\documentclass[12pt]{article}
\usepackage{a4,epsfig,amsmath,amssymb,graphicx,multirow,oldgerm,euscript,mathrsfs}

\newcommand{\SLASH}[1]{/\!\!\! #1}

\begin{document}

\begin{titlepage}
\begin{flushright}
PCCF RI 0601 \\
 ${\rm ECT}^{*}$-05-15\\
 LPNHE/2006-01\\
\end{flushright}
\renewcommand{\thefootnote}{\fnsymbol{footnote}}
\vspace{1.0em}

\begin{center}
{\bf \LARGE{ Testing Fundamental Symmetries with $\boldsymbol{\Lambda_b \to 
\Lambda}$-Vector Decays}}
\end{center}
\vspace{1.0em}
\begin{center}
\begin{large}
O. Leitner$^{1}$\footnote{leitner@ect.it},
Z.J. Ajaltouni$^{2}$\footnote{ziad@clermont.in2p3.fr}, 
E. Conte$^{3}$\footnote{conte@clermont.in2p3.fr}
 \\
\end{large}
\vspace{2.3em}
$^1$ ${\rm{ECT}}^{*}$, Strada delle Tabarelle, 286, 
38050 Villazzano (Trento), Italy \\
$I.N.F.N.$, Gruppo Collegato di Trento, Trento, Italy\\
and\\
Laboratoire de Physique Nucl\'eaire et de Hautes \'Energies\footnote{Unit\'e de Recherche 
des Universit\'es Paris 6 et Paris 7, associ\'ee au CNRS.}, Groupe Th\'eorie\\
Univ. P. $\&$ M. Curie, 4 Pl. Jussieu, F-75252 Paris, France\\
\vspace{0.2cm}
$^{2,3}$ Laboratoire de Physique Corpusculaire de Clermont-Ferrand \\
IN2P3/CNRS Universit\'e Blaise Pascal \\
F-63177 Aubi\`ere Cedex France \\ 
\end{center}
\vspace{5.0em}
\begin{abstract}
\vspace{1.0em}
Putting together kinematical and dynamical analysis, a complete study of the 
decay channels $\Lambda_b \to {\Lambda} V(1^-)$, with $\Lambda \to p {\pi}^-$ and $V (J/\psi) 
\to {\ell}^+ {\ell}^-$ or $V ({\rho}^0) \to {\pi}^+ {\pi}^-,$  is performed. An 
intensive use of the helicity formalism is involved on the kinematical side, while on the 
dynamical side, Heavy Quark Effective Theory (HQET) is applied for an accurate determination of 
the hadronic matrix elements between the baryons $\Lambda_b$ and $\Lambda$. Emphasis is put on 
the major role of the $\Lambda_b$ polarization for constructing T-odd observables and the standard
${\rho}^0-{\omega}$ mixing has the benefit effect of amplifying the process of direct 
$CP$ violation between $\Lambda_b \ \mathrm{and} \ {\bar \Lambda}_b$ decays. 
\vskip 2.8cm
\end{abstract}
%
%
\end{titlepage}
\newpage
%
\section{Introduction}\label{section1}
%
Huge statistics of beauty hadrons are expected to be produced at the 
CERN-LHC proton-proton collider starting in 2007. Obviously this will lead to 
a thorough study of discrete symmetries, $C, P$ and $T$ in $b$-quark physics, 
in the framework of the Standard Model (SM) as well as  beyond the Standard 
Model. It is also well known that the violation of $CP$ symmetry via the 
Cabibbo-Kobayashi-Maskawa (CKM) mechanism is one of the corner-stone of the 
Standard Model of particle physics in the electroweak sector. Thanks to the 
foreseen CERN-LHC collider, non-leptonic and leptonic $b$-baryon decays may 
allow us to get informations about the CKM matrix elements, analysis of the  
$C-P-T$ operators may be performed and different non-perturbative aspects of 
QCD may also be investigated. 

Looking for Time Reversal (TR) violation effects in $b$-baryon decays can 
provide us a new field of research. Firstly, this can be seen as  a 
complementary test of $CP$ violation by assuming the correctness of the $CPT$ 
theorem. Secondly, this can also be a path to follow in order 
to search for processes beyond the Standard Model. On the theoretical side, 
most of the time, the Time Reversal violation effect  comes from an additional 
term related to the physics beyond SM and added by hand in the QCD Lagrangian. 
On the experimental side, various observables which are $T$-odd under time 
reversal operations can be measured, so that  $\Lambda_b$-decay  
seems to be one of the most  promising channel to reveal TR violation signal.

In a previous letter quoted as~\cite{Ajaltouni:2004zu}, a general formulation 
based on the M.~Jacob-G.C.~Wick-J.D.~Jackson (JWJ) helicity formalism has 
been set for studying the decay process $\Lambda_b \to \Lambda V(1^-)$, where 
$\Lambda \to p \pi^-$ with  $V \to l^+ l^-$  or $V \to \pi^+ \pi^-$. Emphasis 
was put on the importance of the initial $\Lambda_b$ polarization as well as 
the correlations among the angular distributions of the final decay products. 
On the dynamical side, the Hadronic Matrix Elements (HME) appearing in the decay 
amplitude were computed, at the tree level approximation, in the framework of 
the factorization hypothesis for  two-body non-leptonic weak decay of heavy quark. 
A confirmation in the validity of the factorization hypothesis was the agreement 
found between the theoretical estimate of the $\Lambda_b \to \Lambda J/\psi$ 
decay width and its experimental measurement  given by the Particle Data 
Group (PDG) average~\cite{Eidelman:2004wy}. 

In our present work, calculations are performed in a more exhaustive and 
detailed manner. On the dynamical side,  both tree and penguin diagrams are 
involved in the evaluation of the HME. In our case, the non-leptonic 
$\Lambda_b$-decay proceeding through the $W$-loop involves either the $b \to u$ or $b \to s$ 
transitions, respectively. The Cabibbo-Kobayashi-Maskawa matrix elements, $V_{ub}, V_{us}, 
V_{cb}, V_{cs}, V_{tb}, V_{td}$ and $V_{ts}$ representing the charged current couplings 
between quark transitions take place in the tree and  QCD, electroweak, QCD-electroweak 
penguin diagrams calculated for the decay  $\Lambda_b \to \Lambda V(1^-)$. 
Other diagrams such as the annihilation and box diagrams have been neglected in 
our approach. A key point in the calculation of non-leptonic baryon decays 
is the calculation of hadronic transition form factors between baryons that will 
be derived by making use of Heavy Quark Effective Theory (HQET).

On the kinematical  side, helicity methods are employed with an emphasis put 
on the polarization of the intermediate resonances $\Lambda$ and $V= J/\psi,  
\rho^0, \omega$ and their particular role in testing the fundamental symmetries 
like Parity and Time Reversal operations as well. It is worth to say that 
the present work  mainly based on  the cascade-type analysis is very useful 
in analysing polarization properties and Time Reversal effects. The main advantage 
of the decay cascade-type holds in a treatment where every decay in the decay chain 
is performed in its respective rest frame.  Finally, full Monte-Carlo simulations 
including all the kinematics and dynamics are  performed according to the 
computational model.

The reminder of the paper is organized as  follows. In Section 2, we present 
the kinematical properties of  $\Lambda_b \to \Lambda V(1^-)$ decays where 
the M. Jacob, G.C. Wick and J.D. Jackson helicity formalism is extensively applied. 
In this section, we also emphasize the physical importance of  polarization,  
mainly that of the intermediate resonances, $\Lambda$ and $V$, respectively. 
In Section 3, we derive the decays of the intermediate resonances previously 
mentioned: all the calculations of $\Lambda \to p \pi^-$, $V \to l^+ l^-$ and 
$V \to \pi^+ \pi^-$ are performed  \`a la Jacob-Wick-Jackson where the role of the 
helicity frame is strongly alighted.

After the kinematical analysis of $\Lambda_b \to \Lambda V(1^-)$ decays, 
we focus on the dynamical analysis of  $\Lambda_b$-decay in Section 4. Transition 
form factors are derived in Heavy Quark Effective Theory which is well 
suited for $\Lambda_b$-decay since the $b$ quark mass is large enough to play the  
adequate energy scale in order to make an expansion in $1/m_b$ of the QCD Lagrangian 
written in the HQET formalism. Corrections to the order of $\mathcal{O}(1/m_b)$ 
are included into the weak transition form factors between the baryons $\Lambda_b$ 
and $\Lambda$ and an usual model for describing baryon wave functions  is also 
applied in our calculations. The weak decay amplitude of the analysed process is
derived by following an Effective Field Theory (EFT) approach. Operator product 
expansion and Wilson coefficients are used in order to include long and short distance 
physics, respectively. Moreover, in the special case of $V=\rho^0$ and $\omega$, the 
charge symmetry violation mixing  between these two vectors is included since it may 
rise to a large signal of $CP$ violation due 
to a large strong phase dynamically produced at the $\omega$ resonance.

In Section 5, we list all the numerical inputs, CKM matrix elements, quark and 
meson masses and decay constants that are needed for calculating all the physical 
observables related to our analysis. Section 6 is devoted to results and 
discussions regarding all the simulations we made for $\Lambda_b \to \Lambda V(1^-)$ 
decays: branching ratio, $\mathcal{BR}$, $CP$ violating asymmetry parameter, $a_{CP}$, helicity 
asymmetry parameter, $\alpha_{AS}$, polarization, $\vec{\mathcal{P}}$, and various 
angular distributions.  In this section, we also stress on the main physical observables 
which 
can be measured with the future detectors at the CERN-LHC proton-proton collider. 
In Section 7, we discuss  in detail the search for Time-odd observables and we conclude 
this section by  proposing an observable which is odd under Time Reversal operation: 
the distributions in cosine and sine of the normal-vectors to the decay planes of the 
intermediate resonances, $\Lambda \ \mathrm{and} \ V$, in the $\Lambda_b$ frame. 
Finally, our main results are summarized in the last section where conclusions and 
global search for discrete symmetries are discussed.


\section{Kinematical properties of $\boldsymbol{ \Lambda_b  \to  \Lambda V(1^-) } $ 
decays} \label{section2}
%
The hyperons produced in  proton-proton collisions as well as in other hadron 
collisions are usually polarized in the {\it transverse direction}. The average 
value of the hyperon spin being non equal to zero and, owing to Parity 
conservation in strong interaction, the spin direction is {\it orthogonal
to the production plane} defined by the incident beam momentum, $\vec{p}_1$, and the 
hyperon momentum, $\vec{p}_h$. Usually, the degree of polarization 
depends on the center of mass energy, $\sqrt s$, and the hyperon transverse 
momentum~\cite{Andersson:1998}. In the case of $p p \rightarrow  {\Lambda_b} + X$
 at high $\sqrt s$, the $b-$quark could replace the $s-$quark of an 
ordinary $\Lambda$ and it is expected that the beauty baryon, $\Lambda_b$, 
will be polarized in the same way than the  $\Lambda$ hyperon. 
 We define $\vec{N}_P \ $ as the normal vector to the production plane: 
\begin{eqnarray}
  \vec{N}_P =  \frac{\vec{p}_1 \times \vec{p}_b} {|\vec{p}_1 \times \vec{p}_b|}\ , 
\end{eqnarray} 
where $\vec{p}_1$ and $\vec{p}_b$ are  the vector-momenta of one incident 
proton beam and $\Lambda_b$, respectively.

\begin{figure}[hpt]
\begin{center}
\includegraphics*[width=0.32\columnwidth]{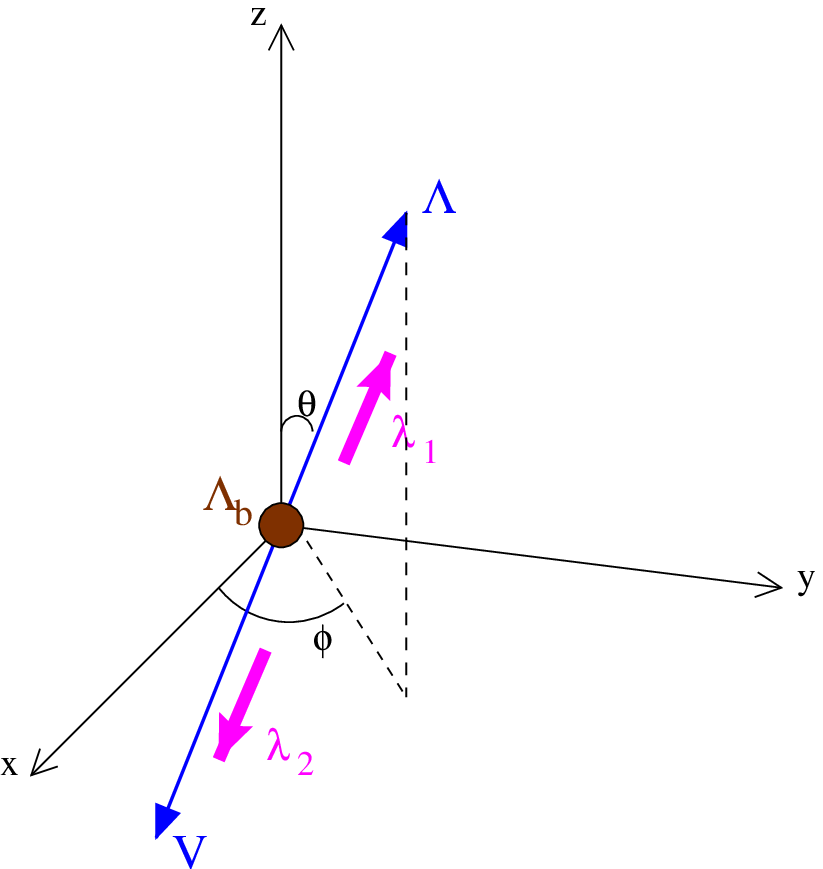}
\includegraphics*[width=0.33\columnwidth]{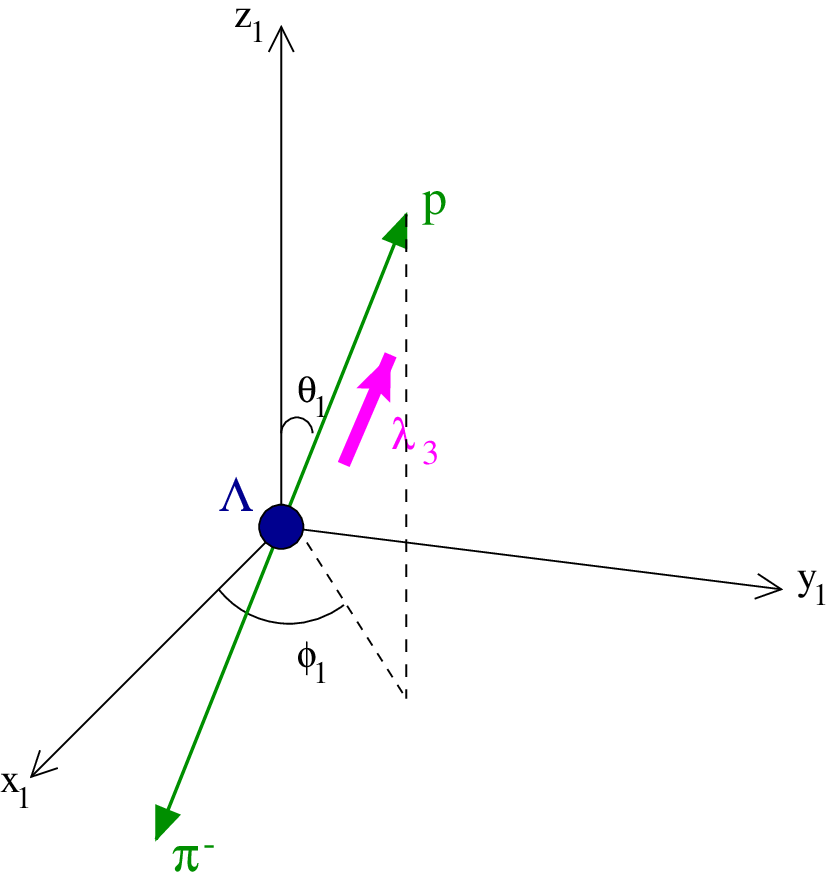}
\includegraphics*[width=0.33\columnwidth]{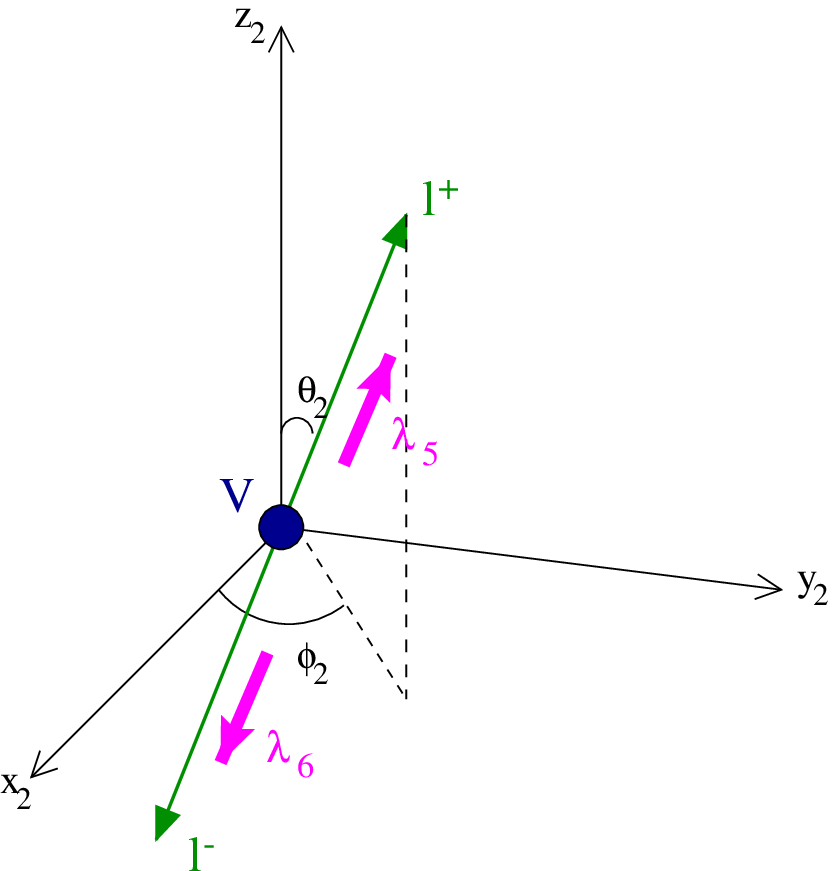}
\end{center}
\caption{Left-handed: $\Lambda_b$ decay in its transversity frame. Right-handed: helicity frames
for the $\Lambda$ and vector meson $V$ decays, respectively.}
\label{frame}
\end{figure}

\vskip 0.2cm  
Let $({\Lambda_b}XYZ)$ be the rest frame (see Fig.~\ref{frame}) of the $\Lambda_b$ 
particle. The quantization axis $({\Lambda_b}Z)$ is chosen to be parallel 
to ${\vec N}_P$. The other orthogonal axis $({\Lambda_b}X)$ and $({\Lambda_b}Y)$ 
are arbitrary in the production plane. In our further analysis, 
the $({\Lambda_b}X)$ axis is taken parallel to the momentum $\vec{p}_1$, which 
constrains the direction of the  $({\Lambda_b}Y)$ axis in the production plane. 
 The spin projection, $M_i$, of the $\Lambda_b$ along the transverse axis 
$({\Lambda_b}Z)$ takes  the values   $\pm {1/2}$. An important physical parameter is 
the $\Lambda_b$ Polarization Density Matrix (PDM), $({\rho}^{\Lambda_b}$). 
It is a $(2 \times 2)$ hermitian matrix with real diagonal elements 
verifying $\sum_{i=1}^{2} {\rho}_{ii}^{\Lambda_b}= 1$. The physical meaning of
the diagonal matrix elements is the following: the  probability of getting $\Lambda_b$  
produced with $M_i = \pm{1/2}$ is 
given by ${\rho}_{11}^{\Lambda_b}$ and ${\rho}_{22}^{\Lambda_b}$, respectively. 
Finally, the initial $\Lambda_b$ polarization, ${\cal P}_{\Lambda_b}$, 
is  given by $\langle \vec{S}_{\Lambda_b} 
\cdot \vec{N}_P \rangle = {\cal P}_{\Lambda_b}
={\rho}_{11}^{\Lambda_b} - {\rho}_{22}^{\Lambda_b}.$

In the framework of the JWJ helicity formalism~\cite{JWJ}, the decay amplitude, 
$A_0{(M_i)}$, for ${\Lambda_b}{(M_i)}  \to  
{\Lambda}{(\lambda_1)} V{(\lambda_2)}$
is obtained by applying the Wigner-Eckart theorem to the $\mathcal{S}$-matrix
element:
\begin{eqnarray}
 A_0{(M_i)} = \langle 1/2,M_i|\mathcal{S}^{(0)}|p,\theta,\phi;\lambda_1,\lambda_2 \rangle 
= \mathcal{A}_{(\lambda_1,\lambda_2)}(\Lambda_b \to \Lambda V) D_{M_i M_f}^{{1/2} \star}
{(\phi,\theta,0)}\ , 
\end{eqnarray}
where $\vec{p} = (p,\theta,\phi)$ is the vector-momentum of the hyperon 
$\Lambda$ in the $\Lambda_b$ frame (Fig.~\ref{frame}). $\lambda_1$ and $\lambda_2$ are the 
respective helicities of $\Lambda$  and $V$ with the possible values $\lambda_1=\pm{1/2}$ 
and  $\lambda_2 =-1,0,+1$. The momentum projection along the 
$({\Delta})$ axis (parallel to $\vec{p}$) is given by $M_f =\lambda_1-\lambda_2=\pm{1/2}$.
The $M_f$ values  constrain those of $\lambda_1$ and $\lambda_2$ since, among six possible 
combinations, only four are physical. If $M_f= +1/2$ then $(\lambda_1,\lambda_2)=(1/2,0)$ 
or $(-1/2,-1)$. If $M_f= -1/2$ then $(\lambda_1,\lambda_2)=(1/2,1)$ or $(-1/2,0)$. On other side, 
the hadronic matrix element, $\mathcal{A}_{(\lambda_1,\lambda_2)}(\Lambda_b \to \Lambda V)$, 
contains  all the  decay dynamics describing, for a set of helicities $\lambda_1$ and $\lambda_2$, 
the hadronic part of the decay $\Lambda_b \to \Lambda V$ and finally the Wigner matrix element, 
\begin{eqnarray}
 D_{M_i M_f}^j{(\phi,\theta,0)} = d_{M_i M_f}^j{(\theta)} {\exp{(-i{M_i}{\phi})}}\ ,
\end{eqnarray} 
is expressed according to the Jackson convention~\cite{JWJ}. According to standard rules, 
the differential cross-section, $d\sigma$, must 
take account of the undetermined helicity initial state and include a summation over 
individual final helicity states, the total angular momentum along the 
helicity $(\Delta)$ axis, $\lambda = M_f = 
\lambda_1 - \lambda_2$, being fixed. We get the expression:

\begin{eqnarray}
d\sigma \propto  \sum_{M_i,M^{\prime}_i} {\sum_{\lambda_1, \lambda_2}}{\rho}_{M_i M^{\prime}_i}^{\Lambda_b}
{|\mathcal {A}_{(\lambda_1,\lambda_2)}(\Lambda_b \to \Lambda V)|}^2 d_{M_i {\lambda}}^{1/2} 
d_{M^{\prime}_i {\lambda}}^{1/2}
{\exp{i(M^{\prime}_i - M_i)\phi}}\ . 
\end{eqnarray}

\noindent As it is known, Parity is not conserved in weak interactions therefore  the hadronic matrix element 
 $\mathcal{A}_{(\lambda_1,\lambda_2)}(\Lambda_b \to \Lambda V)$ is not equal to  
$\mathcal{A}_{(-\lambda_1,-\lambda_2)}(\Lambda_b \to \Lambda V)$. 
In order to handle in an easy manner lengthy calculations, the following mathematical 
expressions are introduced:
\begin{eqnarray}
{\gamma}(\pm 1/2) = \sum_{M_i, M^{\prime}_i}{\rho}_{M_i M^{\prime}_i}^{\Lambda_b} d_{M_i {\pm 1/2}}^{1/2}
d_{M^{\prime}_i {\pm 1/2}}^{1/2} {\exp{i(M^{\prime}_i - M_i)\phi}}\ .
\end{eqnarray}

\noindent Noticing that $\mathcal{P}^{\Lambda_b} = {\rho}_{++}^{\Lambda_b} - 
{\rho}_{--}^{\Lambda_b}$ and
${\rho}_{++}^{\Lambda_b} + {\rho}_{--}^{\Lambda_b} = 1$ (i.e. normalization condition), the 
expression written in Eq.~(5) become simplified:
\begin{eqnarray}
{\gamma}(\pm 1/2) = {\frac{1}{2}}{\left( 1 \pm {\mathcal{P}^{\Lambda_b}}{\cos \theta} \pm 
2 \Re e{({\rho}_{+-}^{\Lambda_b} \exp{i\phi})} {\sin \theta} \right)}\ .
\end{eqnarray} 

\noindent Other expressions are also introduced, like:
\begin{align}
{\bar \omega}(+1/2) = &{|\mathcal {A}_{(1/2,1)}(\Lambda_b \to \Lambda V)|}^2 {\gamma}(-1/2) 
+ {|\mathcal
{A}_{(1/2,0)}(\Lambda_b \to \Lambda V)|}^2 {\gamma}(+1/2)\ , \nonumber \\
{\bar \omega}(-1/2) = & {|\mathcal {A}_{(-1/2,-1)}(\Lambda_b \to \Lambda V)|}^2 
{\gamma}(+1/2) + {|\mathcal
{A}_{(-1/2,0)}(\Lambda_b \to \Lambda V)|}^2 {\gamma}(-1/2)\ , 
\end{align}
which represent respectively the weight of the $\Lambda$ helicity states $+1/2$ and 
$-1/2$ along the $(\Delta)$ axis. With such notations, the differential cross-section can be 
rewritten in the simple form:
\begin{eqnarray}
  \frac{d\sigma}{d\Omega} = N {\Big( {\bar \omega}(+1/2) + {\bar \omega}(-1/2) \Big)}\ ,
\end{eqnarray}  
\noindent
where $N$ is a normalization constant. It is worth to underline  that, apart from the 
four hadronic matrix elements, $\mathcal{A}_{(\lambda_1,\lambda_2)}(\Lambda_b \to \Lambda V)$, 
which have a dynamical origin, the above cross-section needs {\it three real 
parameters} in  order to be fully determined: $\mathcal{P}^{\Lambda_b}, \  
\Re e({\rho}_{+-}^{\Lambda_b}) \  \mathrm{and} \  \Im m({\rho}_{+-}^{\Lambda_b})$.
\noindent
The relation shown in Eq.~(8) can be put in a more compact form by introducing the {\it helicity 
asymmetry parameter}, ${\alpha}_{AS}$, which is related to the final helicity 
value $M_f = \lambda = \lambda_1 -\lambda_2$. By means of the following relations,
\begin{align}
{|{\Lambda_b}(+)|}^2 = &{|\mathcal {A}_{(1/2,0)}(\Lambda_b \to \Lambda V)|}^2  
+{|\mathcal {A}_{(-1/2,-1)}(\Lambda_b \to \Lambda V)|}^2\ , 
\nonumber \\
{|{\Lambda_b}(-)|}^2 = &{|\mathcal {A}_{(-1/2,0)}(\Lambda_b \to \Lambda V)|}^2 
+{|\mathcal {A}_{(1/2,1)}(\Lambda_b \to \Lambda V)|}^2\ , 
\end{align}
and defining the  helicity asymmetry parameter, ${\alpha}_{AS}$, as
\begin{eqnarray}
 \alpha_{AS} = \frac{{|{\Lambda_b}(+)|}^2-{|{\Lambda_b}(-)|}^2}{{|{\Lambda_b}(+)|}^2 
+{|{\Lambda_b}(-)|}^2}\ ,
\end{eqnarray}

\noindent
the final expression of the differential cross-section reads as
\begin{eqnarray}\label{qe01}
\frac{d\sigma}{d\Omega} \propto  1 +  {\alpha_{AS}}{{\cal P}^{\Lambda_b}}{\cos \theta} 
+ 2{\alpha_{AS}}{\Re e{({\rho}_{+-}^{\Lambda_b} \exp{i\phi})}} {\sin \theta}\ .
\end{eqnarray}

\noindent Then, by averaging over the azimuthal angle, $\phi$, a standard relation is 
obtained for the {\it polar} angular distribution:
\begin{eqnarray}\label{qe02}
\frac{d\sigma}{d\cos\theta} \propto  1 + {\alpha_{AS}}{{\cal P}^{\Lambda_b}}{\cos \theta}\ ,
\end{eqnarray}
\noindent
where it can be noticed that polar angular dissymetries are intimately related to the 
initial polarization of the $\Lambda_b$ resonance.

\subsection{Physical importance of the polarization}

Parity violation in $\Lambda_b$ weak decays into $\Lambda V$ necessarily leads  
to a polarization process of the two intermediate resonances $\Lambda$ and $V$. 
In order to determine the {\it vector-polarization} of each 
resonance, a new set of axis is needed according to which properties of the
vector-polarization $\vec{\mathcal{P}}^{(i)}$   with $i=\Lambda, V$ by Parity
and Time-Reversal (TR) transformations will be more obvious. Let $\vec{e}_Z$ be the unit 
vector which is parallel to the preceding vector $\vec{N}_P$, $\vec{N}_P$ being 
transverse to $\Lambda_b$ production plane,  
and $\vec p$ the momentum of a resonance in the $\Lambda_b$ rest frame. 
The following unit vectors are defined (the index $(i)$ being dropped):
\begin{eqnarray}\label{qe03}
 \vec{e}_L =  \frac{\vec{p}}{p}\ , \ \ \    
 \vec{e}_T =  \frac{\vec{e}_Z \times \vec {e}_L}{|\vec{e}_Z \times \vec{e}_L|}\ , \ \ \   
 \vec{e}_N = \vec{e}_L \times \vec{e}_T\ . 
\end{eqnarray}

\noindent
In this new frame, the vector-polarization of any resonance  defined in the
original $\Lambda_b$ frame can be rewritten like:
\begin{eqnarray}
\vec{\mathcal{P}}^{(i)} = {P_L}^{(i)} \vec{e}_L + {P_N}^{(i)} \vec{e}_N + {P_T}^{(i)} 
\vec{e}_T\ , 
\end{eqnarray}
\noindent 
such that each component can be easily computed, $P_j^{(i)} = \vec{\mathcal{P}}^{(i)} \cdot
\vec {e}_j \  \mathrm{with} \ j = L,N,T$. We recall that $P_L, P_N, P_T$ are named  
{\it longitudinal, normal and transverse} polarizations, respectively.
Using the properties of the spin, $\vec s$, and the
vector-polarization, $\vec{\mathcal{P}}$, under Parity and TR operations, 
we can deduce those of $P_j^{(i)}$ displayed in Table~\ref{tab1trans}.

It is worth noticing that the normal polarization (of any resonance) $P_N$ is odd
by Time-Reversal operation. It allows us to say that the knowledge of the 
vector-polarization, $\vec {\cal P}$,  and particularly its
normal component, $P_N$, are essential observables for {\it performing any test of 
validity of Time-Reversal symmetry}.
 
\subsection{Polarization of the intermediate resonances}

Intermediate resonance states, $\Lambda \ {\mathrm{and}} \ V$, can be described by a 
density-matrix named ${\rho}^f$  which analytic expression is given 
by  standard quantum-mechanical relations: 
\begin{eqnarray}
  {\rho}^f = \mathcal{T}^{\dagger}  {\rho}^{\Lambda_b}\mathcal{T}\ , \  \mathrm{with}  
\   Tr{({\rho}^f)}  = 
  \frac{d\sigma}{d\Omega} =  N  W(\theta,\phi)\ ,
\end{eqnarray}
where $\mathcal{T}$ is the transition-matrix related to the $\mathcal{S}$-matrix by 
$\mathcal{S} = 1 + 
i\mathcal{T}$. Its elements being the hadronic elements, $\mathcal {A}_{(\lambda_1,\lambda_2)}(\Lambda_b 
\to \Lambda V)$, mentioned beforehand. With the help of the density-matrix  ${\rho}^f$, the 
vector-polarization of the global system made out of
$\Lambda  \ {\mathrm{and}} \  V$ can be estimated according to the relation:  
\begin{eqnarray}
\vec{\mathcal{P}}  =   \langle \vec{S} \rangle  =   \frac{Tr{\big( {\rho}^f {\vec S} 
\big)}}{Tr({\rho}^f)}\ , 
\end{eqnarray}
where $Tr$ is the trace operator and $\vec S$ the spin of the global system defined by 
$\vec S = |\vec {S_1} + \vec {S_2}|$, $S_1 \ 
\mathrm{and} \ S_2$ being the spin of the baryon $\Lambda$ and the vector meson $V$, respectively.
\vskip 0.2cm
The striking point is that the final system is composed of two {\it correlated subsystems} 
because only four spin-states instead of six are required to describe the quantum system 
$(\Lambda \oplus V)$. So, in order to determine the kinematics features of each resonance,
 namely $\Lambda$ and $V$, and  
particularly there vector-polarization, $\vec{\mathcal{P}}^{\Lambda}$ and $\vec{\mathcal{P}}^V$, 
some modifications of the previous relations are requested. Our task 
will consist in estimating individual density-matrix and vector-polarization  
for each resonance by strictly applying  quantum-mechanical rules~\cite{Ballentine:1998, Werle:1966}. 

\subsection{Determination of $\boldsymbol{\vec{\mathcal P}^{\Lambda}}$}

$\Lambda$ helicity states and its vector-polarization can be extracted from the above 
relations by summing over the kinematics variables of the vector-meson states, 
especially its helicity states. Departing from 
formulas~(15) and~(16),  the following relation can be inferred:
\begin{eqnarray}\label{eqz1}
\vec{\mathcal{P}}^{\Lambda}   W(\theta,\phi) =   N {\Sigma}_{\lambda} \langle \theta,\phi;\lambda|
{\rho}^f {\vec \sigma}| \theta,\phi;\lambda \rangle\ .
\end{eqnarray}
This clearly shows that $\Lambda$ polarization depends a priori on its emission angles. 
In Eq.~(\ref{eqz1}), we have also defined:\\
$\bullet \  \lambda = {\lambda}_1 - {\lambda}_2$, the vector-meson helicity 
$\lambda_2$ which varies with the kinematical constraint $\lambda = M_f = \pm {1/2}$. \\
$\bullet  \  \vec\sigma$ are the Pauli spin $1/2$ matrices and they are defined 
in the $({\Lambda}xyz)$ rest-frame where the axis $\vec x, \vec y \ \mathrm{and} 
\ \vec z$ are respectively identified with 
the normal, transverse and longitudinal axis previously defined.

\noindent The matrix elements given in the right-handed side of Eq.~(\ref{eqz1}) 
can be explicitly calculated~\cite{Jackson:1965} and 
the three components of $\vec{\mathcal{P}}^{\Lambda}$ get the following expressions:
\begin{align}
P_L^{\Lambda} \ {W(\theta,\phi)} \  & \propto  \  {\bar \omega}(+1/2) -  
{\bar \omega}(-1/2)\ , \nonumber \\  
P_N^{\Lambda} \ {W(\theta,\phi)} \  & \propto  \  2 \Re e{(\langle\theta,\phi,1/2|{\rho}^f| 
\theta,\phi,-1/2 \rangle)}\ , \nonumber  \\
P_T^{\Lambda} \ {W(\theta,\phi)} \  & \propto  \ -2 \Im m{(\langle \theta,\phi,1/2|{\rho}^f| 
\theta,\phi,-1/2 \rangle)}\ . 
\end{align}
\noindent
Explicit calculations lead to:
\begin{multline}
P_L^{\Lambda} \ W(\theta,\phi) \  \propto  \  \gamma(+1/2) \Bigl( 
|\mathcal{A}_{(1/2,0)}(\Lambda_b 
\to \Lambda V)|^2 -
|\mathcal{A}_{(-1/2,-1)}(\Lambda_b \to \Lambda V)|^2 \Bigr) \\
 + \gamma(-1/2) \Bigl( |\mathcal{A}_{(1/2,1)}(\Lambda_b \to \Lambda V)|^2 -
|\mathcal{A}_{(-1/2,0)}(\Lambda_b \to \Lambda V)|^2 \Bigr)\ , 
\end{multline}
\begin{multline}
 P_N^{\Lambda} \ W(\theta,\phi) \  \propto  \ 
\Re e \Biggl(\mathcal{A}_{(1/2,0)}(\Lambda_b \to \Lambda V)\mathcal{A}_{(-1/2,0)}^{*}
(\Lambda_b \to \Lambda V)\\
\Bigl(-{\cal P}^{\Lambda_b}\sin\theta
 + 2 \Re e(\exp{(i\phi)}{\rho}^{\Lambda_b}_{+-})\cos\theta 
 + 2i\Im m(\exp{(i\phi)}{\rho}^{\Lambda_b}_{+-})\Bigr) \Biggr)\ ,
\end{multline}
and 
\begin{multline}
P_T^{\Lambda} \ W(\theta,\phi) \ \propto  \  
-\Im m \Biggl({\mathcal{A}_{(1/2,0)}(\Lambda_b \to \Lambda V) 
\mathcal{A}_{(-1/2,0}^{*}(\Lambda_b \to \Lambda V)} \\
\Bigl(-{\cal P}^{\Lambda_b}\sin\theta
 + 2 \Re e(\exp{(i\phi)}{\rho}^{\Lambda_b}_{+-})\cos\theta  
 + 2i\Im m(\exp{(i\phi)}{\rho}^{\Lambda_b}_{+-})\Bigr) \Biggr)\ .
\end{multline}

\noindent
In these detailed formulas given in Eqs.~(19-21), we underline the importance of two {\it input 
parameters}:\\
 (i) the initial $\Lambda_b$ polarization, $\mathcal{P}^{\Lambda_b}$, and (ii) 
the non-diagonal matrix element, ${\rho}_{+-}^{\Lambda_b}$, which appears 
essentially in the interference terms.
 
\subsection{Longitudinal polarization of $\boldsymbol{V}$}

Vector meson has spin $S_{(2)}= 1$ and therefore three helicity states. The components of its 
vector-spin operator, $\vec{S}_j, j = x,y,z $, are $(3 \times 3)$ matrices and they are defined in a new set of 
longitudinal, normal and transverse axis. The longitudinal one being the spin quantization axis and 
verifying the standard relation: $S_z{|{\lambda_2}\rangle} = {\lambda_2}{|{\lambda_2}\rangle} $.
\vskip 0.3cm

Computation of the normal and transverse components, respectively $P_N^V \ \mathrm{and} \ P_T^V$,
as well as the analytical method of extracting these values from experimental data will not be exposed here. 
In this paper, in a first aim, only the longitudinal component, $P_L^V$, will be estimated.
According to the basic relation:
\begin{eqnarray}  
\vec{\mathcal{P}}^V  \ W(\theta,\phi)  =   N \
{\Sigma}_{\lambda_2} \Bigl({\Sigma}_{\lambda_1}\langle\theta,\phi,{\lambda_1},{\lambda_2}|{\rho}^f {\vec
S}|\theta,\phi,{\lambda_1},{\lambda_2}\rangle \Bigr)\ ,
\end{eqnarray}
\noindent
$P_L^V$ can be deduced in a relatively easy way and one gets,
\begin{eqnarray}
 P_L^V  \  W(\theta,\phi)   =  {\gamma}(-1/2)|\mathcal{A}_{(1/2,1)}(\Lambda_b \to \Lambda V)|^2 -
 {\gamma}(1/2)|\mathcal{A}_{(-1/2,-1)}(\Lambda_b \to \Lambda)|^2\ ,
\end{eqnarray} 
\noindent
where the kinematics functions ${\gamma}(\pm {1/2}) \ $ have been  defined in Section 2. 
It may be noticed that the mathematical form of $P_L^V$ is very similar to that of $P_L^{\Lambda}$ 
(same longitudinal axis) so that the helicity value $\lambda_2 = 0 \ $ does not play any role in the 
longitudinal polarization of the vector-meson.

\section{Decays of the intermediate resonances}

Performing appropriate rotations and Lorentz boosts, the decay products 
of each resonance can be studied in its own helicity frame (see Fig.~1). The 
quantization axis of each frame is chosen parallel to the corresponding resonance 
momentum in the $\Lambda_b$ frame i.e. $\overrightarrow{O_1z_1} || \vec{p}_{\Lambda}$ and 
$\overrightarrow{O_2z_2} || \vec{p}_{V}=-\vec {p}_{\Lambda}$.  Again, conservation 
of the total angular momentum (in each frame) constrains the helicities of the 
final particles. For the decays 
${\Lambda}{(\lambda_1)} \to  P{(\lambda_3)} {\pi}^-{(\lambda_4)}$  and 
$V{(\lambda_2)} \to  {\ell}^-{(\lambda_5)} {\ell}^+{(\lambda_6)}$  
or $V{(\lambda_2)} \to h^-{(\lambda_5)} h^+{(\lambda_6)},$ the respective 
helicities are $(\lambda_3,\lambda_4)=(\pm{1/2},0)$ 
and $(\lambda_5,\lambda_6)=(\pm{1/2},\pm{1/2})$ in case of 
leptons or $(\lambda_5,\lambda_6)=(0,0)$ in case of  $0^-$ mesons.

In the $\Lambda$ helicity frame, the projection of the total angular momentum, 
$m_i$, along the proton momentum, $\vec{p}_P$, is given by $m_1=\lambda_3-\lambda_4 
=\pm{1/2}$. In the vector meson helicity frame, this projection is equal to $m_2=\lambda_5-\lambda_6
= -1,0,+1$ if leptons and $m_2=0$ if pions. The decay amplitude, $A_i(\lambda_i)$, of 
each resonance can be written similarly as in Eq.~(2), requiring only that the 
kinematics of its decay products are fixed. Thus, we obtain,
\begin{align}
A_1(\lambda_1) = & \langle \lambda_1, m_1|S^{(1)}|p_1,{\theta}_1,{\phi}_1;\lambda_3,\lambda_4 
\rangle =  
 \mathcal{A}_{(\lambda_3,\lambda_4)}(\Lambda \to p \pi^-) D^{{1/2} \star}_{\lambda_1 m_1}
{(\phi_1,\theta_1,0)}\ , \nonumber \\
A_2(\lambda_2) = & \langle  \lambda_2, m_2|S^{(2)}|p_2,{\theta}_2,{\phi}_2;\lambda_5,\lambda_6 \rangle  =  
 \mathcal{A}_{(\lambda_5,\lambda_6)}(V \to l^+ l^-, h^+ h^-) D^{1 \star}_{\lambda_2 m_2}{(\phi_2,\theta_2,0)}\ , 
\end{align}
where  $\theta_1$ and  $\phi_1$ are respectively the polar and azimuthal angles 
of the proton momentum in the $\Lambda$ rest frame while $\theta_2$ and $\phi_2$ 
are those of ${\ell}^- {(h^-)}$ in the $V$ rest frame. Finally, $\mathcal{A}_{(\lambda_3,\lambda_4)}
(\Lambda \to p \pi^-)$ and $\mathcal{A}_{(\lambda_5,\lambda_6)}(V \to l^+ l^-, h^+ h^-)$ are the dynamical
amplitudes of the strong interaction part of the decay processes $\Lambda \to p \pi^-$ and 
$V \to l^+ l^-, h^+ h^-$, respectively. 
%
\subsection{Analytical form of the decay probability}
%
The general decay amplitude, ${\mathcal A}_I^{g}(M_i,\lambda_1,\lambda_2)$, for the process 
of sequence decays $\Lambda_b{(M_i)} \to \Lambda{(\lambda_1)}  V{(\lambda_2)} \to  p {\pi}^- {\ell}^+ {\ell}^-  
(h^+ h^-) \ $ can be factorized according to the three decay amplitudes
 $A_0(M_i), A_1(\lambda_1)$ and $A_2(\lambda_2)$. 
It must include all the possible intermediate states, so that a sum 
over the helicity states ${(\lambda_1, \lambda_2)}$ is performed:
\begin{eqnarray}
 {\mathcal A}_I^g(M_i,\lambda_1,\lambda_2) = \sum_{\lambda_1,\lambda_2}{A_0{(M_i)} 
A_1{(\lambda_1)} A_2{(\lambda_2)}}\ .
\end{eqnarray} 
The decay probability, $d\sigma$, depending on the amplitude, ${\mathcal A}_I^g(M_i,\lambda_1,\lambda_2)$, takes 
the form,
\begin{eqnarray}
d\sigma  \propto   \sum_{M_i,M^{\prime}_i}{\rho}_{M_i M^{\prime}_i}^{\Lambda_b} 
{\mathcal A}_I^g(M_i,\lambda_1,\lambda_2) {\mathcal A}_I^{g*}(M_i,\lambda_1,\lambda_2)\ ,
\end{eqnarray}
where  the polarization density matrix, ${\rho}_{M_i M^{\prime}_i}^{\Lambda_b}$, 
is used to take into account  the unknown initial $\Lambda_b$ spin states, 
$M_i$. Since the helicities of the final particles  are not measured, a summation over 
the helicity values  $\lambda_3, \lambda_4, \lambda_5$ and  $\lambda_6$  is performed 
as well. In a final step, the decay probability, $d\sigma$, can be written in a such way 
that only the intermediate resonance helicities appear:
\begin{multline}\label{eq11}
d\sigma  \propto  
\sum_{\lambda_1,\lambda_2,{\lambda}^{\prime}_1,{\lambda}^{\prime}_2} D_{\lambda_1-\lambda_2,
{\lambda}^{\prime}_1 -{\lambda}^{\prime}_2}{(\theta,\phi)} 
\mathcal{A}_{(\lambda_1,\lambda_2)}(\Lambda_b \to \Lambda V)
\mathcal{A}_{({\lambda}^{\prime}_1,{\lambda}^{\prime}_2)}^*(\Lambda_b \to \Lambda V)\\
F^{\Lambda}_{\lambda_1{\lambda}^{\prime}_1}{(\theta_1,\phi_1)} G^V_{\lambda_2 {\lambda}^{\prime}_2}
{(\theta_2,\phi_2)}\ ,
\end{multline}
where $F^{\Lambda}_{\lambda_1 {\lambda}^{\prime}_1}{(\theta_1,\phi_1)}$ and $G^V_{\lambda_2 
{\lambda}^{\prime}_2}{(\theta_2,\phi_2)}$  are the hadronic tensor elements describing
the decay dynamics of the intermediate 
resonances $\Lambda \to p {\pi}^-, V \to {\ell}^+ {\ell}^- (h^+ h^-)$ and $D_{\lambda,
{\lambda}^{\prime}}$ is a kinematical factor. Their expressions are:
\begin{multline}
F^{\Lambda}_{\lambda_1{\lambda}^{\prime}_1}{(\theta_1,\phi_1)} =
\Big({|\mathcal{A}_{(1/2,0)}(\Lambda \to p \pi^-)|}^2 d_{{\lambda_1} 1/2}^{1/2}{(\theta_1)}
d_{{\lambda}^{\prime}_1 1/2}^{1/2}{(\theta_1)} + 
\\
{|\mathcal{A}_{(-1/2,0)}(\Lambda \to p \pi^-)|}^2 d_{{\lambda_1}
-1/2}^{1/2}{(\theta_1)} d_{{\lambda}^{\prime}_1 -1/2}^{1/2}{(\theta_1)}\Big) 
\exp{i({\lambda}^{\prime}_1-{\lambda_1})}{\phi_1}\ ,
\end{multline}
and,
\begin{align}
G^V_{\lambda_2 {\lambda}^{\prime}_2}{(\theta_2,\phi_2)} = & \sum_{\lambda_5,
\lambda_6} {|{\mathcal{A}_{(\lambda_5,\lambda_6)}(V \to l^+l^-, h^+ h^-)}|}^2 d_{{\lambda_2}
m_2}^1{(\theta_2)} d_{{\lambda}^{\prime}_2 m_2}^1{(\theta_2)}\exp{i({\lambda}^{\prime}_2-{\lambda_2})}{\phi_2}\ , 
\nonumber \\
D_{\lambda_1-\lambda_2,{\lambda}^{\prime}_1 -{\lambda}^{\prime}_2}{(\theta,\phi)} = &
 \sum_{M_i M^{\prime}_i}{\rho}_{M_i M^{\prime}_i}^{\Lambda_b} d_{M_i,
 {\lambda_1}-{\lambda_2}}^{1/2}{(\theta)} d_{M^{\prime}_i,{\lambda}^{\prime}_1
 -{\lambda}^{\prime}_2}^{1/2}{(\theta)}{\exp{i(M'_i - M_i){\phi}}}\ ,
\end{align} 
\noindent 
with $m_2 = \lambda_5 - \lambda_6$. As far as the vector-meson $V$ is concerned, in 
case of lepton pair in the final state and because of parity
conservation,  two hadronic matrix elements, $\mathcal{A}_{(\pm \frac12,\pm \frac12)}(V \to l^+l^-, h^+h^-)$, 
are necessary whereas only one, $\mathcal{A}_{(0,0)}(V \to l^+l^-, h^+h^-)$, is required 
in case of pseudo-scalar mesons.

From the above relations, angular distributions of the decay products of both the two 
resonances can be deduced. Although these angular distributions are well 
established in the literature, the method outlined 
before allows us to compute the Polarization Density Matrix (PDM) elements of each 
resonance. We underline that these angular distributions are  
important ingredients for the determination of its vector-polarization, ${\vec{\mathcal P}}^{(i)} \ 
( i= \Lambda, V)$.

\subsection{$\boldsymbol{\Lambda \to  p {\pi}^-}$ decay}

Departing from Eq.~(\ref{eq11}), integrating over the angles $\theta, \phi, \theta_2 \ \mathrm{and} \  \phi_2,$
and summing over the vector helicity states, the general formula for proton angular 
distributions, $W_1{(\theta_1, \phi_1)}$, in the $\Lambda$ rest-frame can be obtained:
 \begin{multline}
W_1{(\theta_1,\phi_1)}    \propto  \\
\frac12 \Biggl\{ ({\rho}_{++}^{\Lambda}+{\rho}_{--}^{\Lambda})+
({\rho}_{++}^{\Lambda}-{\rho}_{--}^{\Lambda}) 
\alpha_{AS}^{\Lambda} \cos\theta_1   
 -  \frac{\pi}{2}{\cal P}^{\Lambda_b}\alpha_{AS}^{\Lambda}
  \Re e \Big[{\rho}_{ij}^{\Lambda} \exp{(i\phi_1)} \Big] \sin{\theta_1}\Biggr\} \ ,
\end{multline} 
where the PDM elements, ${\rho}_{ij}^{\Lambda}$, of the $\Lambda$ hyperon are not  
normalized yet and expressed as:
\begin{align}
{\rho}^{\Lambda}_{ii} &=   {\int_{\theta_2,\phi_2}}G^V_{00}{(\theta_2, \phi_2)} 
\ {|\mathcal{A}_{(\pm1/2,0)}(\Lambda_b \to \Lambda V)|}^2 + \\
& \hspace{7.0cm}
{\int_{\theta_2,\phi_2}}G^V_{\pm 1 \pm1}{(\theta_2, \phi_2)} {|\mathcal{A}_{(\pm 1/2,\pm 1)}
(\Lambda_b \to \Lambda V)|}^2\ , \nonumber \\
{\rho}^{\Lambda}_{ij} &= {\int_{\theta_2,\phi_2}}
{G^V_{00}}{(\theta_2, \phi_2)}  
\mathcal{A}_{(-1/2,0)}(\Lambda_b \to \Lambda V) \mathcal{A}_{(1/2,0)}^*(\Lambda_b \to \Lambda V)\ .
\end{align}

Then, integrating over the azimuthal angle, $\phi_1$, the standard expression for the polar 
angular distribution can be recovered:
\begin{eqnarray}
\frac{d\sigma}{d\cos{\theta_1}}  \propto    1 + {\alpha}^{\Lambda}_{AS} {\cal P}^{\Lambda} 
{\cos{\theta_1}}\ ,
\end{eqnarray}
\noindent
where ${\alpha}^{\Lambda}_{AS}$ is the usual $\Lambda$ helicity asymmetry parameter and 
${\cal P}^{\Lambda}$ is the
$\Lambda$ polarization according to the helicity axis. Integrating the formula given in Eq.~(\ref{eq11}) over 
the polar angle $\theta_1$, it allows us to deduce the proton {\it azimuthal
angular} distribution: 

\begin{eqnarray}
\frac{d\sigma}{d\phi_1}  \  \propto  \  1 + \frac{{\pi}^2}{8}{\cal P}^{\Lambda_b}\alpha_{AS}^{\Lambda}
 \Re e \Big[{\rho}_{12}^{\Lambda} \exp{(i\phi_1)} \Big]\ . 
 \end{eqnarray}
It will be shown later that the PDM element, ${\rho}_{12}^{\Lambda}$, is real in the 
framework of the {\it factorization hypothesis} used to compute the hadronic matrix element $
{\cal A}_{\lambda_1,\lambda_2}(\Lambda_b \to \Lambda V)$.

\subsection{$\boldsymbol{V \to {\ell}^+ {\ell}^-, h^+ h^-}$ decays}

Vector meson, V, decaying into a lepton pair or a hadronic one is described by the 
$(3 \times 3)$ hermitian matrix $G^V_{\lambda_i {\lambda}^{\prime}_i}{(\theta_2,\phi_2)}$. 
The angular distributions, $W_2{(\theta_2,\phi_2)}$, in the $V$ rest-frame, are obtained by 
integrating Eq.~(\ref{eq11}) over the angles $\theta,\phi,\theta_1$ and $\phi_1$.  Summing over the two 
$\Lambda$ helicity states, one gets,
\begin{multline}\label{zz002}
W_2{(\theta_2, \phi_2)}    \propto     
 ({\rho}_{ii}^{V}+{\rho}_{jj}^{V})(G^V_{00}(\theta_2,\phi_2)+G^V_{\pm 1 \pm1}
(\theta_2,\phi_2))
\\
-  \frac{\pi}{4}{\cal P}^{\Lambda_b}  \Re e \Big[{\rho}_{ij}^V \exp{(i\phi_2)} \Big] 
\sin{2\theta_2}\ ,
\end{multline}
where  the PDM elements, ${\rho}_{ij}^{V}$, of the meson $V$ are (to a normalization factor):
\begin{align}
{\rho}^{V}_{ii} &=  {\int_{\theta_1,\phi_1}}  F^{\Lambda}_{\lambda_1 
{\lambda}^{\prime}_1}{(\theta_1,\phi_1)}
\Biggl[  \delta_{\lambda_2 \lambda_2^{\prime}}  {|\mathcal{A}_{(\pm1/2,0)}(\Lambda_b 
\to \Lambda V)|}^2 + 
\delta_{\lambda_2 \pm\lambda_2^{\prime}}  {|\mathcal{A}_{(\pm 1/2,\pm 1)}(\Lambda_b 
\to \Lambda V)|}^2 
\Biggr]\ , \nonumber \\
{\rho}^{V}_{ij} &= 
{\int_{\theta_1,\phi_1}}  F^{\Lambda}_{\lambda_1 {\lambda}^{\prime}_1}{(\theta_1,\phi_1)}
 \Biggl[\Bigl\{\mathcal{A}_{(1/2,0)}(\Lambda_b 
\to \Lambda V) \mathcal{A}_{(1/2,1)}^*(\Lambda_b 
\to \Lambda V) + h.c.\Bigr\}
\nonumber
\\ & \hspace{3.cm} - \Bigl\{\mathcal{A}_{(-1/2,0)}(\Lambda_b \to \Lambda V) 
\mathcal{A}_{(-1/2,-1)}^*
(\Lambda_b \to \Lambda V) 
+ h.c.\Bigr\}\Biggr]\mathcal{M}_{V \to hh(ll)} \ ,
\end{align}
where, $\mathcal{M}_{V\to hh(ll)}$, takes the following form according to 
the given decay:
\begin{align}
\mathcal{M}_{V \to hh(ll)}=&|\mathcal{A}_{(0,0)}(V \to h^+h^-)|^2\ , \;\; {\rm for} \;\;V \to h^+ h^-\ , 
\nonumber \\
\mathcal{M}_{V \to hh(ll)}=& |\mathcal{A}_{(1/2,-1/2)}(V \to l^+l^-)|^2 - 2 
|\mathcal{A}_{(+1/2,+1/2)}(V \to l^+l^-)|^2\ , 
\;\; {\rm for} \;\; V \to l^+ l^-\ .
\nonumber
\end{align}

\noindent Taking advantage of {\it parity conservation} in strong or electromagnetic 
$V$-decays, a relatively simple polar angular distribution can be obtained 
after averaging over the azimuthal angle $\phi_2$. Thus, in case of
{\it leptonic decays}:
\begin{eqnarray}
 \frac{d\sigma}{d\cos{\theta_2}} \  \propto  \  \big({1 - 3 {\rho}_{00}^V }\big) 
{\cos}^2{\theta_2} +
 (1+{\rho}_{00}^V)\ ,
\end{eqnarray}
\noindent 
 while, for {\it hadronic decays}, one gets:
\begin{eqnarray}
 \frac{d\sigma}{d\cos{\theta_2}} \  \propto  \  \big({3 {\rho}_{00}^V -1}\big) 
{\cos}^2{\theta_2} +
 {(1-{\rho}_{00}^V)}\ ,
\end{eqnarray}
\noindent 
where ${\rho}_{00}^V$ is the PDM element indicating the probability for the 
vector-meson $V$ to be {\it longitudinally polarized}. In both the two 
angular distributions $W_1 \ \mathrm{and} 
\ W_2$ displayed in Eqs.~(30) and~(35), it is worth noticing 
the major role played by the $\Lambda_b$ polarization, ${\cal P}^{\Lambda_b}$, 
especially in the azimuth angular  $\phi_1$ and $\phi_2$ distributions,  which 
arise from {\it interference terms}.

\section{Hadronization}\label{section4}

In this section, the Heavy Quark Effective Theory~\cite{Korner:1994nh, Hussain:1994zr,
Neubert:1993mb} (HQET) formalism 
will be used to evaluate the hadronic form factors involved in $\Lambda_b$-decay. 
Weak transitions including heavy quarks can be safely described when the mass
of a heavy quark is large enough compared to the QCD scale, $\Lambda_{QCD}$. 
Properties such as flavour and spin symmetries can be exploited in such way 
that corrections of the order of $1/m_Q$ are systematically  calculated within an 
effective field theory. Then, the hadronic amplitude of the weak decay is 
investigated by means of the  effective Hamiltonian, $\Delta B=1$, where the 
Operator Product Expansion formalism separates the soft and hard regimes. 

\subsection{Transition form factors}

\noindent The decay, $\Lambda_b \to \Lambda V$, is usually described by the 
following general amplitude, $\mathcal{A}(\Lambda_b \to \Lambda V)$, 
\begin{eqnarray}
\mathcal{A}(\Lambda_b \to \Lambda V)= i \bar{U}_{\Lambda}(p_{\Lambda},s_{\Lambda}) 
\epsilon^{\star}_{\mu} 
\Bigl[A_1 \gamma^{\mu} \gamma_5 + A_2 \; p_{\Lambda}^{\mu}\gamma_5 + V_1 
\gamma^{\mu} 
+V_2 \; p_{\Lambda}^{\mu} \Bigr] U_{\Lambda_b}(p_{\Lambda_b},s_{\Lambda_b})\ , 
\end{eqnarray}
where, $V_i$ and $A_i$, are respectively the vector and axial components of 
the transition form factors characterizing the decay $\Lambda_b \to 
\Lambda V$. 
The  momentum and spin of the baryons $\Lambda$ and $\Lambda_b$ are given by 
$p_{\Lambda_{(b)}}$ and $s_{\Lambda_{(b)}}$,  respectively.  
$\epsilon^{\star\mu}$ is the vector polarization and $U_{\Lambda_b}$ and 
$\bar{U}_{\Lambda}$ are the Dirac spinors. Based on Lorentz decomposition, the 
hadronic matrix element,  
$\langle\Lambda(p_{\Lambda},s_{\Lambda})|\bar{s}\gamma_{\mu}(1-\gamma_5)b|
\Lambda_b(p_{\Lambda_b},
s_{\Lambda_b}) \rangle$, may also be written  as,
\begin{multline}
\langle\Lambda(p_{\Lambda},s_{\Lambda})|\bar{s}\gamma_{\mu}(1-\gamma_5)b|
\Lambda_b(p_{\Lambda_b},
s_{\Lambda_b}) \rangle=\\
\bar{U}_{\Lambda}(p_{\Lambda},s_{\Lambda}) \Bigl[\bigl(f_1(q^2)\gamma_{\mu}
+if_2(q^2)\sigma_{\mu\nu}q^{\nu}
+f_3(q^2)q_{\mu}\bigr) \\
-\bigl(g_1(q^2)\gamma_{\mu}+ig_2(q^2)\sigma_{\mu\nu}q^{\nu}+
g_3(q^2)q_{\mu}\bigr)\gamma_5\Bigr] U_{\Lambda_b}(p_{\Lambda_b},
s_{\Lambda_b})\ ,
\label{eq1}
\end{multline}
where, $q^2=(p_{\Lambda_b}-p_{\Lambda})^2$, defines the momentum transfer in 
the hadronic transition. The form factors, $f_i(q^2)$ and $g_i(q^2)$, refer 
to the vector and axial parts of the transition, respectively.

Another way of parameterizing the electroweak amplitude in decays of baryons is the 
following:
\begin{multline}
\langle\Lambda(p_{\Lambda},s_{\Lambda})|\bar{s}\gamma^{\mu}(1-\gamma_5)b|
\Lambda_b(p_{\Lambda_b},
s_{\Lambda_b}) \rangle= \\
\bar{U}_{\Lambda}(p_{\Lambda},s_{\Lambda}) \Bigl[\Bigl(F_1(q^2)\gamma^{\mu}
+F_2(q^2)v^{\mu}_{\Lambda_b}
+F_3(q^2) \frac{p^{\mu}_{\Lambda}}{m_{\Lambda}}\Bigr) \\
-\Bigl(G_1(q^2)\gamma^{\mu}+ G_2(q^2) v^{\mu}_{\Lambda_b}+
G_3(q^2) \frac{p^{\mu}_{\Lambda}}{m_{\Lambda}}\Bigr)\gamma_5\Bigr] 
U_{\Lambda_b}(p_{\Lambda_b},s_{\Lambda_b})\ .
\label{eq2}
\end{multline}
By comparing the two sets of form factors given in Eqs.~(\ref{eq1}) 
and~(\ref{eq2}), one gets the 
relations between the $f_i(q^2)$'s ($g_i(q^2)$'s) and $F_i(q^2)$'s 
($G_i(q^2)$'s) as  follows,
\begin{align} 
f_1(q^2) & =  F_1(q^2) + (m_{\Lambda_b} + m_{\Lambda})
    \left [ \frac{F_2(q^2)}{2m_{\Lambda_b}} + \frac{F_3(q^2)}{2m_{\Lambda}} 
\right ]\ ,
\nonumber \\
f_2(q^2) & =   \frac{F_2(q^2)}{2m_{\Lambda_b}} + \frac{F_3(q^2)}
{2m_{\Lambda}}\ ,
\nonumber \\
f_3(q^2) &= \frac{F_2(q^2)}{2m_{\Lambda_b}} - \frac{F_3(q^2)}{2m_{\Lambda}}\ ,
\end{align}
and,
\begin{align} 
g_1(q^2) &= G_1(q^2) - (m_{\Lambda_b} - m_{\Lambda}) \left [
\frac{G_2(q^2)}{2m_{\Lambda_b}} + \frac{G_3(q^2)}{2m_{\Lambda}} \right ]\ ,
\nonumber \\
g_2(q^2) & =   \frac{G_2(q^2)}{2m_{\Lambda_b}} + \frac{G_3(q^2)}
{2m_{\Lambda}}\ ,
\nonumber \\
g_3(q^2) &= \frac{G_2(q^2)}{2m_{\Lambda_b}} - \frac{G_3(q^2)}{2m_{\Lambda}}\ .
\end{align}

\noindent 
In case of working in the HQET formalism, 
the matrix element of the weak transition, $\Lambda_b \to \Lambda$, takes the 
following form,
\begin{eqnarray}
\langle \Lambda(p_{\Lambda},s_{\Lambda})|\bar{s}\gamma_{\mu}(1-\gamma_5)b|
\Lambda_b(p_{\Lambda_b},
s_{\Lambda_b})\rangle = 
\bar{U}_{\Lambda}(p_{\Lambda},s_{\Lambda})
\Bigl[\theta_1(q^2) + \SLASH v_{\Lambda_b} \theta_2(q^2) \Bigr]U_{\Lambda_b}
(p_{\Lambda_b},s_{\Lambda_b})\ .
\label{eqhqet}
\end{eqnarray}
In Eq.~(\ref{eqhqet}), $v_{\Lambda_b}$, defines the velocity of the baryon 
$\Lambda_b$. 
Writing the momentum, $p_{\Lambda_b}$, of the heavy baryon, $\Lambda_b$, such 
as,
\begin{eqnarray}
p_{\Lambda_b}=m_b v_{\Lambda_b} + k\ ,
\end{eqnarray}
where, $k$, is the residual momentum, the velocity, $v_{\Lambda_b}$, of the 
heavy quark, $b$, is almost that of the heavy baryon, $\Lambda_b$. Because of 
$m_b \gg \Lambda_{QCD}$, the parameterization of the hadronic amplitude, 
$\mathcal{A}(\Lambda_b \to \Lambda V)$,  
in term of the velocity, $v_{\Lambda_b}$,  gives us a reasonable picture where 
corrections only arise in the $1/m_b$ expansion.

The Stech's approach~\cite{Guo:1995zb} mainly assumes that the off-shell 
energy of a constituent diquark
is close to its constituent mass (without dependence of its space momentum) 
in the rest frame of the baryon. Moreover, the spectator quark retains its original momentum and 
spin before final hadronization. Therefore, the energy carried
by the spectator quark is equal to that of the spectator in the rest frame 
of the final state particle and the
relevant $b$-quark space momenta are much smaller than the $b$ quark mass: indeed, it is 
assumed to be of the order of the confinement scale, $\Lambda_{QCD}$.
This approach firstly used in the meson case can be  generalized to a heavy 
baryon considered as a bound state of 
a $b$ quark and a scalar diquark. Thus, in the baryon case, the hadronic matrix, 
$\langle \Lambda(p_{\Lambda},s_{\Lambda})|\bar{s}\gamma_{\mu}(1-\gamma_5)b|
\Lambda_b(p_{\Lambda_b},
s_{\Lambda_b})\rangle$, written in terms of components of Dirac spinors such 
as, 
\begin{eqnarray}
\bar{U}_s({\vec{p}}_{\Lambda},m_s)\gamma_{\mu}(1-\gamma_5)U_{b}
({\vec{p}}_{\Lambda_b}=\vec{0},m_b)\ , \nonumber
\end{eqnarray}
leads to the following expressions for the form factors, $\theta_1(q^2)$ 
and $\theta_2(q^2)$, when the heavy quark mass goes to infinity:
\begin{align}
\theta_1(q^2)=& 
2(E_{\Lambda}+m_{\Lambda}+m_s)\Gamma(E_{\Lambda},m_s,m_{\Lambda})\ ,  
\nonumber \\
\theta_2(q^2)=&
(m_s-m_{\Lambda})\Gamma(E_{\Lambda},m_s,m_{\Lambda})\ ,
\end{align}
and where the function, $\Gamma(E_{\Lambda},m_s,m_{\Lambda})$, is,
\begin{eqnarray}
\Gamma(E_{\Lambda},m_s,m_{\Lambda})=\frac{1}{2(E_{\Lambda}+m_s)}
\sqrt{\frac{(E_{\Lambda}+m_s)m_{\Lambda}}{(E_{\Lambda}+m_{\Lambda})m_{s}}}\ .
\end{eqnarray}
\noindent Thus, the ratio, $\theta_2(q^2)/\theta_1(q^2)$, gives, 
\begin{eqnarray}
\frac{\theta_2(q^2)}{\theta_1(q^2)}=
\frac{m_s-m_{\Lambda}}{2 (E_{\Lambda}+m_s+m_{\Lambda})}\ ,
\end{eqnarray}
where, $E_{\Lambda}$, the energy of $\Lambda$ in the $\Lambda_b$ rest-frame reads as,
\begin{eqnarray}
E_{\Lambda}=\frac{1}{2 m_{\Lambda_b}}(m_{\Lambda_b}^{2}+m_{\Lambda}^{2}-q^2)\ ,
\end{eqnarray}
with $q^2$ being the  momentum transfer as previously defined. Working in 
terms of velocity, it may also be 
useful to express the invariant velocity transfer, $\omega(q^2)$, as a 
function of the lambda masses, 
$m_{\Lambda_b}$ and $m_{\Lambda}$. Therefore, one obtains, 
\begin{eqnarray}
\omega(q^2)=v_{\Lambda_b} \cdot v_{\Lambda} =\frac{m^2_{\Lambda_b}+
m^2_{\Lambda}-q^2}{2 m_{\Lambda_b} 
m_{\Lambda}}\ , \;\; \omega_{min}(q^2)=1\ , \;\;\omega_{max}(q^2)=\frac{m^2_{\Lambda_b}+
m^2_{\Lambda}}{2 m_{\Lambda_b} m_{\Lambda}}\;\; \ ,
\end{eqnarray}
where, $v_{\Lambda_b}$ and $v_{\Lambda}$, are the velocities of the baryon 
$\Lambda_b$ 
and $\Lambda$, respectively. 
The effective QCD Lagrangian for heavy quark can be expanded both in powers 
of $\alpha_s(m_b)$ 
and $1/m_b$. The radiative corrections will not be taken into account since they 
are not relevant in our analysis whereas the corrections proportional to 
$\Lambda_{QCD}/m_b$ 
will be systematically calculated. These latter nonperturbative corrections 
are computed in the next section. In the following, all the form factors will be 
defined as a function of the invariant velocity transfer, $\omega$, instead of 
the momentum transfer, $q^2$.

\subsection{The $\boldsymbol{1/m_b}$ corrections}

\noindent The QCD Lagrangian in HQET is written as
\begin{eqnarray}
\mathcal{L}_{HQET}=\bar{h}_{v_{\Lambda_b}}^b i v_{\Lambda_b} \cdot D 
h_{v_{\Lambda_b}}^b\ ,
\end{eqnarray}
where, in the limit of infinite heavy quark mass, the quark field, 
$h_{v_{\Lambda_b}}^b(x)$, 
replaces the heavy quark field, $b(x)$, acting on $x$. Thus, the relation 
between the two quark 
fields reads as,
\begin{eqnarray}
h_{v_{\Lambda_b}}^b(x)= \exp\biggl[i m_b \; v_{\Lambda_b}\cdot x \; P_{+} 
b(x)\biggr]\ ,
\end{eqnarray}
with $P_+$ being the positive energy projection operator and $D^{\alpha}$ 
being the gauge 
covariant derivative $(=\partial^{\alpha} - ig_s t^a A_a^{\alpha})$. 
Including the corrections $1/m_b$ to the effective Lagrangian previously 
written, one has:
\begin{eqnarray}
\mathcal{L}_{HQET}= \bar{h}_{v_{\Lambda_b}}^b i v_{\Lambda_b} \cdot D 
h_{v_{\Lambda_b}}^b
+\frac{1}{2 m_b} \biggr[ \bar{h}_{v_{\Lambda_b}}^b (iD)^2  h_{v_{\Lambda_b}}^b
+ \frac{g_s}{2}  \bar{h}_{v_{\Lambda_b}}^b \sigma_{\alpha \beta}   
G^{\alpha \beta} h_{v_{\Lambda_b}}^b \biggl]\ ,
\end{eqnarray}
where the gluon field strength is defined as usual by $G^{\alpha \beta}=
[iD^{\alpha},iD^{\beta}]=ig_s t^a G^{\alpha \beta}_a$.
In case of heavy to light quark mass transitions, the weak current will have  
the following general structure where the corrections $1/m_b$ are taken into 
account:
\begin{eqnarray}
\bar{q}\Gamma b \rightarrow \bar{q} \Gamma h_{v_{\Lambda_b}}^b + 
\frac{1}{2 m_b}\bar{q} i \; 
\SLASH \hspace{-0.1cm} D 
h_{v_{\Lambda_b}}^b+ ...\  ,
\end{eqnarray}
where $\Gamma$ reads as $\gamma_{\mu}$ or $\gamma_{\mu}\gamma_5$. The dots 
define the 
higher order corrections which will be neglected in our analysis.

By including the covariant derivative, $D$, as well as the corrections 
at the order of $1/m_b$ to the effective Lagrangian, it leads, respectively, 
to the local correction~\cite{Datta:1994ij} given by,
\begin{eqnarray}
\delta \mathcal{L}^{lo,1} = \frac{1}{2m_b} \bar{q} \Gamma i \; \SLASH 
\hspace{-0.1cm} D h_{v_{\Lambda_b}}^b\ ,  
\end{eqnarray}
and to the nonlocal corrections~\cite{Datta:1994ij} given by,
\begin{align}
\delta \mathcal{L}^{nlo,2} =&\frac{1}{2m_b} \bar{h}_{v_{\Lambda_b}}^b 
(i v_{\Lambda_b} \cdot D)^2  
h_{v_{\Lambda_b}}^b\ , \nonumber \\
\delta \mathcal{L}^{nlo,3} =&\frac{1}{2m_b} \bar{h}_{v_{\Lambda_b}}^b 
(i D)^2  h_{v_{\Lambda_b}}^b\ , \nonumber \\
\delta \mathcal{L}^{nlo,4} =&\frac{1}{2m_b} \frac{-g_s}{2} 
\bar{h}_{v_{\Lambda_b}}^b 
\sigma_{\alpha \beta} G^{\alpha \beta} h_{v_{\Lambda_b}}^b\ ,
\end{align}
where, $q$, holds for a light quark such as $u,d$ and $s$, $\Gamma$ is 
an arbitrary Dirac 
structure and, $g_s$ is the field coupling constant. As usual, 
$\sigma_{\alpha \beta}$ is given 
by $\frac{i}{2}[\gamma_{\alpha},\gamma_{\beta}]$.

\subsubsection{The local correction}

\noindent Let us first start with the local term correction, 
$\delta \mathcal{L}^{lo,1}$, 
to the effective Lagrangian, $\mathcal{L}_{HQET}$, written in the
HQET formalism. It usually takes the form,
\begin{eqnarray}\label{eqq1}
\langle \Lambda(p_{\Lambda},s_{\Lambda}) | \bar{q} \; \Gamma i 
D h_{v_{\Lambda_b}}^{b} | 
\Lambda_b(p_{\Lambda_b},s_{\Lambda_b}) \rangle
= \bar{u}_{\Lambda}(p_{\Lambda},s_{\Lambda}) \phi^\nu(\omega) \Gamma 
\gamma_\nu u_{\Lambda_b}
(p_{\Lambda_b},s_{\Lambda_b})\ , 
\end{eqnarray}
where one of the most general form of $\phi^\nu(\omega)$ is,
\begin{multline}
\phi^\nu(\omega)=\Bigl[\phi_{11}(\omega) v^{\nu}_{\Lambda_b} + 
\phi_{12}(\omega) v^{\nu}_{\Lambda} 
+ \phi_{13}(\omega) \gamma_\nu \Bigr]+ 
\\
\SLASH v_{\Lambda_b} \Bigl[\phi_{21}(\omega) v^{\nu}_{\Lambda_b}+ 
\phi_{22}(\omega) v^{\nu}_{\Lambda} 
+ \phi_{23}(\omega) \gamma_\nu\Bigr]\ ,
\end{multline}
because of spin symmetry properties. On one hand, the equation of motion for  heavy quark,
\begin{eqnarray}\label{eq8}
v_{\Lambda_b} \cdot Dh_{v_{\Lambda_b}}^b=0\ ,
\end{eqnarray}
applied to Eq.~(\ref{eqq1}), gives
\begin{eqnarray}
v_{\Lambda_b} \cdot\langle \Lambda(p_{\Lambda},s_{\Lambda}) | \bar{q} \; 
\Gamma i D h_{v_{\Lambda_b}}^{b} 
| \Lambda_b(p_{\Lambda_b},s_{\Lambda_b}) \rangle = 0\ ,
\end{eqnarray}
and leads therefore to the following constraints where $\Gamma=1$ and 
$\Gamma= 
\gamma_5$ have been  used, respectively:
\begin{align}
\bar{u}_{\Lambda}(p_{\Lambda},s_{\Lambda}) & \bigl[ v_{\Lambda_b} \cdot 
\phi(\omega)\bigr]
u_{\Lambda_b}(p_{\Lambda_b},s_{\Lambda_b})=0\ , \nonumber\\
\bar{u}_{\Lambda}(p_{\Lambda},s_{\Lambda}) & \bigl[ v_{\Lambda_b} \cdot 
\phi(\omega) \gamma^5 \bigr]
u_{\Lambda_b}(p_{\Lambda_b},s_{\Lambda_b}) =0\ . 
\end{align}
Thus, two relations between the $\phi_{ij}(\omega)$'s can be obtained and read as,
\begin{align}
\phi_{11}(\omega)+ \omega \phi_{12}(\omega)&=-\phi_{23}(\omega)\ , \nonumber\\
\phi_{21}(\omega)+ \omega \phi_{22}(\omega)&=-\phi_{13}(\omega)\ . 
\end{align}
On the other hand, the momentum conservation also implies that, 
\begin{multline}
\langle \Lambda(p_{\Lambda},s_{\Lambda}) | i \partial_{\mu} (\bar{q}\;  
\Gamma h_{v_{\Lambda_b}}^{b}) | 
\Lambda_b(p_{\Lambda_b},s_{\Lambda_b}) \rangle= \\
\langle \Lambda(p_{\Lambda},s_{\Lambda}) | i D_{\mu}^{\star}\bar{q}\Gamma 
h_{v_{\Lambda_b}}^b 
|\Lambda_b(p_{\Lambda_b},s_{\Lambda_b}) \rangle+
\langle \Lambda(p_{\Lambda},s_{\Lambda}) | \bar{q} \Gamma i D_{\mu} 
h_{v_{\Lambda_b}}^b 
|\Lambda_b(p_{\Lambda_b},s_{\Lambda_b}) \rangle = \\
\biggr\{(m_{\Lambda_b}-m_b)v_{\mu}^{\Lambda_b}- m_{\Lambda} v_{\mu}^{\Lambda}
\biggr\}
\langle \Lambda(p_{\Lambda},s_{\Lambda}) |\bar{q}\Gamma h_{v_{\Lambda_b}}^b
|\Lambda_b(p_{\Lambda_b},s_{\Lambda_b}) \rangle\ .
\end{multline} 
By using the equation of motion for  light quark,
\begin{eqnarray}
(i \;  \SLASH \hspace{-0.1cm} D - m_q)q = 0\ ,
\end{eqnarray}
then changing $\Gamma$ into $\gamma_{\mu} \Gamma$ and contracting the index 
$\mu$, one gets,
\begin{multline}\label{eq10}
\langle \Lambda(p_{\Lambda},s_{\Lambda}) | \bar{q} \; \gamma^{\mu} \Gamma 
i D_{\mu} h_{v_{\Lambda_b}}^b 
|\Lambda_b(p_{\Lambda_b},s_{\Lambda_b}) 
\rangle= \\
 \biggr\{(m_{\Lambda_b}-m_b)v_{\mu}^{\Lambda_b}- (m_{\Lambda}-m_q)
v_{\mu}^{\Lambda}\biggr\} 
\langle \Lambda(p_{\Lambda},s_{\Lambda}) |\bar{q}\gamma^{\mu}
\Gamma h_{v_{\Lambda_b}}^b
|\Lambda_b(p_{\Lambda_b},s_{\Lambda_b}) \rangle\ .
\end{multline}
By choosing $\Gamma=1$ and $\Gamma=\gamma_5$, Eq.~(\ref{eq10}) gives, 
respectively, to
\begin{multline}
\label{eq07a}
\biggl[\frac{\omega-1}{\omega}\biggr](\phi_{11}(\omega)-\phi_{21}(\omega))
+\biggl[\frac{2 \omega+1}{\omega}\biggr] 
\phi_{13}(\omega)+
\biggl[\frac{4\omega -1}{\omega}\biggr]\phi_{23}(\omega)= \\ 
(m_{\Lambda_b}-m_b)(F_1^0(\omega)+F_2^0(\omega))-(m_{\Lambda}-m_q)
(F_1^0(\omega)+\omega F_2^0(\omega))\ ,
\end{multline} 
and,
\begin{multline}
\label{eq07b}
\biggl[ \frac{\omega+1}{\omega}\biggr](\phi_{11}(\omega)+\phi_{21}(\omega))
- \biggl[\frac{2 \omega -1}{\omega}\biggr]
\phi_{13}(\omega)
-\biggl[\frac{4\omega -1}{\omega}\biggr]\phi_{23}(\omega)= \\ 
(m_{\Lambda_b}-m_b)(G_1^0(\omega)-G_2^0(\omega))+(m_{\Lambda}-m_q)
(G_1^0(\omega)+\omega G_2^0(\omega))\ .
\end{multline} 
In Eqs.~(\ref{eq07a}) and~(\ref{eq07b}) the form factors, $F_i^0(\omega)$, 
and, $G_i^0(\omega)$, are
 the zeroth order form factors. The relations
between them and the form factors, $\theta_i(\omega)$, in the limit of 
$m_b$ going to infinity, read as  follows,
\begin{align}
F_1^0(\omega)=& \theta_1(\omega)-\theta_2(\omega)\ , \nonumber\\
F_2^0(\omega)=& 2 \theta_2(\omega)\ , \nonumber\\
F_3^0(\omega)=& 0\ , 
\end{align}
and, 
\begin{align}
G_1^0(\omega)=& \theta_1(\omega)+\theta_2(\omega)\ , \nonumber\\
G_2^0(\omega)=& 2 \theta_2(\omega)\ , \nonumber\\
G_3^0(\omega)=& 0\ .
\end{align}

\noindent From Eqs.~(\ref{eq07a}) and (\ref{eq07b}), it is  possible to 
express the $\phi_{ij}(\omega)$'s in 
terms of the zeroth order of the vector and axial form factors, 
$F_i^0(\omega)$ and $G_i^0(\omega)$ and  one has,
\begin{multline}\label{eq3}
\phi_{11}(\omega)= \frac{\omega(\omega+1)}{2(\omega^2 -1)}
\biggl[-(m_b-m_{\Lambda_b})(F_1^0(\omega)+F_2^0(\omega))
-(m_{\Lambda}-m_q)(F_1^0(\omega)+\omega F_2^0(\omega))\biggr]+ \\
\frac{\omega(\omega-1)}{2(\omega^2 -1)}
\biggl[(m_b-m_{\Lambda_b})(G_2^0(\omega)-G_1^0(\omega))
+(m_{\Lambda}-m_q)(G_1^0(\omega)+\omega G_2^0(\omega))\biggr] 
-\frac{(7 \omega-1)}{(\omega^2-1)}\phi_{123}(\omega)\ ,
\end{multline}
\begin{multline}\label{eq3a}
\phi_{12}(\omega)= \frac{\omega-1}{2(\omega^2 -1)}
\biggl[(m_b-m_{\Lambda_b})(F_1^0(\omega)+F_2^0(\omega))
+(m_{\Lambda}-m_q)(F_1^0(\omega)+\omega F_2^0(\omega))\biggr]+ \\
\frac{\omega-1}{2(\omega^2 -1)}
\biggl[(m_b-m_{\Lambda_b})(G_1^0(\omega)-G_2^0(\omega))
-(m_{\Lambda}-m_q)(G_1^0(\omega)+\omega G_2^0(\omega))\biggr] 
-\frac{(\omega-7)}{(\omega^2-1)}\phi_{123}(\omega)\ ,
\end{multline}
\begin{multline}\label{eq3b}
\phi_{21}(\omega)= \frac{\omega(\omega+1)}{2(\omega^2 -1)}
\biggl[(m_b-m_{\Lambda_b})(F_1^0(\omega)+F_2^0(\omega))
+(m_{\Lambda}-m_q)(F_1^0(\omega)+\omega F_2^0(\omega))\biggr]+\\ 
\frac{\omega(\omega-1)}{2(\omega^2 -1)}
\biggl[(m_b-m_{\Lambda_b})(G_2^0(\omega)-G_1^0(\omega))
+(m_{\Lambda}-m_q)(G_1^0(\omega)+\omega G_2^0(\omega))\biggr]+\\
\frac{(6\omega^2-\omega+1)}{(\omega^2-1)}\phi_{123}(\omega)\ ,
\end{multline}
and,
\begin{multline}\label{eq4}
\phi_{22}(\omega)= -\frac{\omega+1}{2(\omega^2 -1)}
\biggl[(m_b-m_{\Lambda_b})(F_1^0(\omega)+F_2^0(\omega))
+(m_{\Lambda}-m_q)(F_1^0(\omega)+\omega F_2^0(\omega))\biggr]-\\
\frac{(\omega-1)}{2(\omega^2 -1)}\biggl[(m_b-m_{\Lambda_b})
(G_2^0(\omega)-G_1^0(\omega))
+(m_{\Lambda}-m_q)(G_1^0(\omega)+\omega G_2^0(\omega))\biggr]\\
+\frac{(1-7\omega)}{(\omega^2-1)}\phi_{123}(\omega)\ .
\end{multline}
In Eqs.~(\ref{eq3}-\ref{eq4}), the assumption of $\phi_{13}(\omega)
\simeq \phi_{23}(\omega)=
\phi_{123}(\omega)$ has been made for simplicity. 
It is now straightforward to derive the local corrections, $\delta F_i^{lo,1}
(\omega)$ and 
$\delta G_i^{lo,1}(\omega)$, to the 
transition form factors, $F_i(\omega)$ and $G_i(\omega)$ and one gets,
\begin{align}
\delta F_1^{lo,1}(\omega) & = -\frac{1}{2m_b} \biggl[
\phi_{11}(\omega)+\phi_{12}(\omega)2 (\omega+1)-\phi_{21}(\omega)
+\phi_{22}(\omega)\biggr]\ , \nonumber\\
\delta F_2^{lo,1}(\omega) & = \frac{1}{m_b}\biggl[
2 \phi_{11}(\omega)+ 2 \omega \phi_{12}(\omega)+\phi_{21}(\omega)
+\phi_{22}(\omega)\biggr]\ , \nonumber\\
\delta F_3^{lo,1}(\omega) & =\frac{1}{m_b}
\biggl[\phi_{11}(\omega)+\phi_{21}(\omega)\biggr]\ , 
\end{align}
and,
\begin{align}
\delta G_1^{lo,1}(\omega) & = \frac{1}{2m_b}
\biggl[\phi_{11}(\omega)+\phi_{12}(\omega) (2 \omega-1)
+\phi_{21}(\omega)+\phi_{22}(\omega)\biggr]\ , \nonumber\\
\delta G_2^{lo,1}(\omega) & = \frac{1}{m_b}
\biggl[2 \phi_{11}(\omega)+ 2  \omega \phi_{12}(\omega)
-\phi_{21}(\omega)+\phi_{22}(\omega)\biggr]\ ,   \nonumber\\
\delta G_3^{lo,1}(\omega) & =\frac{1}{m_b}
\biggl[\phi_{12}(\omega)-\phi_{22}(\omega)\biggr]\ .
\end{align}

\noindent Moreover, assuming that,
\begin{eqnarray}
\delta F_1^{lo,1}(\omega)+\delta F_2^{lo,1}(\omega)+\delta F_3^{lo,1}(\omega)= 
\frac{\epsilon}{2 m_b}\ ,
\end{eqnarray}
it allows us to derive the  expression of $\phi_{123}(\omega)$ and one obtains the
following solution:
\begin{multline}
\phi_{123}(\omega)=
\frac18 \Biggl[ (m_b-m_{\Lambda_b})(G_1^0(\omega)-G_2^0(\omega))
-(m_{\Lambda}-m_q)(G_1^0(\omega)+ \omega G_2^0(\omega))+ \frac{\epsilon (\omega+1)}{2 (\omega-1)}\Biggr]\\
-\frac{\omega+1}{16(\omega-1)}\Biggl[-\epsilon+(m_b-m_{\Lambda_b})(F_1^0(\omega)+F_2^0(\omega))
+(m_{\Lambda}-m_q)(F_1^0(\omega)+\omega F_2^0(\omega))\Biggr]\ ,
\end{multline}
where the form factor, $\phi_{123}(\omega)$, is expressed in terms of the zeroth 
order form factors, $F_i^0(\omega)$ and $G_i^0(\omega)$, respectively.

\subsubsection{The non-local corrections}

\noindent Next is the first non local correction, $\delta 
\mathcal{L}^{nlo,2}$, given by,
\begin{eqnarray}
\delta \mathcal{L}^{nlo,2} = \frac{1}{2m_b} \bar{h}_{v_{\Lambda_b}}^b 
(i v_{\Lambda_b} \cdot D)^2 
h_{v_{\Lambda_b}}^b \ ,
\end{eqnarray}
which does not contribute because of the equation of motion, $v_{\Lambda_b} 
\cdot Dh_{v_{\Lambda_b}}^b=0$, 
of the heavy quark, $b$. Thus, it directly leads 
to $\delta F_i^{nlo,2}(\omega)=\delta G_i^{nlo,2}(\omega)=0$. The second 
non local correction, 
$\delta \mathcal{L}^{nlo,3}$, given by,
\begin{multline}
\langle \Lambda (p_{\Lambda},s_{\Lambda})| i \int dy T \{\bar{q}
h_{v_{\Lambda_b}}^b(0),\delta \mathcal{L}^{nlo,3}(y)\}| 
\Lambda_b (p_{\Lambda_b},s_{\Lambda_b})\rangle  = \\ \frac{\lambda_{i}(\omega)}
{2 m_b}
\langle \Lambda (p_{\Lambda},s_{\Lambda})| \bar{q} \Gamma 
h_{v_{\Lambda_b}}^b(0) | 
\Lambda_b (p_{\Lambda_b},s_{\Lambda_b})\rangle\ ,
\end{multline}
and being a singlet under spin symmetry,  leads  
to two form factors, $\lambda_i^F(\omega)$ and $\lambda_i^G(\omega)$, which 
only renormalize the zeroth order form factors, $F_i^0(\omega)$ and $G_i^0(\omega)$. 
Namely and independently of $\Gamma$, one has,
\begin{align}
\bar{F}_i^0(\omega) & = F_i^0(\omega) + \frac{\lambda_i^F(\omega)}{2 m_b}\ , \nonumber \\
\bar{G}_i^0(\omega) & = G_i^0(\omega) + \frac{\lambda_i^G(\omega)}{2 m_b}\ .
\end{align}
For simplicity, it will be assumed that  $\lambda_i^F(\omega)=
\lambda_i^G(\omega)=\lambda_i(\omega)$. Moreover, these corrections can be safely 
neglected as they will only appear at the order of $1/m_b^2$.

\noindent Finally, by means of  spin symmetry, the magnetic operator leads to
the last non local correction, $\delta \mathcal{L}^{nlo,4}$, 
which reads as,
\begin{multline}
\langle \Lambda(p_{\Lambda},s_{\Lambda}) | i \int dy T \biggl\{\bar{q}
h_{v_{\Lambda_b}}^b(0),
\frac{g_s}{2} \bar{h}_{\Lambda_b}^b
 \sigma_{\alpha \beta} G^{\alpha \beta}
h_{\Lambda_b}^b(y) \biggr\} | \Lambda_b(p_{\Lambda_b},s_{\Lambda_b}) \rangle = \\
\frac{1}{2m_b}\bar{u}_{\Lambda}(p_{\Lambda},s_{\Lambda}) 
\varphi_{\alpha \beta}(\omega) 
\Gamma \biggl(\frac{1+\SLASH v_{\Lambda_b}}{2}\biggr) s^{\alpha \beta} 
u_{\Lambda_b}(p_{\Lambda_b},s_{\Lambda_b})\ ,
\end{multline}
where,
\begin{eqnarray}
\varphi_{\alpha \beta}(\omega) = \Bigl[\varphi_{11}(\omega)+ \SLASH 
v_{\Lambda_b} \varphi_{12}(\omega)\Bigr] 
s_{\alpha \beta} 
+ \Bigl[\varphi_{21}(\omega)+ \SLASH v_{\Lambda_b} \varphi_{22}(\omega) 
\Bigr] t_{\alpha \beta}\ , 
\end{eqnarray}
with the tensors, $s_{\alpha \beta}$ and $t_{\alpha \beta}$, defined as,
\begin{eqnarray}
s^{\alpha \beta}=\frac14 \Bigl\{\gamma^{\alpha} \gamma^{\beta} - 
\gamma^{\beta} \gamma^{\alpha}\Bigr\} \;\; {\rm and} \;\;
t^{\alpha \beta}=\frac14 \Bigl\{\gamma^\alpha v_{\Lambda}^\beta - 
\gamma^\beta v_{\Lambda}^\alpha \Bigr\}\ .
\end{eqnarray}
Assuming that $\varphi_{11}(\omega)=\varphi_{21}(\omega)$ and 
$\varphi_{12}(\omega)=\varphi_{22}(\omega)$, 
the corrections to the form factors $F_i(q^2)$ and $G_i(q^2)$ are 
given by,
\begin{align}
\delta F_1^{nlo,4}(\omega) & = \frac{1}{4m_b} \biggl[\varphi_{11}(\omega)
-\varphi_{12}(\omega)\biggr]\ , \nonumber\\
\delta F_2^{nlo,4}(\omega) & = -\frac{1}{4m_b}\biggl[3\varphi_{11}(\omega)+
\varphi_{12}(\omega)(3-2\omega)\biggr]\ , \nonumber\\
\delta F_3^{nlo,4}(\omega) & =-\frac{1}{4m_b} \biggl[\varphi_{11}(\omega)
+\varphi_{12}(\omega)\biggr]\ , 
\end{align}
and,
\begin{align}
\delta G_1^{nlo,4}(\omega) & =\frac{1}{4m_b} \biggl[\varphi_{11}(\omega)
+\varphi_{12}(\omega)\biggr]\ , \nonumber\\
\delta G_2^{nlo,4}(\omega) & =\frac{1}{4m_b} \biggl[3\varphi_{11}(\omega)-
\varphi_{12}(\omega)(3+2\omega)\biggr]\ , \nonumber\\
\delta G_3^{nlo,4}(\omega) & =\frac{1}{4m_b} \bigl[\varphi_{12}(\omega)
-\varphi_{11}(\omega)\bigr]\ .
\end{align}
Next and finally is the determination of the form factors, $\varphi_{ij}(\omega)$,
coming from the magnetic operator corrections. By solving Eq.~(\ref{eq20}) where the 
form factors, $F_i(\omega)$ and $G_i(\omega)$, have been replaced by their expressions 
given in Eqs.~(\ref{eqqq01}-\ref{eqqq02}), it yields to the expressions of  the form factors,
$\varphi_{ij}(\omega)$, as a function of $F_i^0(\omega)$ and $G_i^0(\omega)$, respectively.

\subsubsection{The form factors at the order of $\boldsymbol{1/m_b}$}

After having derived in the HQET formalism the local, $\delta 
\mathcal{L}^{lo,1}$,  and nonlocal, $\delta \mathcal{L}^{nlo,j}$, corrections 
to the effective Lagrangian, $\mathcal{L}_{HQET}$, the full form factors, $F_i(\omega)$ and 
$G_i(\omega)$, read as,
\begin{align}
F_i(\omega) &=F_i^0(\omega)+\delta F_i^{lo,1}(\omega)+ \sum_{j=2}^4 
\delta F_i^{nlo,j}(\omega)\ , \nonumber \\
G_i(\omega) &=G_i^0(\omega)+\delta G_i^{lo,1}(\omega)+ \sum_{j=2}^4 
\delta G_i^{nlo,j}(\omega)\ .
\end{align} 
Explicitly, the expressions of the form factors $F_i(\omega)$ and 
$G_i(\omega)$ up to the corrections $1/m_b$ can be written as follows,
\begin{align}
F_1(\omega) &= F_1^0(\omega)-\frac{1}{2 m_b} \biggl[ \phi_{11}(\omega)
+2 \phi_{12}(\omega)(1+2\omega)-\phi_{21}(\omega)+\phi_{22}(\omega)
\nonumber \\
&  \hspace{10.4cm} -\frac12 (\varphi_{11}(\omega)-\varphi_{12}(\omega))\biggr]\ , 
\nonumber 
\end{align}
\begin{align}\label{eqqq01}
F_2(\omega) &= F_2^0(\omega)+\frac{1}{2 m_b}\biggl[4 \phi_{11}(\omega)+2\phi_{21}(\omega)
+2  \phi_{22}(\omega)+ \omega(4 \phi_{12}(\omega) +\varphi_{12}(\omega))
\nonumber \\
&  \hspace{10.4cm} -\frac32 (\varphi_{11}(\omega)-\varphi_{12}(\omega))\biggr]\ ,\nonumber \\
F_3(\omega) &= F_3^0(\omega) 
+\frac{1}{m_b}\biggl[\phi_{12}(\omega)+\phi_{22}(\omega)-\frac14
(\varphi_{11}(\omega)+\varphi_{12}(\omega))\biggr]\ , 
\end{align}
and,
\begin{align}\label{eqqq02}
G_1 (\omega)&= G_1^0(\omega)+\frac{1}{2 m_b} \biggl[ \phi_{11}(\omega)
+2 \phi_{12}(\omega)(2\omega-1)+\phi_{21}(\omega)+\phi_{22}(\omega)
\nonumber \\
&  \hspace{10.4cm} +\frac12 (\varphi_{11}(\omega)+\varphi_{12}(\omega))\biggr]\ , \nonumber \\
G_2(\omega) &= G_2^0(\omega)+\frac{1}{2 m_b}\biggl[4 \phi_{11}(\omega)-2\phi_{21}(\omega)
+2  \phi_{22}(\omega)+ \omega(4 \phi_{12}(\omega) -\varphi_{12}(\omega))
\nonumber \\
&  \hspace{10.4cm} +\frac32 (\varphi_{11}(\omega)-\varphi_{12}(\omega))\biggr]\ , \nonumber \\
G_3(\omega) &= G_3^0(\omega)+\frac{1}{m_b}\biggl[\phi_{12}(\omega)-\phi_{22}(\omega)-\frac14
(\varphi_{11}(\omega)-\varphi_{12}(\omega))\biggr]\ . 
\end{align}
The evolution of the form factors $F_i(\omega)$ and $G_i(\omega)$ as a function 
of the invariant velocity transfer, $\omega$, in case of $\Lambda_b \to 
\Lambda V$ where $V$ holds for the vectors
 $\rho^0, \omega$ and $J/\psi$ is plotted in Figs.~\ref{figffF} and~\ref{figffG}. 
It can also be shown that the invariant velocity transfer, $\omega$, to which the form factors 
$F_i(\omega)$ and $G_i(\omega)$ have to be evaluated are equal to $\omega=2.573$ for $\Lambda_b \to 
\Lambda \rho^0$,  $\omega=2.572$ for $\Lambda_b \to \Lambda \omega$ and $\omega=1.856$ 
for $\Lambda_b \to \Lambda J/\psi$, respectively.

\subsubsection{The integral overlap}

The calculation of the overlap between the initial and final baryon
wave functions will be done by making use of the Drell-Yann approach  
where the current matrix elements are calculated in the impulse 
approximation. The two relations to leading order in momentum, $P_{\Lambda_b}$, are,
\begin{align}\label{eq20}
F_1(\omega) + &\frac12 (\omega+1)(m_{\Lambda_b}+m_{\Lambda})\biggl( 
\frac{F_2(\omega)}{m_{\Lambda} + m_{\Lambda_b} \omega}+ \frac{F_3(\omega)}{m_{\Lambda_b}
+ m_{\Lambda} \omega}
\biggr) = c_s \mathcal{I}(\omega)\ , \nonumber \\
G_1(\omega)-& \frac12 (\omega-1)(m_{\Lambda_b}-m_{\Lambda})\biggl( 
\frac{G_2(\omega)}{m_{\Lambda}+m_{\Lambda_b} \omega}+\frac{G_3(\omega)}{m_{\Lambda_b}
+\omega m_{\Lambda}}
\biggr)= -c_s \mathcal{I}(\omega)\ , 
\end{align}
where the overlap integral, $\mathcal{I}(\omega)$, is,
\begin{eqnarray}
\mathcal{I}(\omega)=\int_0^1 d x_1 \Phi_s(x_1)^{\star}\Phi_b(x_1) \equiv 
\mathcal{I}(\omega,m_{\Lambda_b} )\ .
\end{eqnarray}
The Clebsch-Gordan coefficient, $c_s$, suppresses the integral overlap since 
the s(ud) state is the only one which
may contribute to the transition $\Lambda_b$ to $\Lambda$. As a first approximation, 
the $\Lambda$ may be seen as a superposition of various 
quark-diquark configurations.  In our model, one will follow the assumption 
of taking $c_s$ equal to $1/\sqrt{3}$.
Then, let us  expand the overlap function $\mathcal{I}(\omega,m_{\Lambda_b})$
into inverse powers of the heavy baryon mass $1/m_{\Lambda_b}$, i.e.
\begin{eqnarray}\label{Itilde}
\mathcal{I}(\omega,m_{\Lambda_b} ) &=& \mathcal{I}^{(0)}(\omega) +
\left[\frac{1}{m_{\Lambda_b}}\right] \mathcal{I}^{(1)}(\omega) + 
\mathcal{O}\biggl(\frac{1}{m_{\Lambda_b}^2}\biggr)\mathcal{I}^{(2)}(\omega)\ ,
\nonumber\\
&=& \mathcal{I}^{(0)}(\omega)
\left\{
1+\left[
\frac{1}{m_{\Lambda_b}}\right]\tilde{\mathcal{I}}^{(1)}(\omega) + 
\mathcal{O}\biggl(\frac{1}{m_{\Lambda_b}^2}\biggr)\tilde{\mathcal{I}}^{(2)}
(\omega)
\right\}\ ,
\end{eqnarray}
where the leading term overlap integral, $\mathcal{I}^{(0)}(\omega)$, is 
\begin{eqnarray}
\mathcal{I}^{(0)}(\omega) &=&
\sqrt{\frac{1}{\omega}}^{2}
\exp{\{ - \kappa^2 \frac{\omega-1}{2\omega} \}}
\cdot
\frac{H_{1}(\bar{\kappa})}{H_{1}(\kappa)}\ ,
\end{eqnarray}
and the next-to-leading order correction~\cite{Konig:1993ze}, $\tilde{\mathcal{I}}^{(1)}
(\omega)$, reads as,
\begin{multline}
\tilde{\mathcal{I}}^{(1)}(\omega) = \;\frac{\omega -1}{\omega}
   \left[\frac{\bar{b}}{b}
 -\sqrt{2}\kappa (\bar{\alpha} b + \alpha \bar{b})
 + \kappa \alpha \bar{b}/(\sqrt{2}\omega)\right] \\
 +\frac{1}{2\sqrt{2}b}
  \left[ \frac{H_{2}(\kappa)}{H_{1}(\kappa)} -
  \sqrt{\frac{2}{\omega(\omega+1)}} \frac{
  H_{2}(\bar{\kappa})}
  {H_{1}(\bar{\kappa})} \right]   
  - \frac{1}{\sqrt{2}} (\bar{\alpha} b + \alpha \bar{b}) \left[
  \frac{H'_{1}(\kappa)}{H_{1}(\kappa)} -
  \sqrt{\frac{2}{\omega(\omega+1)}}\frac{
  H'_{1}(\bar{\kappa})}
  {H_{1}(\bar{\kappa})} \right] \\
  + \frac{1}{2} \alpha \bar{b} \frac{\omega-1}{\sqrt{\omega^3(\omega+1)}}
  \frac{H'_{1}(\bar{\kappa})}
  {H_{1}(\bar{\kappa})}\ ,
\end{multline}
with, $H_{l}(x)$,  a function defined as,
\begin{eqnarray}
H_{l}(x) = \int_{- x }^{\infty} {\rm d}z \left(z + x \right)^{l} 
e^{- z^2}\ ,\;\; {\rm and}\;\;\;
H'_l(x) = {\rm d} H_l(x)/{\rm d}x\ , 
\end{eqnarray}
where $\bar{\kappa}$ and $\kappa$ are
\begin{eqnarray}
\bar{\kappa} = \kappa \sqrt{\frac{\omega+1}{2\omega}}\ , \;\; {\rm with} \;\;
\kappa = \sqrt{2} \alpha b\ .
\end{eqnarray}
The parameters, $\alpha, \bar{\alpha}$ and $b, \bar{b}$ are defined in Section 4.3. Note also that the 
scaling functions, $\mathcal{I}^{(0)}(\omega)$, and,
$\tilde{\mathcal{I}}^{(1)}(\omega)$, obey the normalization conditions
$\mathcal{I}^{(0)}(1) = 1$ and $\tilde{\mathcal{I}}^{(1)}(1) = 0$.
 From Eq.~(\ref{eq20}), one can calculate the normalization condition, 
which gives 
\begin{align}
\sum_{i=1}^3 F_i(1) & = c_s \mathcal{I}(1)+ \mathcal{O}\biggl(\frac{1}{m_b^2} 
\biggr) , \nonumber \\
G_1(1) & = -c_s \mathcal{I}(1) + \mathcal{O}\biggl(\frac{1}{m_b^2} \biggr)\ ,
\end{align}
at $\omega=1$. In a similar way of Eq.~(\ref{eq20}), another relation between the asymptotical
form factors, $\theta_1(\omega)$ and $\theta_2(\omega)$, is given in terms of the overlap integral, 
$\mathcal{I}(\omega)$,  between  the $\Lambda_b$ and $\Lambda$ 
hadronic wave functions. Therefore, by working in the appropriate infinite momentum frame
and evaluating the good current components~\cite{Guo:1995zb},
the form factors, $\theta_1(\omega)$ and $\theta_2(\omega)$, are obtained as the following,
\begin{eqnarray}
\theta_{1}&=& \frac{2E_{\Lambda}+m_{\Lambda}+m_{s}}{2(E_{\Lambda}+m_{s})}c_{s}
\mathcal{I}(\omega),\nonumber \\
\theta_{2}&=& \frac{m_{s}-m_{\Lambda}}{2(E_{\Lambda}+m_{s})}c_{s}\mathcal{I}(\omega)\ .
\label{3c}
\vspace{2mm}
\end{eqnarray}

\subsection{The baryon wave function}

\noindent Working within a quark-diquark approach, the baryon wave-function 
is proposed as
\begin{eqnarray}
\Psi_i(x,{\bf k}^2_{\perp})=N_i^{\Psi} \Phi_i(x) \exp\Bigl[-b^2_i 
{\bf k}^2_{\perp}\Bigr]\ ,
 \label{eqwfb1}
\end{eqnarray}
where the function, $\Phi_i(x)$, is  defined as
\begin{eqnarray}
 \Phi_i(x) = N_i^{\Phi} \sqrt{x} \sqrt{(1-x)}\exp\Bigl[-b^2_i M^2_i (x
-x_{0i})^2\Bigr]\ , 
\label{eqwfb2}
\end{eqnarray}
with $x_{0i}$ being the peak of the baryon distribution function that 
represents the momentum fraction, $x$, carried by the heavy quark in the 
quark-diquark picture of the baryon. The parameter, $b_i$, is 
related to the root of the average square of the transverse momentum of 
$\bf{k}_{\perp}$. The index, $i$, holds for a given baryon, namely $\Lambda_b$ 
or $\Lambda$. We also impose the following condition of normalization for both of the 
functions, 
$\Psi_i(x,{\bf k}^2_{\perp})$ and $\Phi(x)$, such as,
\begin{align}
\label{eqwfb3}
\int_0^1 & \int_0^{\infty} {\rm d} x \; {\rm d^2} {\bf k}_{\perp}
\; \Psi_i(x,{\bf k}^2_{\perp})^{\star} 
\Psi_i(x,{\bf k}^2_{\perp})   = 1\ , \nonumber \\
\int_0^1 & {\rm d} x \;  \Phi(x)^{\star}\Phi(x)  = 1 \ .
\end{align}
The constants of normalization, $N_i^{\Psi}$ and $N_i^{\Phi}$, are 
directly given by  making use of Eq.~(\ref{eqwfb3}). 
The parameters, $b_i$ and $x_{0i}$, read as,
\begin{align}
b_i&=b+ \frac{\bar{b}}{M_i}\ , \nonumber \\
x_{0i}&=1- \frac{\alpha_i}{M_i}\ , \;\; {\rm and} \;\; \alpha_i=\alpha+
\frac{\bar{\alpha}}{M_i}\ ,
\end{align}
where $M_i$ is  the baryon mass. From the average square of the 
transverse momentum of $\bf{k}_{\perp}$ 
given by,
\begin{eqnarray}
\langle {\bf k}^2_{\perp} \rangle = \int_0^1 \int_0^{\infty} 
{\rm d^2} {\bf k}_{\perp}
\; |\Psi_i(x,{\bf k}^2_{\perp})|^2 {\bf k}^2_{\perp}\ ,
\label{eqwfb4}
\end{eqnarray}
one can effectively constrain the parameter, $b_i$, since on gets,
\begin{eqnarray}
2 b^2_i \langle {\bf k}^2_{\perp} \rangle = 1\ .
\end{eqnarray}
Regarding numerical values, we will follow Ref.~\cite{Konig:1993ze}:  
assuming that the root of the average transverse momentum,
$\sqrt{{\bf k}^2_{\perp}}$, is of the order of a few
hundred MeV, one will choose an average of $500$ MeV that gives $b_i=\sqrt{2}$ 
${\rm GeV}^{-1}$. The parameters, $b$ and $\bar{b}$, are equal to
$1.4$ ${\rm GeV}^{-1}$ and $0.1$ GeV, respectively.
The parameters, $\alpha$ and $\bar{\alpha}$, are equal to
$1.0$ ${\rm GeV}$ and $0.3$ ${\rm GeV}^{2}$, respectively.
\noindent The baryon quark distribution, $D_{i}(x)$,  
takes the following form,
\begin{eqnarray}
D_{i}(x)=\int_0^{\infty} {\rm d^2} {\bf k}_{\perp}
\; \Psi_i(x,{\bf k}^2_{\perp})^2=
\frac{ (N_i^{\Phi} N_i^{\Psi})^2 \pi}{2 b^2_i} (1-x)^6 x^2 \exp\Bigl[-2 b^2_i
 M^2_i (x-x_0)^2\Bigr]\ ,
\end{eqnarray}
and is plotted in Fig.~\ref{figbaryons} in case of $\Lambda_b$ and $\Lambda_s$. 
Finally, it  will be assumed that the model of the baryon wave function of $\Lambda_b$ is also
valid for $\Lambda_s$.

\subsection{The weak decay amplitude}

In tree approximation, the effective interaction, $\mathcal{H}^{eff}$, 
written as,
\begin{eqnarray}
\mathcal{H}^{eff}=\frac{G_{F}}{\sqrt{2}} V_{qb}V^{\star}_{qs} \sum_{i=1}^{10} 
c_i(m_b) O_i(m_b)\ , 
\end{eqnarray} 
gives the weak following amplitude, $\mathcal{A}_{(\lambda_1,\lambda_2)}(\Lambda_b \to \Lambda V$, 
factorized into,
\begin{multline}\label{eqoli}
\mathcal{A}_{(\lambda_1,\lambda_2)}(\Lambda_b \to \Lambda V)= \frac{G_{F}}{\sqrt{2}} 
f_V E_V {\langle \Lambda | \bar{s} \gamma_{\mu} (1-\gamma_{5}) b| \Lambda_b 
\rangle}_{(\lambda_{\Lambda},\lambda_V)} \\
 \biggl\{ \mathcal{M}^T_{\Lambda_b}(\Lambda_b \to \Lambda V)-
\mathcal{M}^P_{\Lambda_b}(\Lambda_b \to \Lambda V) \biggr\}\ ,
\end{multline} 
where the intermediate amplitudes, $\mathcal{M}^{T,P}_{\Lambda_b}(\Lambda_b 
\to \Lambda V)$, are,
\begin{eqnarray}
\mathcal{M}^{T,P}_{\Lambda_b}(\Lambda_b \to \Lambda V)= V_{ckm}^{T,P} 
A^{T,P}_{V}(a_{i})\ . 
\end{eqnarray}  
The CKM matrix elements, $V_{ckm}^{T}(V_{ckm}^{P})$, read as $V_{ub}V^{\star}_{us}
(V_{tb}V^{\star}_{ts})$ 
and $V_{cb}V^{\star}_{cs}(V_{tb}V^{\star}_{ts})$, in case of $\Lambda_b \to 
\Lambda \rho$, $\Lambda_b \to 
\Lambda \omega$ and  $\Lambda_b \to \Lambda J/\psi$, respectively. $f_V$ and $G_F$ are
the decay constant of vector meson, $V$ and Fermi constant. The  amplitude, $A^{T,P}_{V}(a_{i})$, 
expressed in term of Wilson coefficients, are the tree and
penguin electroweak amplitudes  which respect quark interactions in  $\Lambda_b$-decay. They are 
listed in Section 4.6. Finally, the baryonic matrix element, 
${\langle \Lambda | \bar{s} \gamma_{\mu} (1-\gamma_{5}) b| \Lambda_b 
\rangle}_{(\lambda_{\Lambda},\lambda_V)} (\equiv \mathcal{B}_{(\lambda_{\Lambda},
\lambda_V)}^{\Lambda_b})$, 
depending on the helicity state, $(\lambda_{\Lambda},\lambda_V)$, reads as,
\begin{equation}\label{eq18}
  \mathcal{B}_{(\lambda_{\Lambda},
\lambda_V)}^{\Lambda_b}
   = \left\{\,\,
   \begin{array}{ll}
    {\displaystyle -\frac{P_V}{E_V} \Biggl( \frac{m_{\Lambda_b}+m_{\Lambda}}
{E_{\Lambda}+m_{\Lambda}} \mathcal{\zeta}^-(\omega) + 2 \mathcal{\zeta}_2(\omega) \Biggl)} \,; & 
(\lambda_{\Lambda},\lambda_V)=(\frac12,0)\ ,
        \\[0.4cm] 
    {\displaystyle  \frac{1}{\sqrt{2}} \Biggl( \frac{P_{V}}
{E_{\Lambda}+m_{\Lambda}}\mathcal{\zeta}^-(\omega)+   \mathcal{\zeta}^+(\omega) \Biggl)} \,; & 
(\lambda_{\Lambda},\lambda_V)=(-\frac12,-1)\ ,
         \\[0.4cm] 
    {\displaystyle \frac{1}{\sqrt{2}}  \Biggl( \frac{P_{V}}
{E_{\Lambda}+m_{\Lambda}}\mathcal{\zeta}^-(\omega) - \mathcal{\zeta}^+(\omega) \Biggl)   } \,; & 
(\lambda_{\Lambda},\lambda_V)=(\frac12,1)\ ,
        \\[0.4cm] 
     {\displaystyle  \Biggl(\mathcal{\zeta}^+(\omega) + \frac{P^2_{V}}
{E_V (E_V+m_{\Lambda})} \mathcal{\zeta}^-(\omega) \Biggl)  } \,; & (\lambda_{\Lambda},
\lambda_V)=(-\frac12,0)\ ,
       \\ [0.4cm]      
   \end{array}\right.
\end{equation}
where, $P_V$ and $E_V$ are the momentum and energy of $V$ in the rest frame of $\Lambda_b$.
 The form factors $\mathcal{\zeta}^{\pm}(\omega)=\mathcal{\zeta}_1(\omega)\pm \mathcal{\zeta}_2(\omega)$ are 
defined for convenience. The baryonic matrix element calculated in terms of the form factors, 
$\mathcal{\zeta}_i(\omega)$, can be converted in terms of the 
form factors, $F_i(\omega)$ and/or $G_i(\omega)$ as defined in Eq.~(\ref{eq2}). As a function 
of $F_i(\omega)$, the form factors $\zeta_i(\omega)$ are written as follows,
\begin{align}
\mathcal{\zeta}_1(\omega)&=\frac12 \Biggl[2 F_1(\omega)+ F_2(\omega)+ F_3(\omega) 
\biggl(1+\frac{m_{\Lambda_{b}}}{m_{\Lambda}} \biggr)\Biggl]\ , \nonumber \\
\mathcal{\zeta}_2(\omega)&=\frac{F_2(\omega)}{2}\ ,
\end{align}
and they are plotted in Fig.~\ref{teta12} as a function of $\omega$, the invariant velocity transfer. Results
are similar to those given in Ref.~\cite{Huang:1998ek} in case of $\Lambda_b \to \Lambda J/\psi$ 
but they are different in case of $\Lambda_b \to \Lambda \rho^0$ or  $\Lambda_b \to \Lambda \omega$.

In our numerical analysis, the experimental values of the mass, 
$m_V$, and the width, $\Gamma_V$, of the vector meson, $V$,
will be used to generate the vector mass resonance, $m_{vr}$. Therefore, 
the hadronic matrix element,  $\langle  \Lambda(p_{\Lambda},s_{\Lambda}) V (p_V,s_V)| \mathcal{H}_{eff} 
|\Lambda_b(p_{\Lambda_b},s_{\Lambda_b})\rangle$ will be modified as  follows:  
\begin{multline}
\langle  \Lambda(p_{\Lambda},s_{\Lambda}) V(p_V,s_V) | \mathcal{H}_{eff} 
|\Lambda_b(p_{\Lambda_b},s_{\Lambda_b})\rangle = 
\frac{m_V \Gamma_V}{m_{vr}^2-m^2_V + i \Gamma_V m_V}\times  \\
\langle m_V |J_2|0\rangle 
\langle \Lambda(p_{\Lambda},s_{\Lambda})|J_1|\Lambda_b(p_{\Lambda_b},
s_{\Lambda_b})\rangle \biggl[ 1 
+ \sum_n r_n \alpha_s^n + 
\mathcal{O}(\Lambda_{QCD}/m_b)\biggr]\ ,
\end{multline}
where $J_i$ denote the quark currents and $r_n$ refers to the radiative corrections
in $\alpha_s$.
\noindent
Thus, the branching ratio, $\mathscr{BR}(\Lambda_b  \rightarrow V \Lambda)$, will read as,
\begin{multline}\label{eq4.1567}
\mathscr{BR}(\Lambda_b  \rightarrow V \Lambda) \propto
\Bigg| \frac{m_V \Gamma_V}{m_{vr}^2-m^2_V + i \Gamma_V m_V} \sum_{\lambda_{\Lambda},\lambda_V}
\mathcal{B}_{(\lambda_{\Lambda},\lambda_V)}^{\Lambda_b}
 \bigg[\mathcal{M}^T_{\Lambda_b}(\Lambda_b \to \Lambda V)- \\
  \mathcal{M}^P_{\Lambda_b}(\Lambda_b \to \Lambda V)    \Bigg|^{2}\ .
\end{multline}
Furthermore, when we calculate the branching ratio for 
$\Lambda_b \rightarrow \rho^{0} \Lambda$,  we should also take into account 
the $\rho^0-\omega$ 
mixing contribution  since  we are working to the first order of  
isospin violation. The application is straightforward and we obtain:
\begin{multline}\label{eq4.15}
\mathscr{BR}(\Lambda_b  \rightarrow \rho^{0} \Lambda) \propto
\Bigg| 
\sum_{\lambda_{\Lambda},\lambda_V}
\mathcal{B}_{(\lambda_{\Lambda},\lambda_V)}^{\Lambda_b} \biggl\{
\frac{m_{\rho^0} \Gamma_{\rho^0}}{m_{vr}^2-m^2_{\rho^0} + i \Gamma_{\rho^0} m_{\rho^0}}
 \bigg[\mathcal{M}^T_{\Lambda_b}(\Lambda_b \to \Lambda \rho^0)- \\
  \mathcal{M}^P_{\Lambda_b}(\Lambda_b \to \Lambda \rho^0)  \bigg]    
+\frac{m_{\omega} \Gamma_{\omega}}{m_{vr}^2-m^2_{\omega} + i \Gamma_{\omega} m_{\omega}}
\bigg[\mathcal{M}^T_{\Lambda_b}(\Lambda_b \to \Lambda \omega)- \\
  \mathcal{M}^P_{\Lambda_b}(\Lambda_b \to \Lambda \omega)  \bigg]
\frac{
\tilde{\Pi}_{\rho \omega}}{(s_{\rho}-m_{\omega}^{2})+im_{\omega}
\Gamma_{\omega}}\biggr\}  \Bigg|^{2}\ .
\end{multline}

\subsection{Operator Product Expansion}
%

The Operator Product Expansion (OPE)~\cite{Buras:1999rb, Buras:1998ra, 
Buchalla:1996vs} is  used to separate the calculation of a baryonic decay 
amplitude, into two distinct physical regimes.
One is called {\it hard} or short-distance physics, represented by Wilson 
Coefficients  and the other is called {\it soft} or long-distance physics. This part is described 
by $O_{i}(\mu)$, and is derived by using a non-perturbative approach. The operators, $O_{n}$, 
entering from the Operator Product Expansion (OPE) to reproduce 
the weak interaction of quarks, can be understood as local operators 
which govern a given decay. They can be written, in a generic form, as,
\begin{equation}\label{eq3.2}
O_{n} = ({\bar q}_{i} \Gamma_{n1} q_{j})({\bar q}_{k} \Gamma_{n2} q_{l})\ ,
\end{equation}
where $\Gamma_{ni}$ denotes a combination of gamma matrices. They  should 
respect the Dirac structure, the
colour structure and the type of quark relevant for the  decay being studied. 
Two kinds of topology contributing 
to the decay can be defined: there is the tree diagram of which  the 
operators are  $O_{1}, O_{2}$ and the 
penguin diagram expressed by the operators  $O_{3}$ to $O_{10}$. The 
operators related to these diagrams mentioned previously are the following,
\begin{align}
O_{1}^{u}& = \bar{q}_{\alpha} \gamma_{\mu}(1-\gamma{_5})u_{\beta}
\bar{u}_{\beta} \gamma^{\mu}(1-\gamma{_5})
b_{\alpha}\ , & \!\!O_{2}^{u}& = \bar{q} \gamma_{\mu}(1-\gamma{_5})u\bar{u} 
\gamma^{\mu}(1-\gamma{_5})b\ ,  \nonumber \\
O_{3}& = \bar{q} \gamma_{\mu}(1-\gamma{_5})b \sum_{q\prime}\bar{q}^{\prime}
\gamma^{\mu}(1-\gamma{_5})
q^{\prime}\ , & \!\!O_{4}& =\bar{q}_{\alpha} \gamma_{\mu}(1-\gamma{_5})b_{\beta}
\sum_{q\prime}\bar{q}^{\prime}_{\beta}\gamma^{\mu}(1-\gamma{_5})
q^{\prime}_{\alpha}\ , \nonumber \\
O_{5}& =\bar{q} \gamma_{\mu}(1-\gamma{_5})b \sum_{q'}\bar{q}^
{\prime}\gamma^{\mu}(1+\gamma{_5})q^{\prime}\ , & \!\!O_{6}& =\bar{q}_{\alpha} 
\gamma_{\mu}(1-\gamma{_5})b_{\beta}
\sum_{q'}\bar{q}^{\prime}_{\beta}\gamma^{\mu}(1+\gamma{_5})
q^{\prime}_{\alpha}\ ,  \nonumber \\
O_{7}& =\frac{3}{2}\bar{q} \gamma_{\mu}(1-\gamma{_5})b \sum_{q'}e_{q^{\prime}}
\bar{q}^{\prime} \gamma^{\mu}(1+\gamma{_5})q^{\prime}\ , & \!\!O_{8}& 
=\frac{3}{2}\bar{q}_{\alpha}
\gamma_{\mu}(1-\gamma{_5})b_{\beta}
\sum_{q'}e_{q^{\prime}}\bar{q}^{\prime}_{\beta}\gamma^{\mu}(1+\gamma{_5})
q^{\prime}_{\alpha}\ , \nonumber \\
O_{9}& =\frac{3}{2}\bar{q} \gamma_{\mu}(1-\gamma{_5})b \sum_{q'}e_{q^{\prime}}
\bar{q}^{\prime} \gamma^{\mu}(1-\gamma{_5})q^{\prime}\ , & \!\!O_{10}& =
\frac{3}{2}\bar{q}_{\alpha}
 \gamma_{\mu}(1-\gamma{_5})b_{\beta} \sum_{q'}e_{q^{\prime}}
\bar{q}^{\prime}_{\beta}
\gamma^{\mu}(1-\gamma{_5})q^{\prime}_{\alpha}\ .
\end{align}

In the above equations, $\alpha$ and $\beta$ are the colour indices. $e_{q}$ 
denotes the quark electric charge and $q^{\prime}$, the quarks 
$(u,d,c,s)$ which may contribute in the penguin loop.
%

\subsection{Wilson coefficients}

The Wilson coefficients~\cite{Buras:1998ra}, $C_{i}(\mu)$, represent the
physical contributions from scales higher than $\mu$ (of the order 
of $O(m_{b})$ in $b$-quark decay) and 
since  QCD has the property of  asymptotic freedom, they can be calculated in perturbation theory. 
They include contributions of  all heavy particles,
and  are calculated to the next-to-leading order (NLO) in such a way that 
one can get some  corrections $O(\alpha_{s})$ from 
the leading-log-order (LO). By definition, $C(\mu)$
 (we remove for convenience the index $i$) is given by~\cite{Buras:1999rb, 
Buras:1998ra, Buchalla:1996vs},
\begin{equation}\label{eq3.3}
C(\mu)= U(\mu,M_{W})C(M_{W})\ ,
\end{equation}
where $U(\mu,M_{W})$  describes the QCD evolution and  reads as,
\begin{equation}\label{eq3.4}
U(\mu,M_{W})= \biggl[ 1+ \frac{\alpha_{s}(\mu)}{4 \pi}J \biggr] \biggl[ 
\frac{\alpha_{s}(M_{W})}{\alpha_{s}(\mu)}
\biggr]^{d} \biggl[1- \frac{\alpha_{s}(M_{W})}{4 \pi}J \biggr] \ ,
\end{equation}
with $J$ the matrix element including the leading order and  the next-to-leading 
 order corrections. $d$ is  the anomalous dimension. The final 
expression for $C(\mu)$ in the NLO,
with $U^{0}(\mu,M_{W})=  \bigl(\alpha_{s}(M_{W})/\alpha_{s}(\mu)\bigr)^{d}$  
is,
\begin{equation}\label{eq3.5}
C(\mu)= \biggl[ 1+ \frac{\alpha_{s}(\mu)}{4 \pi}J \biggr] U^{0}(\mu,M_{W})
 \biggl[1+ \frac{\alpha_{s}(M_{W})}{4 \pi} (B-J)\biggr]  \ ,
\end{equation}
where $B$ is a constant term which depends on the factorization scheme. 
To be consistent, the matrix elements of the operators, $O_{i}$, should 
also be renormalized to the one-loop
order. This results in the effective Wilson coefficients, $C_{i}^{\prime}$, 
which  satisfy the constraint,
\begin{eqnarray}\label{eq3.6}
C_{i}(m_{b})\langle O_{i}(m_{b})\rangle=C_{i}^{\prime}{\langle O_{i}
\rangle}^{tree}\ ,
\end{eqnarray}
where ${\langle O_{i}\rangle}^{tree}$ are the matrix elements at the tree 
level. These matrix elements   will be
 evaluated
in the factorization approach. From  Eq.~(\ref{eq3.6}), the relations 
between $C_{i}^{\prime}$ and $C_{i}$
are~\cite{Deshpande:1995pw, Fleischer:1994gr, Fleischer:1993gp, 
Fleischer:1997bv},
\begin{align}\label{eq3.7}
C_{1}^{\prime}& =C_{1}\ ,\; \nonumber &
C_{2}^{\prime}& =C_{2}\ , \nonumber \\
C_{3}^{\prime}& =C_{3}-P_{s}/3\ ,\; \nonumber &
C_{4}^{\prime}& =C_{4}+P_{s}\ , \nonumber \\
C_{5}^{\prime}& =C_{5}-P_{s}/3\ ,\; \nonumber &
C_{6}^{\prime}& =C_{6}+P_{s}\ , \nonumber \\
C_{7}^{\prime}& =C_{7}+P_{e}\ ,\; \nonumber &
C_{8}^{\prime}& =C_{8}\ , \nonumber \\
C_{9}^{\prime}& =C_{9}+P_{e}\ ,\;  &
C_{10}^{\prime}& =C_{10}\ ,
\end{align}
where
\begin{eqnarray}
P_{s}  =\frac{\alpha_{s}}{8\pi} C_{2}\biggl(\frac{10}{9}+G(m_{c},\mu,q^{2})
\biggr)\ , \;{\rm and}\;\;
P_{e}  =\frac{\alpha_{em}}{9\pi}(3C_{1}+C_{2})\biggl(\frac{10}{9}+
G(m_{c},\mu,q^{2})\biggr)\ ,
\end{eqnarray}
with
\begin{eqnarray*}
 G(m_{c},\mu,q^{2})=4\int_{0}^{1}dx \; x(x-1){\rm ln} \frac{m_{c}^{2}
-x(1-x)q^{2}}{\mu^{2}}\ .
\end{eqnarray*}
Here $q^{2}$ is   the typical  momentum transfer of the gluon or photon 
in the penguin diagrams and
 the expression of $G(m_{c},\mu,q^{2})$ can be found in 
Ref.~\cite{Kramer:1994yu},
Finally, the values for $C_{i}^{\prime}$  are given in Table~1 where we 
have  taken  $\alpha_{s}(m_{Z})=0.112$, 
$\alpha_{em}(m_{b})=1/132.2$,  $m_{b}=4.9$ GeV, and $m_{c}=1.35$ GeV.

%
\subsubsection{Explicit Wilson coefficient amplitudes for $
\boldsymbol{\Lambda_b \to \Lambda V}$}
%
Finally, in the following one lists  the tree and penguin amplitudes 
which appear in the given transition:
\newline
\noindent for the decay  $\Lambda_b \rightarrow \Lambda J/\Psi$,
\begin{multline}
A^{T}_{J/\Psi}(a_{1},a_{2})  = a_1\ ,
\hspace{26.75em}
\end{multline}
\begin{multline}
A^{P}_{J/\Psi}(a_{3},  \cdots,   a_{10})  = a_3+a_5+a_7+a_9\ ;
\hspace{19em}
\end{multline}
\noindent for the decay  $\Lambda_b \rightarrow \Lambda \omega$, 
\begin{multline}
\sqrt{2}A^{T}_{\omega}(a_{1},a_{2})  = a_1\ ,
 \hspace{26.2em}
\end{multline}
\begin{multline}
\sqrt{2}A^{P}_{\omega}(a_{3},  \cdots,   a_{10})  = \frac12 \Bigl( 
4(a_3+a_5)+a_7+a_9\Bigr)\ ;
\hspace{15em}
\end{multline}
\noindent for the decay  $\Lambda_b \rightarrow \Lambda \rho^{0}$, 
\begin{multline}
\sqrt{2}A^{T}_{\rho^0}(a_{1},a_{2})  =a_1\ ,
 \hspace{26.0em}
\end{multline}
\begin{multline}\label{eq5.21}
\sqrt{2} A^{P}_{\rho^0}(a_{3},  \cdots,   a_{10})  =\frac32 (a_7+a_9)\ ;
\hspace{20em}
\end{multline}
%

\subsection{$\boldsymbol{\rho^{0}-\omega}$ mixing scheme}

The direct  $CP$  violating asymmetry parameter, $a_{CP}$, integrated 
over all the available range of energy of the $\pi^+ \pi^-$ invariant mass is 
found to be small for most of
the non-leptonic exclusive $b$-decays when either the factorization framework
is applied. However,  it appears that the asymmetry may be large in the vicinity 
of a given resonance, namely the $\omega$ meson in our case. To obtain a  large signal
for direct  $CP$  violation requires some mechanism to make both 
$\sin\delta$  and  $r$ large.
We stress that   $\rho^0-\omega$ mixing has the dual advantages that 
the strong phase difference
is large (passing rapidly through $90^{o}$ at the $\omega$ resonance) 
and well known~\cite{Enomoto:1996cv,Gardner:1998yx,
Guo:1998eg,Guo:1999ip,Gardner:1998za,Gardner:1997qk}.
In the vector meson dominance model~\cite{Sakurai:1969ju}, the photon 
propagator is dressed by coupling
to  the vector mesons $\rho^0$ and $\omega$. In this regard, the 
$\rho^0-\omega$ mixing mechanism~\cite{O'Connell:1997wf,O'Connell:1996ns} 
has been developed. Let $A$ be the
amplitude for the decay $\Lambda_b \rightarrow \rho^{0} ( \omega ) 
\Lambda \; \rightarrow  
\pi^{+}  \pi^{-} \; \Lambda$, then one has,
\begin{equation}\label{eq35}
A=\langle  \Lambda \; \pi^{-} \pi^{+}|H^{T}|\Lambda_b \rangle + \langle  
\Lambda \;  
\pi^{-} \pi^{+}|H^{P}|\Lambda_b  \rangle\ ,
\end{equation}
with $H^{T}$ and $H^{P}$ being the Hamiltonians for the tree and 
penguin operators. We can define the relative magnitude and phases between these two 
contributions as follows,
\begin{align}\label{eq36}
A &= \langle  \Lambda \;  \pi^{-} \pi^{+}|H^{T}| \Lambda_b \rangle [ 
1+re^{i\delta}e^{i\phi}]\ , \nonumber \\
\bar {A} &= \langle \overline{\Lambda} \;  \pi^{+}  \pi^{-}|H^{T}|
\bar{\Lambda}_b \rangle [ 1+re^{i\delta}e^{-i\phi}]\ ,
\end{align}
where $\delta$ and $\phi$ are strong and weak phases, respectively. 
The phase $\phi$ arises
from the appropriate combination of CKM matrix elements. In case of $b 
\to d$ or $b \to s$
transitions, $\phi$ is given by $ \phi={\rm arg}[(V_{tb}
V_{td}^{\star})/(V_{ub}V_{ud}^{\star})]$
or  ${\rm arg}[(V_{tb}V_{ts}^{\star})/(V_{ub}V_{us}^{\star})]$, 
respectively. As a result,
$\sin \phi$ is equal to $\sin \alpha \; (\sin \gamma)$ for $b \to d 
\; (b \to s)$,  with
$\alpha \; (\gamma)$ defined in the standard way~\cite{Eidelman:2004wy}.

\noindent Regarding the parameter, $r$, it represents  the
absolute value of the ratio of tree and penguin amplitudes:
\begin{equation}\label{eq39}
r \equiv \left| \frac{\langle \rho^{0}(\omega) \Lambda|H^{P}|\Lambda_b 
\rangle}{\langle\rho^{0}(\omega)
\Lambda|H^{T}|\Lambda_b \rangle} \right|.
\end{equation}

\noindent With this mechanism, to first  order 
in  isospin violation, we have the following results 
when the invariant mass of $\pi^{+}\pi^{-}$ is near the 
$\omega$ resonance mass,
\begin{align}\label{eq40}
\langle \Lambda \pi^{-} \pi^{+}|H^{T}|\Lambda_b  \rangle & = 
\frac{g_{\rho}}{s_{\rho}s_{\omega}}
 \tilde{\Pi}_{\rho \omega}t_{\omega} +\frac{g_{\rho}}
{s_{\rho}}t_{\rho}\ , \nonumber \\
\langle  \Lambda \pi^{-} \pi^{+}|H^{P}|\Lambda_b  \rangle & = 
\frac{g_{\rho}}{s_{\rho}s_{\omega}} 
\tilde{\Pi}_{\rho \omega}p_{\omega} +\frac{g_{\rho}}
{s_{\rho}}p_{\rho}\ .
\end{align}
Here $t_{V} \; (V=\rho \;{\rm  or} \; \omega) $ is the tree 
 amplitude
and $p_{V}$ the penguin  amplitude for producing a vector meson,
$V$, $g_{\rho}$ is the coupling for $\rho^{0} \rightarrow \pi^{+}\pi^{-}$, 
$\tilde{\Pi}_{\rho \omega}$
is the effective $\rho-\omega$ mixing amplitude, and $s_{V}$  is  from the 
inverse  propagator
of the vector meson $V$,  $s_{V}=s-m_{V}^{2}+im_{V}\Gamma_{V}$ 
(with $\sqrt s$  the
invariant mass of the $\pi^{+}\pi^{-}$ pair). We stress that the direct 
coupling $ \omega
\rightarrow \pi^{+} \pi^{-} $ is effectively absorbed into
$\tilde{\Pi}_{\rho \omega}$~\cite{O'Connell:1994uc,Maltman:1996kj,
O'Connell:1997xy,Williams:1998nj,Gardner:1998ta},
leading  to the explicit $s$ dependence of $\tilde{\Pi}_{\rho \omega}$. 
Making the expansion
$\tilde{\Pi}_{\rho \omega}(s)=\tilde{\Pi}_{\rho \omega}(m_{\omega}^{2})
+(s-m_{w}^{2})
\tilde{\Pi}_{\rho \omega}^{\prime}(m_{\omega}^{2})$, the  $\rho^{0}-\omega$ 
mixing parameters
 were determined in the fit of Gardner and 
O'Connell~\cite{Gardner:1998ie}: we will use $\Re e \;
\tilde{\Pi}_{\rho \omega}(m_{\omega}^{2})=-3500 \pm 300 \; {\rm MeV}^{2}, 
\;\;\; \Im m \;
\tilde{\Pi}_{\rho \omega}(m_{\omega}^{2})= -300 \pm 300 \; {\rm MeV}^{2}$, 
and  $\tilde{\Pi}_{\rho \omega}^{\prime}
(m_{\omega}^{2})=0.03 \pm 0.04$. In practice, the effect of the derivative 
term is negligible.

\section{General inputs}

\subsection{Source of $\boldsymbol{CP}$ violation: CKM matrix}\label{part3.1}
%
In most  phenomenological applications, the widely used  CKM matrix  parametrization is 
the {\it Wolfenstein parametrization}~\cite{Wolfenstein:1983yz,Wolfenstein:1964ks}. 
One of the main advantages in comparison with the standard 
parametrization~\cite{Hocker:2001xe,Chau:1984fp}
is  its easy  analytical derivation at any order of $\lambda$. a
Four independent parameters, $\lambda, A, \rho$ and $\eta$, are usually used to describe 
the CKM matrix and each of these parameters can be (in)directly measured experimentally. 
By expanding each element of the CKM matrix  as a power series in the parameter $\lambda =
 \sin \theta_{c} = 0.2224$ ($\theta_{c}$ is the Gell-Mann-Levy-Cabibbo angle), 
and going beyond the leading order  in terms of $\lambda$ 
in a perturbative expansion of  the CKM matrix, it is found that the CKM matrix takes
 the following form (up to corrections of $O(\lambda^7)$):  
\begin{equation}\label{eq60}
{\hat V}_{CKM}= \left( \begin{array}{ccc}
1-\frac{1}{2} \lambda^{2}- \frac18 \lambda^4 &  \lambda                    & 
A\lambda^{3}(\rho-i\eta) \\
-\lambda + \frac12 A^2 \lambda^5 (1- 2 (\rho + i \eta))    & 1-\frac{1}{2}\lambda^{2}
-\frac18 
\lambda^4 (1+4 A^2)    & A\lambda^{2}             \\
A\lambda^{3}(1-\bar{\rho}-i\bar{\eta})& -A\lambda^{2}+ \frac12 A \lambda^4 (1- 2 (\rho 
+ i \eta)) &   
   1-\frac12 A^2 \lambda^4    \\
\end{array}  \right)\ ,
\end{equation}
where
\begin{equation}\label{eq61}
\bar{\rho}= \rho \biggl(1-\frac{\lambda^2}{2}\biggr)\;\; {\rm and} \;\;\bar{\eta}= 
\eta \biggl(1-\frac{\lambda^2}{2}\biggr)\ .
\end{equation}
and  $\eta$  plays the well-known role of the  $CP$-violating phase in the Standard Model
 framework. From the CKM matrix, expressed in terms of the Wolfenstein parameters and 
constrained with
several experimental data,  we will take in our numerical applications, in case of 
$95\%$ confidence level,
\begin{equation}\label{eq65}
 0.076 < \rho < 0.380  \;\; {\rm and }\;\; 0.280< \eta <0.455\ .
\end{equation}
The values for $A$ and $\lambda$ are assumed to be well determined 
experimentally:
\begin{equation}\label{eq66}
\lambda=0.2265  \;\; {\rm and }\;\; A=0.801\ .
\end{equation}
%

\subsection{Input physical parameters}
%
%
\subsubsection{Quark and hadron  masses}\label{part3.2}
%
%
The constituent quark masses are used in order to calculate the electroweak form factor 
transitions 
between baryons and one has (in GeV),
\begin{eqnarray}\label{eq68}
m_{u}=  m_{d}= 0.350\ ,\; m_{s}= 0.500\ ,\; m_{b}= 4.900\ ,\; m_{c}= 1.300\ .
\end{eqnarray}
For hadron masses, we shall use the following values (in  GeV):
\begin{align}
 m_{\Lambda_b}& =5.624 \ ,  & m_{\Lambda}& =1.115\ , & m_{J/\psi}   &=3.096\ , \nonumber\\
m_{\rho^{0}}& = 0.769\ ,   & m_{\omega} &  = 0.782\ .& 
\end{align}
%
%
\subsubsection{Form factors and decay constants}
%
%
The baryon heavy-to-light form factors, $F_{i}(k^{2})$ and $G_{i}(k^{2})$,
depending on the inner structure of  the hadrons have been calculated in Section 4.

The decay constants for vector mesons, $f_{V}$,  do not 
suffer from uncertainties as large as those for form factors 
since they are well determined experimentally from leptonic and semi-leptonic decays. 
Let us first recall the usual definition for  a vector meson,
\begin{eqnarray}\label{eq73}
c \langle V(q) | \bar{q}_{1} \gamma_{\mu} q_{2} | 0 \rangle & = f_{V} m_{V} \epsilon_{V}\ ,
\end{eqnarray}
where  $m_{V}$ and $\epsilon_{V}$ are respectively the 
mass and polarization 4-vector of the vector meson, and $c$ is a constant depending 
on the given meson: $c=\sqrt{2}$ for the $\rho^0$ and $\omega$ and $c=1$ otherwise. 
\noindent Numerically, in our calculations, for the decay constants we  take
 (in MeV),
\begin{eqnarray}\label{eq75}
   f_{\rho^0}  = 209\ ,  \;\;   f_{\omega}=187\ ,  \;\; f_{J/\psi}  =400\ .
\end{eqnarray}
Finally, for the total $\Lambda_b$ decay width, $\Gamma_{\Lambda_b}(= 1/\tau_{\Lambda_b})$, 
we use $\tau_{\Lambda_b}  = 1.229 \pm 0.080 \; {\rm ps}$.

\section{Physical results and simulations}

According to the preceding sections, all the essential parameters to perform precise 
simulations can be  determined. We will  particularly focuss on the $\Lambda$ 
polarization, ${\mathcal P}^{\Lambda}$, the PDM element, ${\rho}^V_{00}$, 
of the vector-meson, the helicity and $CP$ violating asymmetry parameters, $\alpha_{AS}^{\Lambda_b}$ and $a_{CP}$, 
of $\Lambda_b$-decay and finally the branching ratios $\mathcal{BR}(\Lambda_b \to \Lambda V)$ where $V$ holds 
for $\rho^0, \omega$ and $J/\psi$.

\subsection{Branching ratios}

By means of kinematical analysis developed in Sections 2 and 3 and the factorization procedure 
developed in Section 4, it allows us to compute  the branching ratios of $\Lambda_b \to \Lambda \rho^0, 
\Lambda_b \to \Lambda \omega$ and $\Lambda_b \to \Lambda J/\psi$. The decay width of any process 
like $\Lambda_b \to \Lambda V$ is given by the following 
formula~\cite{Pakvasa:1990if}, 
 \begin{eqnarray}\label{zj}
\Gamma(\Lambda_b \to \Lambda V) 
=\Big(\frac{E_{\Lambda}+m_{\Lambda}}{m_{\Lambda_b}}\Big)\frac{P_V}{16\pi^2} 
\int_{\Omega}\Bigl| \sum_{\lambda_{\Lambda},\lambda_{V}} 
\mathcal{A}_{\lambda_{\Lambda},\lambda_{V}}(\Lambda_b \to \Lambda V) \Bigr|^2 d\Omega\ ,
\end{eqnarray}
where, in the $\Lambda_b$ rest frame, the $V$-momentum takes the following form, 
\begin{eqnarray}
|\vec{p}_V|=\frac{ \sqrt{ [m_{\Lambda_b}^{2}-(m_V+
m_{\Lambda})^{2}][m_{\Lambda_b}^{2}
-(m_V-m_{\Lambda})^{2}]}}
{2m_{\Lambda_b}}\ .
\end{eqnarray} 
In Eq.~(\ref{zj}), $E_{\Lambda}$ and $\Omega$  are, respectively, the energy of the $\Lambda$ baryon 
in the $\Lambda_b$ rest frame and the decay solid angle. The electroweak amplitude, 
$\mathcal{A}_{\lambda_{\Lambda},\lambda_{V}}(\Lambda_b \to \Lambda V)$, is given in Eq.~(\ref{eqoli}).  
In order to take into account the non-factorizable term coming from the color octet contribution, 
calculations have been performed by keeping the effective number of color, $N_c^{eff}$, to vary between 
the values 2 and 3.5. Branching ratios, $\mathcal{BR}$, have been calculated in case 
of $\Lambda_b \to \Lambda V$ where $V$ is $J/\psi, \rho^0$ 
and $\omega$ and one obtains the following results:
\begin{align}
\mathcal{BR}(\Lambda_b \to \Lambda J/\psi) = & (8.95\ , 2.79\ ,0.62\ ,0.03) \times 10^{-4}\ ,  \nonumber \\
\mathcal{BR}(\Lambda_b \to \Lambda \rho^0) = & (1.62\ , 1.89\ , 2.16\ ,2.39) \times 10^{-7}\ ,  \nonumber \\
\mathcal{BR}(\Lambda_b \to \Lambda \omega) = & (22.3\ , 4.75\ , 0.19\ ,0.64) \times 10^{-7}\ ,   
\end{align}
for $N_c^{eff}=2, 2.5, 3.0$ and $3.5$, respectively. It is worth noticing that the 
experimental branching ratio~\cite{Eidelman:2004wy}, 
$\mathcal{BR}^{exp}(\Lambda_b \to \Lambda J/\psi)=(4.7\pm2.1 \pm 1.9)\times 10^{-4} \ $ 
agrees with the theoretical predictions for $2.0 \le {N_c}^{eff} \le 3.0.$ Regarding $\Lambda_b \to \Lambda \rho^0$
and $\Lambda_b \to \Lambda \omega$, no conclusion can be drawn without any experimental data.

\subsection{Special case of  $\boldsymbol{{\rho}^0 - \omega}$ mixing }

It is well known that the $\omega$ meson decays into ${\pi}^+ {\pi}^-$ pair with a 
branching ratio of $2.2 \% \ $ and thus mixes with the ${\rho}^0$ meson by electromagnetic interaction. 
So, this mixing can be taken into account for the computation of the branching ratio $\mathcal{BR}(\Lambda_b 
\to \Lambda \rho^0)$ 
on one hand, and for comparison with the conjugate process ${\bar \Lambda}_b \to  
\bar{\Lambda} \rho^0$ 
on the other hand. This comparison will allow us to check the {\it direct CP violation} 
in the sector of beauty baryons if a notable difference is found between $\Lambda_b$-decay and its charge 
conjugated one. 

Computing the branching ratio of the
charge conjugated process requires that only the CKM matrix elements, $V_{ij}$, must be 
complex conjugated, while the intrinsic tree and penguin amplitudes are left unchanged.  
Owing to the important $\rho^0$ width, ${\Gamma}_{\rho^0} = 150 \ {\rm MeV}/c^2$, by 
comparison to the 
$\omega$ one, ${\Gamma}_{\omega} = 8 \ {\rm MeV}/c^2$, the branching ratios 
$\mathcal{BR}( \Lambda_b ({\bar \Lambda_b})
\to \Lambda ({\bar \Lambda})\rho^0(\omega))$, which matrix element is given by 
relation (112), is estimated by making use of Monte-Carlo techniques. The $\pi^+ \pi^-$ invariant 
squared mass, $s_{\rho}$, 
is generated according to a relativistic Breit-Wigner~\cite{Jackson:1964zd} where the  mass $M_R$ and 
width ${\Gamma}_R$ are
identified  with $M_{\rho}$ and ${\Gamma}_{\rho}$, respectively. For each generated invariant mass, $s_{\rho}$,  
the $CP$ violating asymmetry parameter, $a_{CP}(s_{\rho})$, between the two conjugated channels can be defined as,
\begin{eqnarray} 
a_{CP}(s_{\rho}) \ = \  \frac{\mathcal{BR}(\Lambda_b) - \mathcal{BR}(\bar \Lambda_b)}
{\mathcal{BR}(\Lambda_b)+\mathcal{BR}(\bar \Lambda_b)}\ . 
\end{eqnarray}
In Fig.~7, the $CP$ violating asymmetry is plotted according to 
$s_{\rho}$. A remarkable effect is seen in the mass interval $[720 - 820] MeV/c^2$, 
which includes the $\omega$ meson mass. Whatever is the value of ${N}^{eff}_c$, one observes a strong 
enhancement of $a_{CP}(s_{\rho})$ when the $\pi^+\pi^-$ invariant mass is in the vicinity of $780  MeV/c^2$.

In Table~\ref{tabl3result}, the numerous results obtained from the M-C simulations are gathered in case of 
$\rho^0-\omega$ mixing for $\Lambda_b \to \Lambda \rho^0(\omega) \to \Lambda \pi^+ \pi^-$ decay: 
mainly these are the mean branching ratio, $\mathcal{BR}$, the asymmetry parameter, $a_{CP}$, computed 
in the whole range of $\pi^+ \pi^-$ mass and the same asymmetry, $a_{CP}(\omega)$, limited to the $\omega$ mass 
interval. Because of $\rho^0-\omega$ mixing, the   branching ratio $\mathcal{BR}(\Lambda_b \to \Lambda \rho^0(\omega))$ 
weakly increases in comparison to that of $\mathcal{BR}(\Lambda_b \to \Lambda \rho^0)$. 
We also notice that the $a_{CP}(\omega)$
for $\Lambda_b \to \Lambda  \pi^+ \pi^-$ always reaches its maximum value when $s_{\rho}$ is in the range 
of the omega mass. However, its maximum value strongly varies  according to the effective number of color, 
${N}^{eff}_c$. Due to a lack of experimental data for the branching ratio $\mathcal{BR}(\Lambda_b \to 
\Lambda \rho^0(\omega))$, there is no way of constraining ${N}^{eff}_c$, unfortunately. In any case, 
the actual result can be seen  as a  {\it clear signal of direct $CP$ violation} between beauty baryon 
and beauty anti-baryon.

\subsection{Helicity asymmetry and angular distribution}

\subsubsection{ $\boldsymbol{\Lambda_b  \to  \Lambda  V}$ decay}

With respect to the definition of the helicity asymmetry parameter, ${\alpha}^{\Lambda_b}_{AS}$, 
given in Section 3, 
its numerical value depends on the nature of the vector-meson $V$:
\begin{eqnarray*} 
{\alpha}^{\Lambda_b}_{AS}(\Lambda \rho^0(\omega)) \ = \ 19.4 \%  \ \ , \ \
{\alpha}^{\Lambda_b}_{AS}(\Lambda J/{\psi}) \ =
\ 49.0\%\ , 
\end{eqnarray*}
which leads to a complete determination of the polar angular distribution of the $\Lambda$ 
hyperon in the
$\Lambda_b$ rest-frame.
As far as azimuthal angle ${\phi}_{\Lambda}$ is concerned and owing to the unknown value 
of the parameter ${\rho}_{+-}^{\Lambda_b}\ $, the angle ${\phi}_{\Lambda}$ is generated {\it uniformly}
in the range $[0, 2\pi]$. In Fig.~\ref{figziad2} are plotted the $\cos \theta_{\Lambda}$  and $\phi_{\Lambda}$  
distributions of the $\Lambda$-baryon in $\Lambda_b$ rest-frame, in case of $\Lambda_b \to \Lambda \rho^0(\omega)$. 
 Results are only shown in case of $\rho^0$ vector since they are very similar for $J/\psi$. 
 
\subsubsection{ $\boldsymbol{\Lambda \to  p {\pi}^-}$ decay }

From preceding relations, both polar and azimuthal angular distributions of proton and 
$\pi$  can be obtained
thanks to the complete determination of the 
$\Lambda$  polarization, ${\cal P}^{\Lambda}$, and the PDM element ${\rho}_{12}^{\Lambda}$. Again, 
these values strongly depend on the nature of the vector-meson $V$ coming from $\Lambda_b$ 
decays. One obtains the following results:
\begin{align*}
 {\cal P}^{\Lambda} =  -0.167\ , \ \   {\rho}_{12}^{\Lambda} = & 0.25  \ \   (J/\psi)\ , 
\nonumber \\
 {\cal P}^{\Lambda} =  -0.21\ , \ \   {\rho}_{12}^{\Lambda} = & 0.31  \ \    (\rho^0(\omega))\ .  
\end{align*} 
It is worthy noticing that, in the framework of the  factorization hypothesis 
previously exposed, the non-diagonal matrix element, ${\rho}_{12}^{\Lambda}$, is {\it real}, which makes 
easier the kinematics simulations. In Fig.~\ref{figziad3} are also plotted the $\cos \theta$  and 
$\phi$  distributions of the proton in $\Lambda$ rest-frame in case of  $\Lambda_b \to \Lambda \rho^0(\omega)$. 
Similar results have been obtained in case of $J/\psi$.

\subsubsection{ $\boldsymbol{V \to {\ell}^+ {\ell}^- , \ \ h^+ h^- }$ decays}

The angular distributions of lepton or pseudoscalar meson in the vector-meson rest-frame 
essentially depend on the normalized PDM element ${\rho}_{00}^V$. The latter is related to the probability 
of the vector-meson $V$ to get an helicity value ${\lambda}_2 = 0$. Its numerical values 
are ${\rho}_{00}^{J/\psi} =  0.66  \ \mathrm{and} \  {\rho}_{00}^{\rho} =  0.79$, for $J/\psi$ and 
$\rho^0$, respectively.. 

Despite the complicated form of $W_2(\theta_2,\phi_2)$ diplayed in Eq.~(\ref{zz002}), the distribution of 
the azimuthal angle $\phi_2$ will be {\it uniform} in the angular range $[0,2\pi]$. Finally, in Fig.~\ref{figziad4} are 
shown the $\cos \theta_{\pi}$  and $\cos \theta_{\mu}$ distributions  for $\Lambda_b 
\to \Lambda \rho^0(\omega) \to \Lambda \pi^+\pi^-$ and  $\Lambda_b \to \Lambda J/\psi 
\to \Lambda \mu^+ \mu^-$, respectively.

\section{Time-odd observables}

Previously, it has been shown that, in an appropriate frame related to each intermediate 
resonance, the normal component of the resonance vector-polarization is a Time-odd observable and 
its measure could be a signal of TR violation. However,
experimental determination of $P_N^{(i)} (i = \Lambda, V)$ is a difficult 
task~\cite{Byers:1967tw} and it requires high statistics data. Luckily, other 
possibilities to test TR symmetry exist. They have been developed by 
several authors and they rely on the search for Triple Product Correlations (TPC) 
among physical parameters represented by expressions 
like ${\vec a}_i \cdot ({\vec b}_j \times {\vec c}_k )$, where vectors ${\vec a}_i,
 {\vec b}_j$ and  ${\vec c}_k$ are either momentum or
 spin~\cite{Bensalem:2002pz, Aliev:2004yf, Chen:2003}. It is worth 
recalling that the spin of any particle is not a direct measurable quantity 
and, only polarization with respect to a given direction can be determined.

In the following, emphasis is put on TPC built from final kinematics variables measured 
in the $\Lambda_b$ rest-frame. Needless to say that the detailed calculations performed in this work, both 
the (geometrical) angular distributions and the (dynamical) form factors deduced 
from HQET, represent important ingredients
for precise computations of the TPC or their spectrum. In the recent years, L.~Sehgal~\cite{Sehgal:1999vg} 
and L.~Wolfenstein~\cite{Wolfenstein:1999xb} analyzed very thoroughly experimental data coming from the 
reaction $K_L^0  \to  {\pi}^+ {\pi}^- e^+ e^-$, and they interpreted some kinematics asymmetry as 
T-odd effects. However, these effects do not necessarily indicate a TR violation, provided that complementary
hypothesis (like non-conservation of $CPT$ symmetry) have to be done.

Inspired by the pioneering work of these authors, we give a special care to the analysis of 
some particular angles. First of all, we recall how the basis vectors
of the $\Lambda_b$ rest-frame (defined in Section $1$) transform under TR:
\begin{eqnarray}
\vec {e}_Z \longrightarrow  +\vec{e}_Z\ , \ \ \vec{e}_X \longrightarrow  -\vec{e}_X\ , \ \ \vec{e}_Y = 
\vec{e}_Z \times \vec{e}_X  \  \longrightarrow   -\vec{e}_Y\ . 
\end{eqnarray} 

\noindent
Azimuthal angular distributions like ${\phi}_{\Lambda}$ in  $\Lambda_b$ rest-frame, 
${\phi}_p $ in $\Lambda$ rest-frame and ${\phi}_{\ell} ({\phi}_h)$ in $V$ rest-frame are directly computed
according to the analytical expressions developed beforehand. For instance, 
let us consider the azimuthal angle ${\phi}_{\Lambda}$. Whatever is its
distribution, the two parameters $\cos {\phi}_{\Lambda} \ \mathrm{and} \  \sin {\phi}_{\Lambda}$ are both {\it even}
under TR and their proof is straightforward. According to the mathematical expressions:
\begin{eqnarray}
\vec u = \frac{\vec{p}_{\Lambda} \times \vec{e}_Z} {|\vec{p}_{\Lambda} \times \vec{e}_Z|}\ , \ \ 
\cos {\phi}_{\Lambda} =
\vec{e}_Y \cdot \vec u \ , \ \  \sin {\phi}_{\Lambda} =  \vec{e}_Z \cdot (\vec{e}_Y  \times  \vec{u})\ , 
\end{eqnarray}

\noindent
it is straightforward to see  that $\cos {\phi}_{\Lambda} \  \mathrm{and}  \  \sin {\phi}_{\Lambda}$  
do not change sign under TR. Thus, in order to detect any TR asymmetry, we must look for other 
genuine triple products which can be associated to some specific angles.

Let us define ${\vec n}_{\Lambda} \ \mathrm{and} \  \vec{n}_V$ as unit vectors which are {\it normal}
respectively to $\Lambda$  and vector-meson $V$ decay planes. Their expressions are given by:
\begin{eqnarray}
\vec{n}_{\Lambda} = \frac{\vec{p}_p \times \vec{p}_{\pi}}{|\vec{p}_p \times \vec{p}_{\pi}|}\ , \ \
\vec{n}_V = \frac{ \vec{p}_{l^+} \times  \vec{p}_{l^-}}{ | \vec{p}_{l^+} \times \vec{p}_{l^-}|}\ , \ \
 \mathrm{or} \ \  \vec{n}_V = \frac{ \vec{p}_{h^+} \times  \vec{p}_{h^-}}{ |\vec{p}_{h^+} \times 
\vec{p}_{h^-}|}\ . 
\end{eqnarray}

\noindent
Those vectors are {\it even} under TR operation and their azimuthal angles, respectively ${\phi}_{ {\vec
n}_{\Lambda}} \ \mathrm{and} \ {\phi}_{{\vec n}_V}$, will be referred to as ${\phi}_{(n_i)}$. 
Performing the same calculations as for ${\phi}_{\Lambda}$, it is shown 
that $ \cos {\phi}_{(n_i)} \  \mathrm{and} \  \sin {\phi}_{(n_i)}$ are both {\it odd} under TR. 
Their analytical expressions are given by the following relations:
\begin{eqnarray}
\vec{u}_i = \frac{ \vec{e}_Z \times \vec{n}_i}{ |\vec{e_Z} \times \vec{n}_i|}\ , \ \  \cos
{\phi}_{(n_i)} = \vec{e}_Y \cdot \vec{u}_i\ , \ \  \sin {\phi}_{(n_i)} = \vec{e}_Z \cdot  (\vec{e}_Y
\times \vec{u}_i)\ ,  \ \  \vec{n}_i = \vec{n}_{\Lambda}\  , \  \vec{n}_{V}\ ,  
\end{eqnarray}

\noindent 
and, according to the transformation of each vector by TR, both $\cos {\phi}_{(n_i)} \  \mathrm{and} \ \sin
{\phi}_{(n_i)}$ change sign under Time Reversal. An important question arises:  what could be the order of 
magnitude of these TR violation processes? The only way to
answer this question is the method of M-C simulations which provides us the 
interesting spectra of $ \cos {\phi}_{(n_i)}
 \ \mathrm{and} \  \sin {\phi}_{(n_i)}$ in order to cross-check the TR violation assumption. Exhaustive M-C studies
were performed by using all the allowed values of the input parameters of our model and they reveal 
{\it strong correlations} of the T-odd observables with the azimuthal angular ${\phi}_{\Lambda}$ distribution.

(i) By adopting the simple hypothesis of a  flat distribution for ${d\sigma}/{d{\phi}_{\Lambda}}$,
 no dissymetries are observed in any of $\cos {\phi}_{(n_i)} \ \mathrm{or} \  \sin {\phi}_{(n_i)}$
distributions. Their mean values are very close to zero and  the absolute value of the asymmetry
factor,  $AS(X)$, defined by, 
\begin{eqnarray}
 AS(X)  \ \propto  \  \int_0^1 {\frac{d\sigma}{dX}}dX \  -  \ \int_{-1}^0 {\frac{d\sigma}{dX}}dX\ ,   
\end{eqnarray}
\noindent
with $X = \cos {\phi}_{(n_i)}  \  \mathrm{or}  \  \sin {\phi}_{(n_i)} $ is  less than $0.4\% $. This
only corresponds  to statistical fluctuations.

(ii) Other solution which has not yet been exploited would be  to assign a realistic distribution for the angle 
$\phi_{\Lambda}$. Indeed, the latter directly depends  on the non-diagonal PDM element,
${\rho}_{+-}^{\Lambda_b} \ $, which, a priori, is non-equal to zero.  Departing from the
 relation given in Eq.~(11), the ${\phi}_{\Lambda}$ angular distribution can be inferred and
its analytic expression is:
\begin{eqnarray}
{d\sigma}/{d\phi}  \  \propto  \  1 + \frac{\pi}{2} {\alpha_{AS}}\Big({\Re e({\rho}_{+-}^{\Lambda_b}) \cos \phi
- \Im m({\rho}_{+-}^{\Lambda_b}) \sin \phi } \Big)\ ,
\end{eqnarray}

\noindent
where $ \phi = {\phi_{\Lambda}}; \  \ {\alpha}_{AS} =  49\% \  \ \mathrm{for} \   \ \Lambda J/{\psi} \ \ 
\mathrm{and} \ \  19.4\% \ \mathrm{for}  \  \  \Lambda \rho^0(\omega).$
Normalizing ${d\sigma}/{d\phi}$ provides a probability density function which must be positive in the whole
range of $\phi$ variation: $0 \le \phi  \le {2\pi}$. Furthermore, the (normalized) PDM elements could not
exceed $1$ in absolute value. Taking account of these kinematics constraints and adopting a conservative
point of view, we suppose that ${\Re e({\rho}_{+-}^{\Lambda_b})} \ \mathrm{and} \ {\Im m
({\rho}_{+-}^{\Lambda_b})} \ $   have similar contributions and the following choice is made:
${\Re e({\rho}_{+-}^{\Lambda_b})}  =  -{\Im m({\rho}_{+-}^{\Lambda_b})} =  {\sqrt{2}}/2$.
Interesting results concerning the asymmetry parameters, $AS(X)$,  are 
obtained in both the two channels with a sample of $10^5$ generated events for each one:
\begin{align*}
\Lambda_b \to \Lambda J/{\psi}\ ;   \ \ \ \  & AS(\cos {\phi}_{{\vec n}_{\Lambda}}) = 4.3\%\ , \;\; 
{\rm and} & \!\!\!\!\!\!  
AS(\sin
{\phi}_{{\vec n}_{\Lambda}}) = -5.5\%\ , \nonumber \\
\Lambda_b \to \Lambda {\rho}^0(\omega)\ ;   \ \ \ \  & AS(\cos {\phi}_{{\vec n}_{\Lambda}}) = 2.4\%\ , \;\; {\rm and} 
& \!\!\!\!\!\!  AS(\sin
{\phi}_{{\vec n}_{\Lambda}}) = -2.7\%\ . 
\end{align*}

\noindent
Same calculations have been performed for $\cos {\phi}_{\vec{n}_V} \ \mathrm{and} \  \sin
{\phi}_{\vec{n}_V} \  \mathrm{where} \   \vec{n}_V = \  \vec{n}_{\rho}, \ \vec{n}_{J/{\psi}}$. The
corresponding asymmetries, $AS(X)$, are close to zero: they are $\leq  0.4\% \ $ and compatible 
with statistical fluctuations. What could be the origin of these asymmetry discrepancies, 
despite the fact that the normal vectors 
${\vec n}_{\Lambda} \ \mathrm{and} \ {\vec n}_V \ $ respectively to $ \Lambda \ \mathrm{and} \ V$ decay
planes play similar role?

In our opinion, it is caused by the very difference of the azimuthal angular
distributions ${\phi}_p$ in $\Lambda$ rest-frame and ${\phi}_{\ell} ({\phi}_h)$ in $V$ rest-frame. The former
obeys to Parity violation in $\Lambda$-decay and it is {\it dissymetric} while the latter is {\it flat}
because of Parity conservation so that  it can be inferred that Parity violation in a given process 
like $\Lambda$ hyperon decay could be the origin of Time-odd effect in a subsequent process. 
Other questions arise too: what are the true values of the real and imaginary parts of 
${\rho}_{+-}^{\Lambda_b}$? Are the chosen values,  ${\Re e({\rho}_{+-}^{\Lambda_b})}  =  -{\Im m({\rho}_{+-}^{\Lambda_b})} = 
{\sqrt{2}}/2$, overestimated and be the real source  of the observed T-odd effects? Whatever the 
physical origin of this novel effect is, 
all these interrogations are raised in the framework of the Standard Model because {\it the main hypothesis 
underlying our calculations do not include any TR violation parameter}.

\section{Conclusion}\label{section6}
%
 We have studied the decay process $\Lambda_b \to \Lambda V(1^-)$ where the vector  
 meson $V$ is either $J/\psi, \rho^0$ or $\omega$, with the inclusion of 
 $\rho^0-\omega$ mixing when it was appropriate. When the invariant mass of the
  $\pi^+ \pi^-$ pair is in the vicinity of the $\omega$ resonance, it is found 
  that the $CP$ violating parameter, $a_{CP}$, reaches its maximum value. In our analysis, we have also 
  investigated the branching ratios, $\mathcal{BR}$, polarizations, $\mathcal{P}$, 
and direct $CP$ and helicity violating parameters, $a_{CP}$ and  
 $\alpha_{AS}$, respectively,   for the same channels. 

 Thanks to an intensive use of the Jacob-Wick-Jackson helicity formalism, 
  rigorous and detailed calculations  of the $\Lambda_b$-decays into one baryon and one vector meson have 
  been carried out completely. This helicity formalism allows us to clearly separate the kinematical and 
  dynamical contributions in the computation of the amplitude corresponding to 
  $\Lambda_b \to \Lambda V(1^-)$ decay. The cascade-type analysis\footnote{
  A paper quoted in~\cite{Kadeer:2005aq} and based on the same ''cascade-type'' approach has been released 
  when our paper was in preparation. The authors using Monte-Carlo program and helicity analysis have 
 analysed hyperon decays including lepton mass effects.} is indeed very useful for analysing 
  polarization properties and Time Reversal effects since the analysis of every decay 
 in the decay chain can be performed in its respective rest frame. In order to apply our 
  formalism, all the numerical simulations have been performed thanks to a Monte-Carlo method.  
  We also dealt at length with the uncertainties coming from the input parameters. In particular, 
  these include the Cabibbo-Kobayashi-Maskawa matrix element parameters, $\rho$ and $\eta$, the effective 
 number of colors, $N_c^{eff}$, coming from the factorisation scheme we have followed and the 
 phenomenological model of the baryon wave functions for $\Lambda_b$ and $\Lambda$. Moreover, the heavy 
 quark effective theory has been applied in order to estimate the various form factors, 
 $F_i(q^2)$ and $G_i(q^2)$, which usually describe  dynamics of the electroweak transition 
 between two baryons. Corrections at the order of $\mathcal{O}(1/m_b)$ have been included 
 when the form factors were computed.  In the calculation of $b$-baryon decays, we need 
 the Wilson coefficients, $C(m_b)$, for the tree and penguin operators at the scale $m_b$. 
  We worked with the renormalization scheme independent Wilson coefficients. One  of the 
 major uncertainties  is that the hadronic matrix elements for both tree and penguin operators involve 
 nonperturbative QCD. We have worked in the factorisation approximation, with $N_c$ treated as an effective 
parameter defined as $N_c^{eff}$. Although one must have some doubts about factorisation, 
 it has been pointed out that it may be quite reliable in energetic weak $b$-decays. 
 The recent experimental measurement of branching ratio for  $\Lambda_b \to \Lambda 
 J/\psi$ has confirmed the theoretical calculations we made  for a  value of $N_c^{eff}$ close to 2.5-3.5. 

As regards theoretical results for the branching ratios 
  $\Lambda_b \to \Lambda J/\psi, \  \Lambda_b \to \Lambda \rho^0$ and $\Lambda_b \to \Lambda \omega$, we 
  made a comparison with PDG for  $\Lambda_b \to \Lambda J/\psi$ where an agreement is found 
($\mathcal{BR}^{exp}(\Lambda_b \to \Lambda J/\psi)=4.7 \pm 2.1 \pm 1.9\times 10^{-6}$ and 
  $\mathcal{BR}^{th}(\Lambda_b \to \Lambda J/\psi)=3.1 \times 10^{-6}$). 
For $\Lambda_b \to \Lambda \rho^0$ and $\Lambda_b \to \Lambda \omega$, the lack of experimental results 
does not allow us to draw  conclusions. However, we made two theoretical branching ratio 
  (in unit of $10^{-7}$) predictions, from $1.6$ to $2.4$ and from $0.64$ to $2.23$ for  
  $\Lambda_b \to \Lambda \rho^0$ and $\Lambda_b \to \Lambda \omega$, respectively. We have to keep in mind 
  that, because of the difficulty in dealing with non factorizable effects associated with 
  final state interactions, which are more complex for baryon decays than meson decays,  these latter 
  results are more dependent on the effective number of colors, $N_c^{eff}$.  
 In the special case of $\rho^0-\omega$ mixing, the branching ratio 
  $\Lambda_b \to \Lambda \rho^0(\omega) \to  \Lambda \pi^+ \pi^-$,  has a mean value 
  around $2.15 \times 10^{-7}$. The direct violating $CP$ asymmetry parameter, $a_{CP}$, has also been 
  calculated and a clear signal of direct $CP$ violation between beauty baryon and beauty antibaryon 
  has been observed. We have determined a range for the maximum of direct $CP$ asymmetry, $a_{max}$, as a function of 
  the parameter, $N_c^{eff}$ and the $\pi^+ \pi^-$ invariant mass. Moreover, the mean value of the direct $CP$ 
 violating  asymmetry for $\Lambda_b \to \Lambda \rho^0(\omega) \to \Lambda \pi^+ \pi^-$, has been 
  computed as well.   

By making use  of the helicity asymmetry parameter, $\alpha_{AS}$, determined 
for  $\Lambda_b \to \Lambda J/\psi$, and $\Lambda_b \to \Lambda \rho^0(\omega)$, respectively, it 
allowed us to a complete determination of the polar angular distribution of the $\Lambda$ hyperon in the 
$\Lambda_b$ rest-frame.  In fact, the knowledge of the $\Lambda$ polarization, 
$\mathcal{P}^{\Lambda}$, which depends on the nature of the vector meson produced, in addition to that 
of the PDM elements, $\rho^{\Lambda_b}_{ij}$, made available the polar and azimuthal angular  
distributions of the proton (and pion) in the $\Lambda$ rest frame. In a similar way, the polar and 
azimuthal angular distributions of leptons and pseudo-scalar mesons in the vector meson rest-frame 
have been also computed. From weak decays analysis,  one knows that vector-polarizations of  
outgoing resonances (or some of their components
in appropriate frames) are T-odd observables. In our case, the {\it normal components} of ${\mathcal P}^{\Lambda} 
 \ \mathrm{and} \  {\mathcal P}^V $ are T-odd, respectively  and this might lead to a clear signal of TR violation.
It is also well known that Triple Product Correlations (TPC) between momentum and spin can be used 
to put in evidence TR violation. However, not all TPC's can be exploited on the experimental side, 
because essentially of the difficulties to measure 
spin(s).  In our work, we have shown that new and unsuspected observables can be measured: angular
distributions of the normal vectors to the decay planes of the intermediate resonances in the $\Lambda_b$-frame. 
We found that the magnitude of their effects is 
directly related to the $\Lambda_b$ polarization density matrix (PDM) and more precisely to the 
non-diagonal elements, $\rho^{\Lambda_b}_{+-}$, appearing in the interference terms of the decay amplitude.  
However, despite our ignorance of the real and imaginary parts of $\rho^{\Lambda_b}_{+-}$, sensible effects 
of Time Reversal asymmetry can be measured, provided the latter are not too small with respect to unity. 

In our opinion, new fields of research like direct $CP$ violation and $T$-odd observables indicating a 
possible non-conservation of Time Reversal symmetry can be investigated in the sector of beauty baryons 
and especially the $\Lambda_b$-bayons which can be highly produced  in the future hadronic machine like LHC.
In order to reach this aim, all uncertainties in our calculations still have to be decreased, for example
non-factorizable effects have to be evaluated with more accuracy. Moreover, we strongly need more numerous 
and accurate experimental data in $\Lambda_b$-decays, especially the $\Lambda_b$ polarization. 
We expect that our predictions should provide useful guidance for future investigations and urge our 
experimental colleagues to measure all the observables related to $\Lambda_b$ baryon decays, if we want 
to understand direct $CP$ violation and Time Reversal symmetry better.

%
\subsection*{Acknowledgments}
The authors are indebted to their colleagues of the LHCB Clermont-Ferrand (France) team and to the Theory
Group for their support and the interesting discussions regarding the very promising research subject of 
Time Reversal.   One of us, (O.L.) would like to thank W. Roberts for  stimulating 
discussions at the early stage of the paper. Another author, (Z.J.A.), is very indebted to J.-P. Blaizot, 
Director of the ${\mathrm{ECT}}^*$ in Trento (Italy), for his kind hospitality at the Center
and the very pleasant ambiance in which an important part of this work has been achieved. Finally, a recent 
 graphical interface, so-called JaxoDraw~\cite{Binosi:2003yf} has been used for drawing the diagrams in Fig.~2.
\newpage
%

%
\newpage
%
\section*{Figure captions} 
\begin{figure}[hpt]
\begin{center}
\includegraphics*[width=0.40\columnwidth]{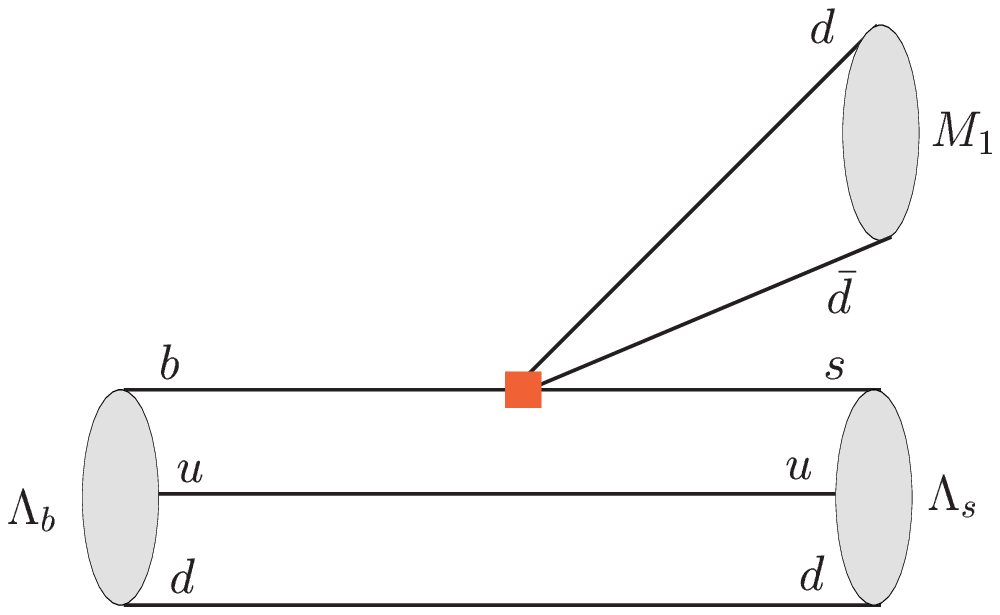}
\includegraphics*[width=0.40\columnwidth]{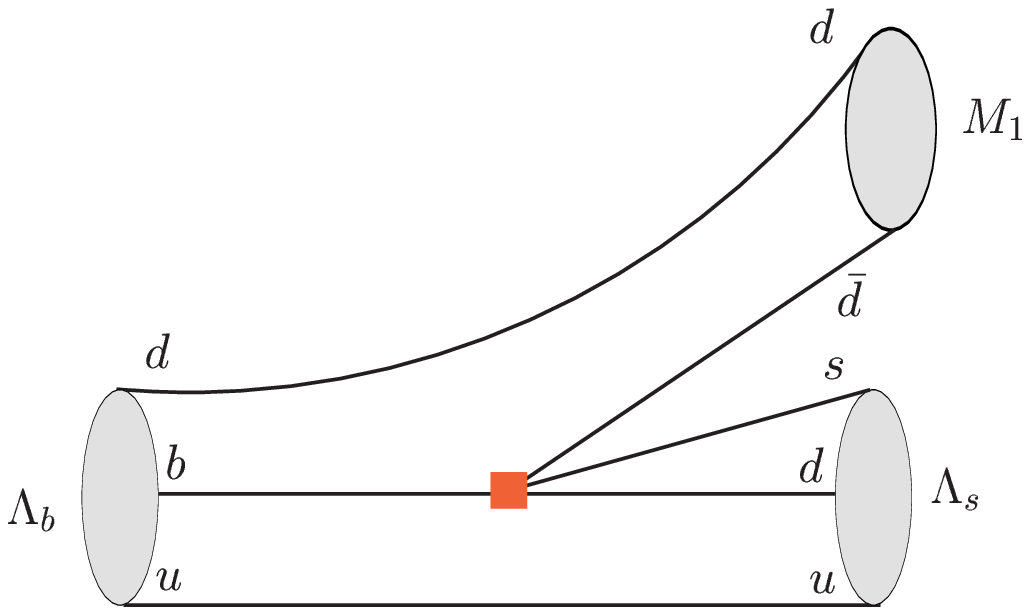}
\vskip 0.5cm
\includegraphics*[width=0.40\columnwidth]{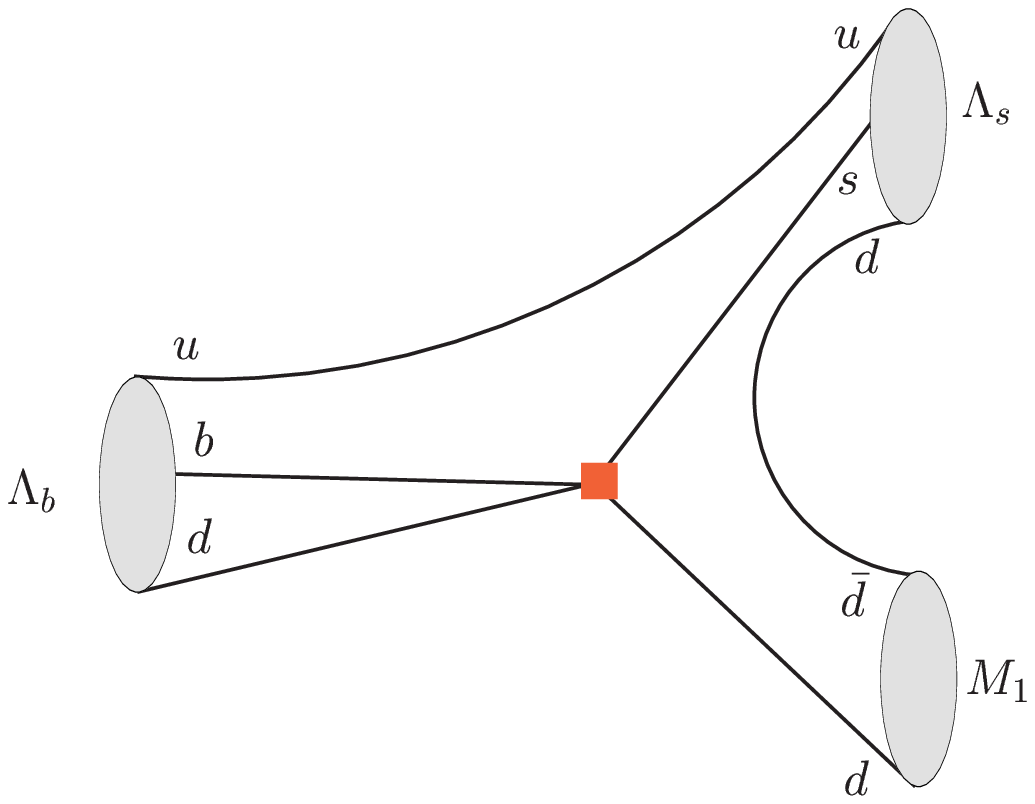}
\includegraphics*[width=0.40\columnwidth]{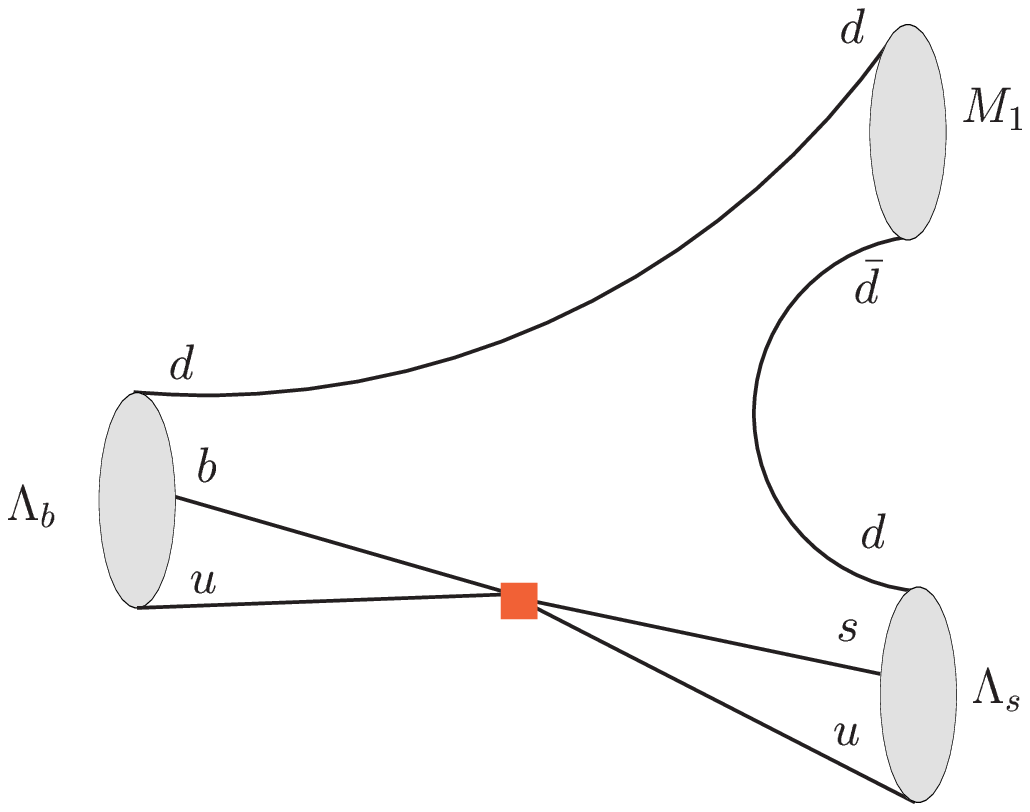}
\end{center}
\caption{From left-up to right-down, tree (T), color-commensurate (C), exchange (E) and 
bow tie (B) diagrams which are allowed in 
 transition $\Lambda_b \to \Lambda M_1.$  The squared dot represents the electroweak 
interaction in the Standard Model. It can be shown that the contribution of each diagram to the 
total amplitude follows  $T \gg C \approx E \gg B$.}
\label{figbaryons01}
\end{figure}

\begin{figure}[hpt]
\begin{center}
\includegraphics*[width=0.8\columnwidth]{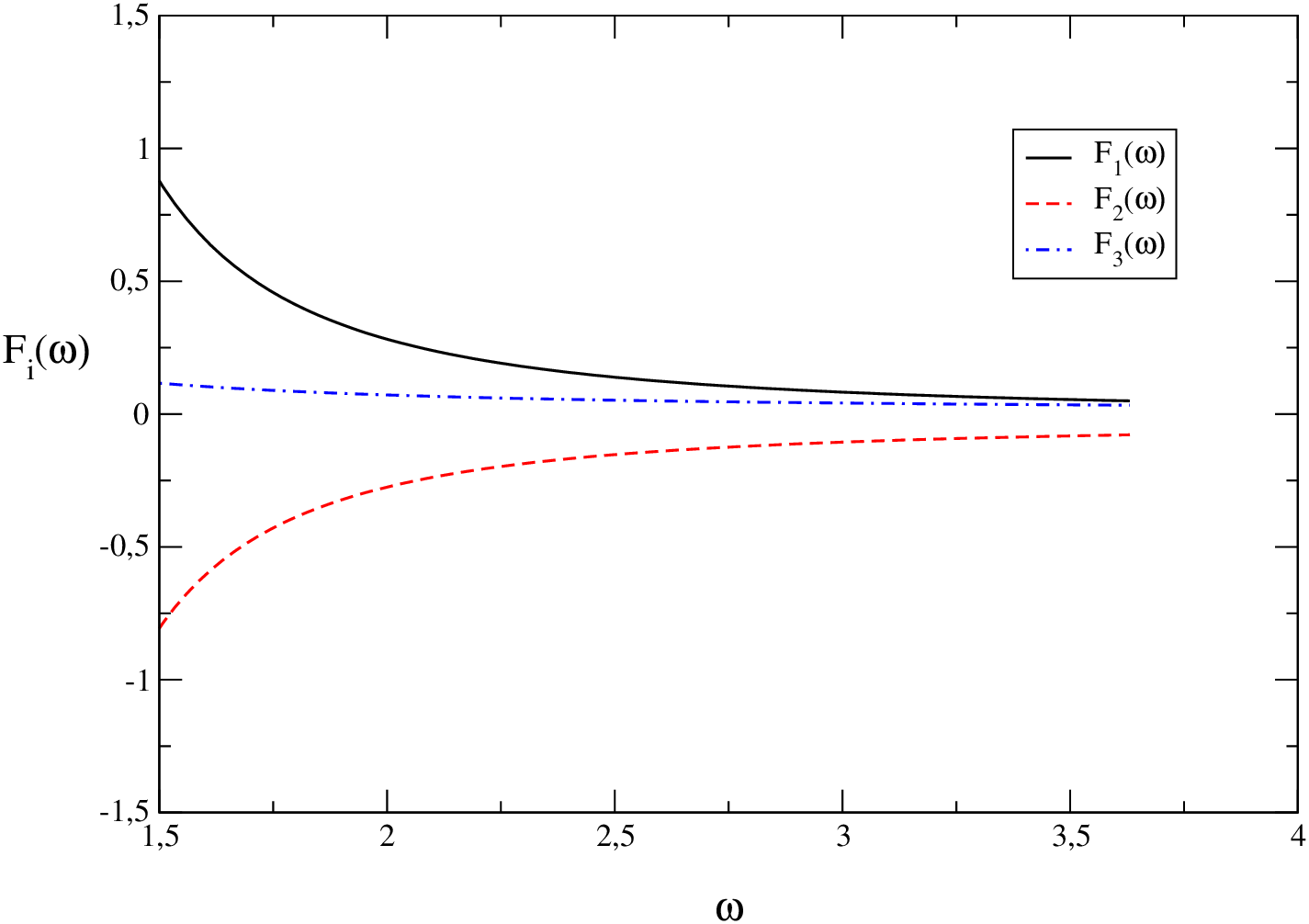}
\end{center}
\caption{Form factor distributions,  $F_i(\omega)$, for $\Lambda_b \to \Lambda V$. The full, dashed and
dot-dashed lines represent $F_1(\omega), F_2(\omega)$ and $F_3(\omega)$, 
respectively.}
\label{figffF}
\end{figure}

\begin{figure}[hpt]
\begin{center}
\includegraphics*[width=0.8\columnwidth]{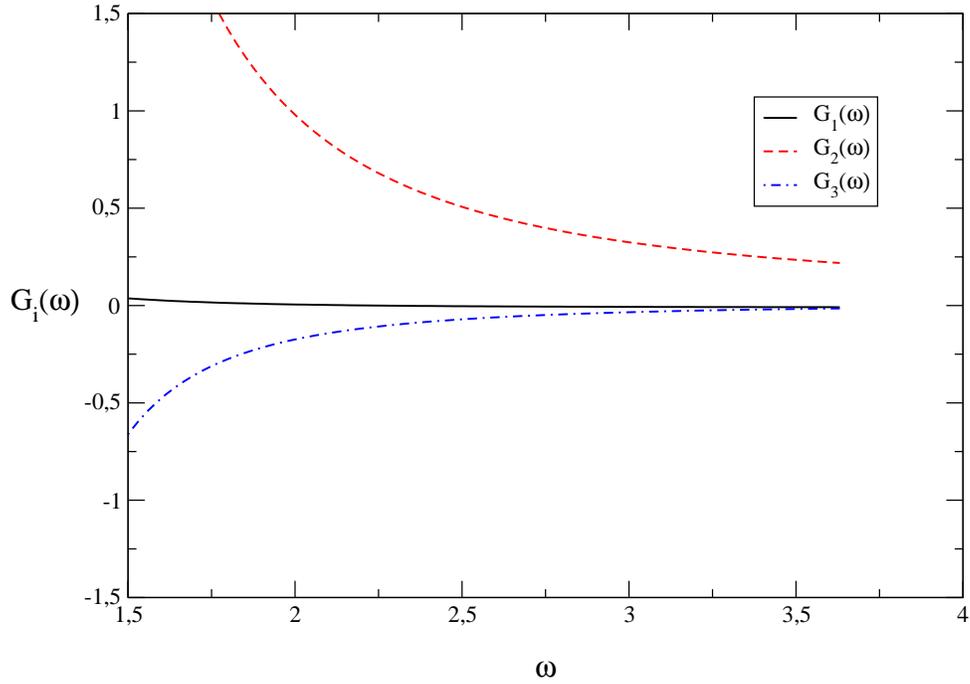}
\end{center}
\caption{Form factor  distributions,  $G_i(\omega)$, for $\Lambda_b \to \Lambda V$. The full, dashed and
dot-dashed lines represent $G_1(\omega), G_2(\omega)$ and $G_3(\omega)$, 
respectively.}
\label{figffG}
\end{figure}

\begin{figure}[hpt]
\begin{center}
\includegraphics*[width=0.785\columnwidth]{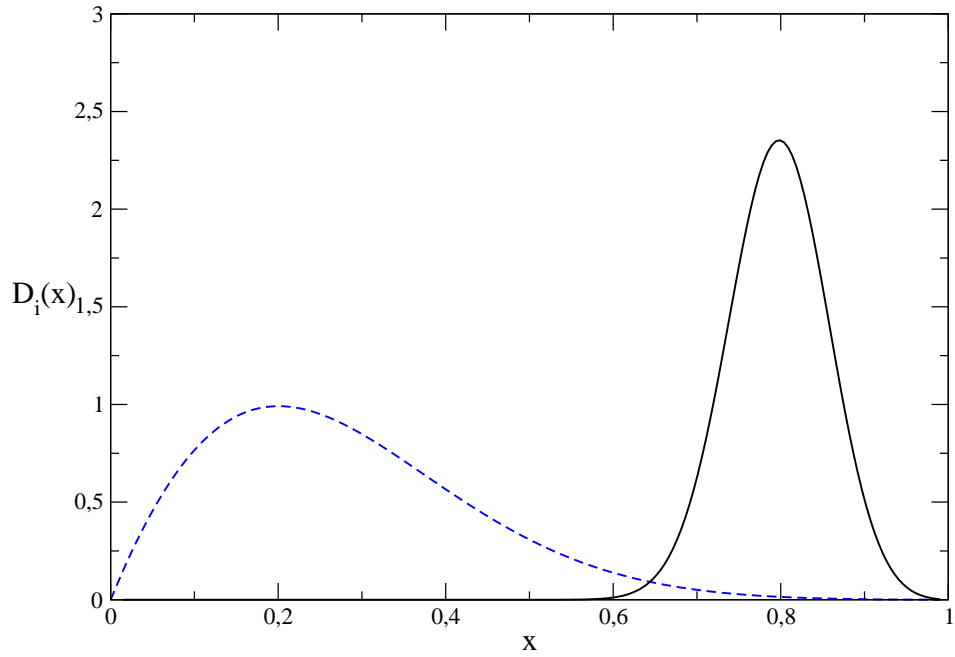}
\end{center}
\caption{Baryon function distributions for $\Lambda_b$ and $\Lambda$. The full and 
dashed lines represent the  $\Lambda_b$ and $\Lambda$ distributions, respectively.}
\label{figbaryons}
\end{figure}

\begin{figure}[hpt]
\begin{center}
\includegraphics*[width=0.8\columnwidth]{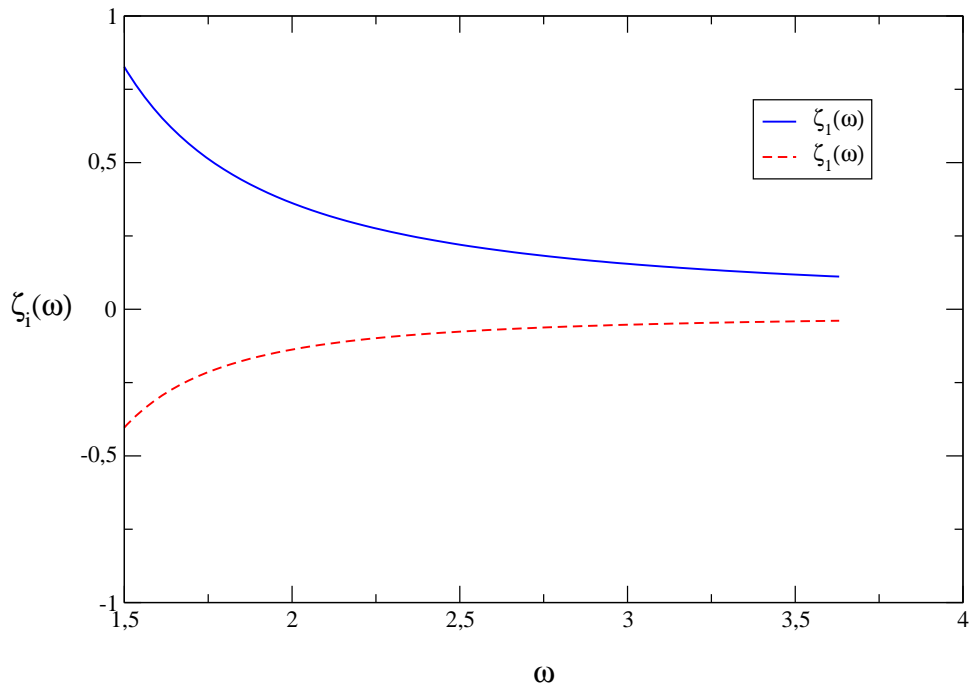}
\end{center}
\caption{Form factor  distributions,  $\zeta_i(\omega)$, for $\Lambda_b \to \Lambda V$. The full 
and dashed  lines represent $\zeta_1(\omega)$ and  $\zeta_2(\omega)$ respectively.}
\label{teta12}
\end{figure}

%
%
\begin{itemize}
\item{Fig.~\ref{frame} $\Lambda_b$ decay in its transversity frame and helicity frames
for the $\Lambda$ and vector meson $V$ decays, respectively.}
\item{Fig.~\ref{figbaryons01} Tree, color-commensurate, exchange and bow tie diagrams which are allowed in 
 transition $\Lambda_b \to \Lambda M_1.$  The squared dot represents the electroweak 
interaction in the Standard Model.}
\item{Fig.~\ref{figffF} Form factor distributions,  $F_i(\omega)$, for $\Lambda_b \to \Lambda V$.}
\item{Fig.~\ref{figffG} Form factor distributions,  $G_i(\omega)$, for $\Lambda_b \to \Lambda V$.}
\item{Fig.~\ref{figbaryons} Baryon wave-function distributions for 
$\Lambda_b$ and $\Lambda$.}
\item{Fig.~\ref{teta12} Form factor  distributions,  $\zeta_i(\omega)$, for $\Lambda_b \to \Lambda V$.}
\item{Fig.~7 Branching ratio asymmetry, $a_{CP}(\omega)$, as a function of  the ${\pi}^+ {\pi}^-$ 
invariant mass in case of $\Lambda_b \to \Lambda \rho^0(\omega) \to \Lambda \pi^+\pi^-$ and for $N_c^{eff}=3$.}
\item{Fig.~\ref{figziad2} $\cos \theta_{\Lambda}$  and $\phi_{\Lambda}$  distributions of the $\Lambda$-baryon 
in $\Lambda_b$ rest-frame, in case of $\Lambda_b \to \Lambda \rho^0(\omega)$.}
\item{Fig.~\ref{figziad3} $\cos \theta_P$  and $\phi_P$  distributions of the proton in $\Lambda$ 
rest-frame in case of  $\Lambda_b \to \Lambda \rho^0(\omega)$.}
\item{Fig.~\ref{figziad4} $\cos \theta_{\pi}$  and $\cos \theta_{\mu}$  distributions in vector rest-frame,  
for $\Lambda_b \to \Lambda \rho^0(\omega) \to \Lambda \pi^+\pi^-$ 
and  $\Lambda_b \to \Lambda J/\psi \to \Lambda \mu^+ \mu^-$, respectively.}
\end{itemize}
%
%
\section*{Table captions} 
%
%
\begin{itemize}
\item{Table~\ref{tab1trans} Vector-polarization under parity and TR operations.}
\item{Table~\ref{tab2wilson} Wilson coefficients related to current-current tree and penguin 
operators.}
\item{Table~\ref{tabl3result} Branching ratio, $\mathcal{BR}$, and  direct $CP$ violating asymmetry 
parameters, $a_{CP}$ (mean value)
and  $a_{CP}(\omega)$ (maximum value), for $\Lambda_b \to \Lambda \rho^0(\omega)$.}
\end{itemize}
\newpage
\begin{figure}[hpt]
\begin{center}
\includegraphics*[width=1.0\columnwidth]{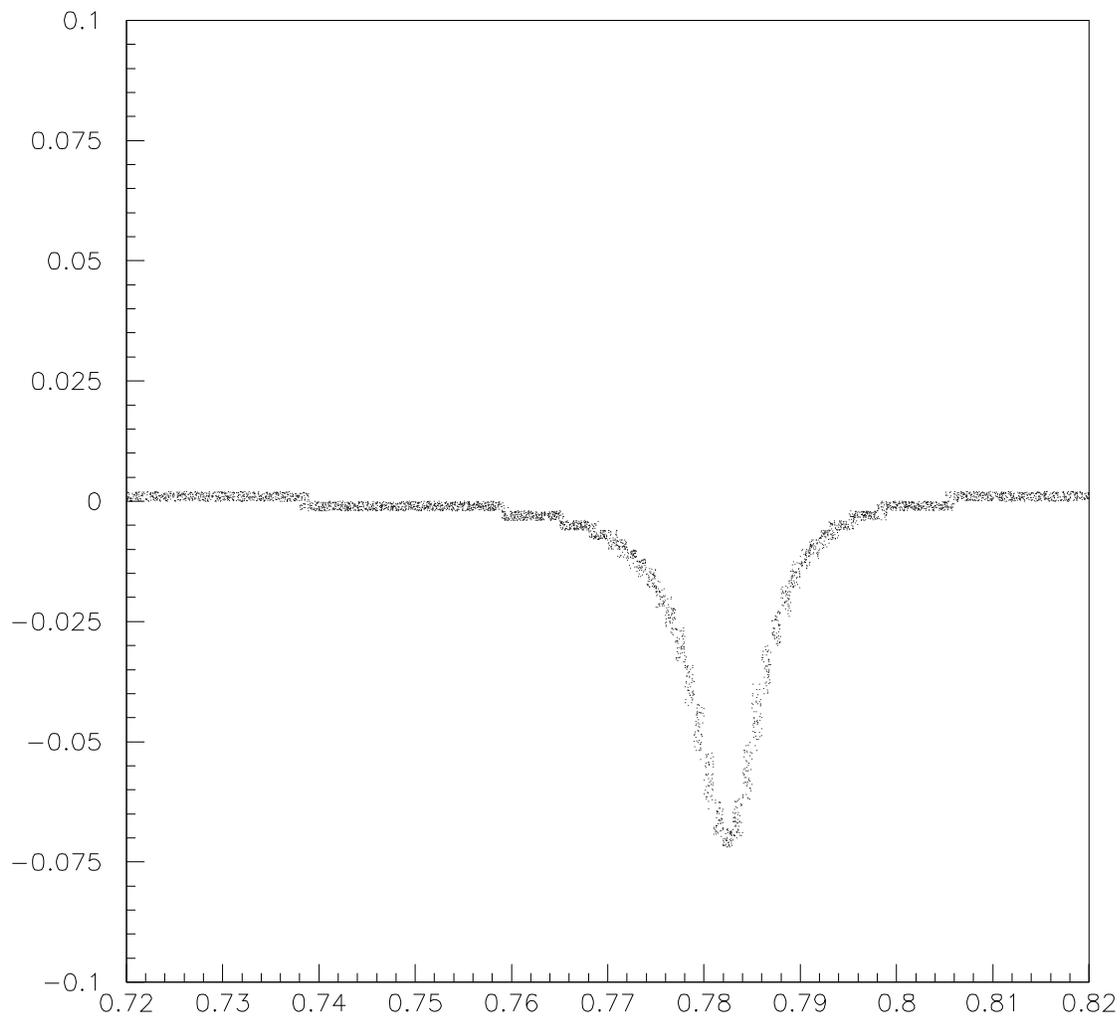}
\end{center}
\caption{Branching ratio asymmetry, $a_{CP}(\omega)$, as a function of  the ${\pi}^+ {\pi}^-$ 
invariant mass in case of $\Lambda_b \to \Lambda \rho^0(\omega) \to \Lambda \pi^+\pi^-$ and 
for $N_c^{eff}=3$.}
\label{figziad0}
\end{figure}
\begin{figure}[hpt]
\begin{center}
\mbox{\epsfig{file = 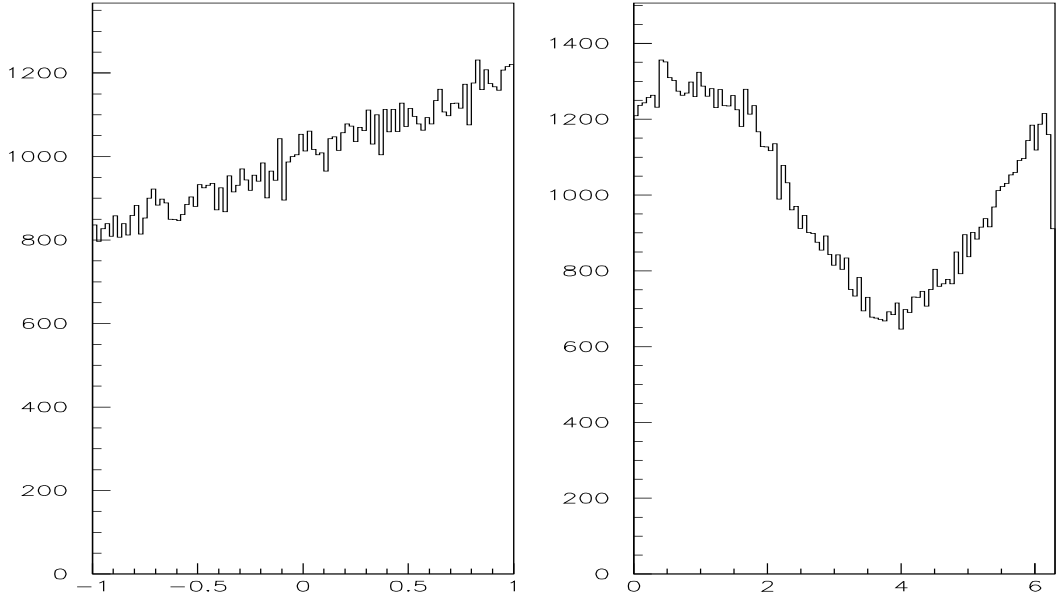, height = 9.5cm, width = 16.0cm}}
\end{center}
\caption{$\cos \theta_{\Lambda}$ (left) and $\phi_{\Lambda}$ (right) distributions of the $\Lambda$-baryon 
in $\Lambda_b$ rest-frame, in case of $\Lambda_b \to \Lambda \rho^0(\omega)$. The $Y$ axis means $\frac{dN}{d \cos 
\theta_{\Lambda}}$ (left) and $\frac{dN}{d \phi_{\Lambda}}$ (right).}
\label{figziad2}
\end{figure}
\begin{figure}[hpt]
\begin{center}
\mbox{\epsfig{file = 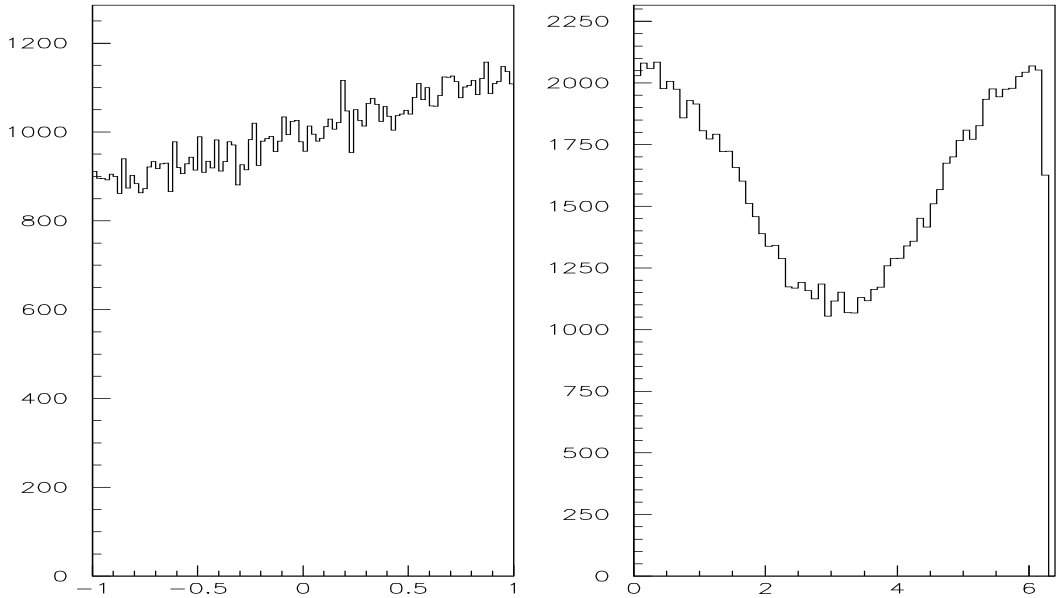, height = 9.5cm, width = 16.0cm}}
\end{center}
\caption{$\cos \theta_P$ (left) and $\phi_P$ (right) distributions of the proton in $\Lambda$ rest-frame in case 
of  $\Lambda_b \to \Lambda \rho^0(\omega)$. The $Y$ axis means $\frac{dN}{d \cos 
\theta_P}$ (left) and $\frac{dN}{d \phi_P}$ (right).}
\label{figziad3}
\end{figure}
\begin{figure}[hpt]
\begin{center}
\mbox{\epsfig{file = 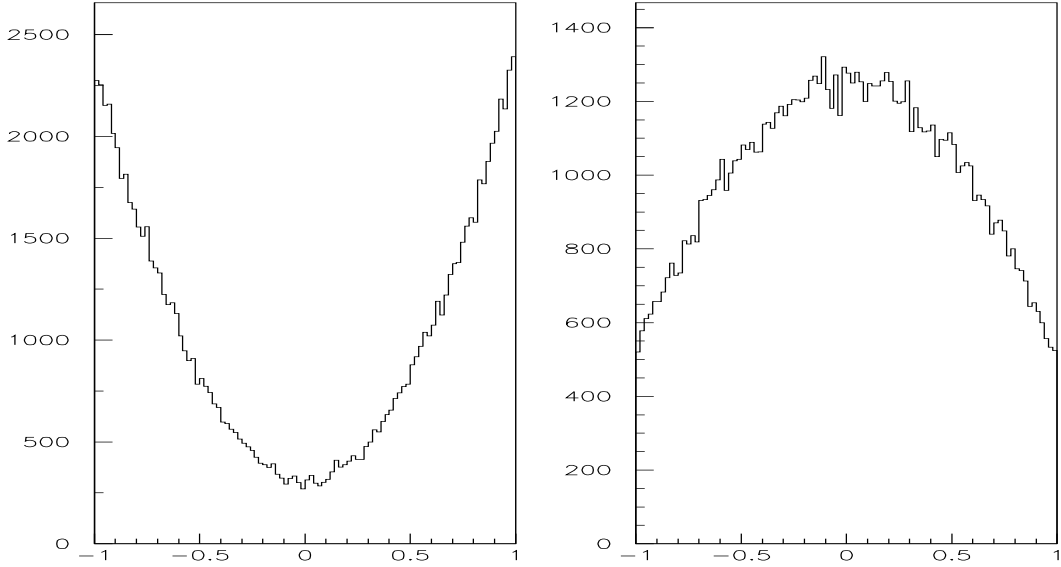, height = 9.cm, width = 16.0cm}}
\end{center}
\caption{$\cos \theta_{\pi}$ (left) and $\cos \theta_{\mu}$ (right) distributions in vector rest-frame,  
for $\Lambda_b \to \Lambda \rho^0(\omega) \to \Lambda \pi^+\pi^-$ 
and  $\Lambda_b \to \Lambda J/\psi \to \Lambda \mu^+ \mu^-$, respectively.  The $Y$ axis means $\frac{dN}{d \cos 
\theta_{\pi}}$ (left) and $\frac{dN}{d \phi_{\mu}}$ (right).}
\label{figziad4}
\end{figure}
\begin{table}[hbp]
\begin{center}
\begin{tabular}{ccc}
\hline
\hline
Observable&Parity&TR\\
\hline
\hline
$\vec s$&Even&Odd\\
$\vec {\cal P}$&Even&Odd\\
\hline
$\vec{e_Z}$&Even&Even\\
$\vec {e_L}$&Odd&Odd\\
$\vec {e_T}$&Odd&Odd\\
$\vec {e_N}$&Even&Even\\
\hline
${P_L}$&Odd&Even\\
${P_T}$&Odd&Even\\
${P_N}$&Even&{\bf ODD}\\
\hline
\hline
\end{tabular}
\end{center}
\caption{Vector-polarization under Parity and TR operations.}\label{tab1trans}
\end{table}
\begin{table}[htpb]
\begin{center}
\begin{tabular}{ccc} \hline \hline
   $C_{i}^{\prime}$          & $q^{2}/m_{b}^{2}=0.3$   & $q^{2}/m_{b}^{2}
=0.5$ \\
\hline
\hline
$C_{1}^{\prime} $   &   $-0.3125$                                     
&$-0.3125 $  \\
$C_{2}^{\prime} $   & $+1.1502$                                        
&$+1.1502$    \\
\hline
$C_{3}^{\prime} $   & $+2.433 \times 10^{-2} + 1.543 \times 10^{-3}i$  
&$+2.120 \times 10^{-2} + 2.174 \times 10^{-3}i$\\
$C_{4}^{\prime} $   & $-5.808 \times 10^{-2} -4.628 \times 10^{-3}i$  
&$-4.869 \times 10^{-2} -1.552 \times 10^{-2}i$\\
$C_{5}^{\prime} $   & $+1.733 \times 10^{-2}+ 1.543 \times 10^{-3}i$   
&$+1.420 \times 10^{-2} + 5.174 \times 10^{-3}i$\\
$C_{6}^{\prime} $   & $-6.668 \times 10^{-2}- 4.628 \times 10^{-3}i$  
&$-5.729  \times 10^{-2}- 1.552 \times 10^{-2}i$\\
\hline
$C_{7}^{\prime} $   & $-1.435 \times 10^{-4} -2.963 \times 10^{-5}i$  
&$-8.340 \times 10^{-5} -9.938 \times 10^{-5}i$\\
$C_{8}^{\prime} $   & $+3.839 \times 10^{-4}$                          
&$ +3.839 \times 10^{-4} $\\
$C_{9}^{\prime} $   & $-1.023 \times 10^{-2} -2.963 \times 10^{-5}i$  
&$-1.017 \times 10^{-2} -9.938 \times 10^{-5}i$\\
$C_{10}^{\prime} $  & $+1.959 \times 10^{-3}$                          
&$+1.959 \times 10^{-3}$\\
\hline
\hline
\end{tabular}
\end{center}
\caption{Wilson coefficients related to current-current tree and penguin 
operators.}\label{tab2wilson}
\end{table}
\begin{table}[htbp]
\begin{center}
\begin{tabular}{ccccc}
\hline
\hline
${N_c}^{eff}$&$2.0$&$2.5$&$3.0$&$3.5$\\
\hline
\hline
$ \langle \mathcal{BR} \rangle \; (10^{-7})$&$2.10 $&$1.98 $&$2.15 $&$2.40 $\\
\hline
$ \langle a_{CP} \rangle$&$6.4\%$&$1.04\%$&$ -0.3\%$&$ -0.2\%$\\
\hline
$ a_{CP}(\omega) $&$96\%$&$15.0\%$&$ -7.5\%$&$ -3.3\%$\\
\hline
\hline
\end{tabular}
\end{center}
\caption{Branching ratio, $\mathcal{BR}$, and direct $CP$ violating asymmetry parameters, $a_{CP}$ (mean value)
and  $a_{CP}(\omega)$ (maximum value), for $\Lambda_b \to \Lambda \rho^0(\omega)$.}\label{tabl3result}
\end{table}

\begin{thebibliography}{10}
\bibitem{Ajaltouni:2004zu}
Z.~J. Ajaltouni, E.~Conte, and O.~Leitner,
\newblock Phys. Lett. {\bf B614}, 165 (2005), hep-ph/0412116.

\bibitem{Eidelman:2004wy}
Particle Data Group, S.~Eidelman {\em et~al.},
\newblock Phys. Lett. {\bf B592}, 1 (2004).

\bibitem{Andersson:1998}
B.~Andersson,
\newblock {\it The LUND Model}, Cambridge Monographs on Particle Physics
  (1998).

\bibitem{JWJ}
M.~Jacob and G.~C. Wick,
\newblock Annals of Physics {\bf 7}, 404 (1959).

\bibitem{Ballentine:1998}
L.~Ballentine,
\newblock {\it Quantum Mechanics: A Modern Development}, Singapore World Sci.
  (1998).

\bibitem{Werle:1966}
J.~Werle,
\newblock {\it Relativistic Theory of Reactions}, North-Holland Publishing
  Company- AMSTERDAM , 419 (1966).

\bibitem{Jackson:1965}
J.~Jackson,
\newblock {\it Resonance Decays}, High Energy Physics, Les Houches lectures,
  edited by C.~DeWitt and M.~Jacob (Gordon and Breach, New York, 1966  (1965).

\bibitem{Korner:1994nh}
J.~G. Korner, M.~Kramer, and D.~Pirjol,
\newblock Prog. Part. Nucl. Phys. {\bf 33}, 787 (1994), hep-ph/9406359.

\bibitem{Hussain:1994zr}
F.~Hussain and G.~Thompson,
\newblock (1994), hep-ph/9502241.

\bibitem{Neubert:1993mb}
M.~Neubert,
\newblock Phys. Rept. {\bf 245}, 259 (1994), hep-ph/9306320.

\bibitem{Guo:1995zb}
X.-H. Guo, T.~Huang, and Z.-H. Li,
\newblock Phys. Rev. {\bf D53}, 4946 (1996), hep-ph/9706402.

\bibitem{Datta:1994ij}
A.~Datta,
\newblock Phys. Lett. {\bf B349}, 348 (1995), hep-ph/9411306.

\bibitem{Konig:1993ze}
B.~Konig, J.~G. Korner, M.~Kramer, and P.~Kroll,
\newblock Phys. Rev. {\bf D56}, 4282 (1997), hep-ph/9701212.

\bibitem{Huang:1998ek}
C.-S. Huang and H.-G. Yan,
\newblock Phys. Rev. {\bf D59}, 114022 (1999), hep-ph/9811303.

\bibitem{Buras:1999rb}
A.~J. Buras,
\newblock Lect. Notes Phys. {\bf 558}, 65 (2000), hep-ph/9901409.

\bibitem{Buras:1998ra}
A.~J. Buras,
\newblock (1998), hep-ph/9806471.

\bibitem{Buchalla:1996vs}
G.~Buchalla, A.~J. Buras, and M.~E. Lautenbacher,
\newblock Rev. Mod. Phys. {\bf 68}, 1125 (1996), hep-ph/9512380.

\bibitem{Deshpande:1995pw}
N.~G. Deshpande and X.-G. He,
\newblock Phys. Rev. Lett. {\bf 74}, 26 (1995), hep-ph/9408404.

\bibitem{Fleischer:1994gr}
R.~Fleischer,
\newblock Z. Phys. {\bf C62}, 81 (1994).

\bibitem{Fleischer:1993gp}
R.~Fleischer,
\newblock Z. Phys. {\bf C58}, 483 (1993).

\bibitem{Fleischer:1997bv}
R.~Fleischer,
\newblock Int. J. Mod. Phys. {\bf A12}, 2459 (1997), hep-ph/9612446.

\bibitem{Kramer:1994yu}
G.~Kramer, W.~F. Palmer, and H.~Simma,
\newblock Nucl. Phys. {\bf B428}, 77 (1994), hep-ph/9402227.

\bibitem{Enomoto:1996cv}
R.~Enomoto and M.~Tanabashi,
\newblock Phys. Lett. {\bf B386}, 413 (1996), hep-ph/9606217.

\bibitem{Gardner:1998yx}
S.~Gardner, H.~B. O'Connell, and A.~W. Thomas,
\newblock Phys. Rev. Lett. {\bf 80}, 1834 (1998), hep-ph/9705453.

\bibitem{Guo:1998eg}
X.~H. Guo and A.~W. Thomas,
\newblock Phys. Rev. {\bf D58}, 096013 (1998), hep-ph/9805332.

\bibitem{Guo:1999ip}
X.~H. Guo and A.~W. Thomas,
\newblock Phys. Rev. {\bf D61}, 116009 (2000), hep-ph/9907370.

\bibitem{Gardner:1998za}
S.~Gardner,
\newblock (1998), hep-ph/9809479.

\bibitem{Gardner:1997qk}
S.~Gardner, H.~B. O'Connell, and A.~W. Thomas,
\newblock (1997), hep-ph/9707414.

\bibitem{Sakurai:1969ju}
J.~J. Sakurai,
\newblock {\it Currents and Mesons}, University of Chicago Press  (1969).

\bibitem{O'Connell:1997wf}
H.~B. O'Connell, B.~C. Pearce, A.~W. Thomas, and A.~G. Williams,
\newblock Prog. Part. Nucl. Phys. {\bf 39}, 201 (1997), hep-ph/9501251.

\bibitem{O'Connell:1996ns}
H.~B. O'Connell, A.~G. Williams, M.~Bracco, and G.~Krein,
\newblock Phys. Lett. {\bf B370}, 12 (1996), hep-ph/9510425.

\bibitem{O'Connell:1994uc}
H.~B. O'Connell, B.~C. Pearce, A.~W. Thomas, and A.~G. Williams,
\newblock Phys. Lett. {\bf B336}, 1 (1994), hep-ph/9405273.

\bibitem{Maltman:1996kj}
K.~Maltman, H.~B. O'Connell, and A.~G. Williams,
\newblock Phys. Lett. {\bf B376}, 19 (1996), hep-ph/9601309.

\bibitem{O'Connell:1997xy}
H.~B. O'Connell, A.~W. Thomas, and A.~G. Williams,
\newblock Nucl. Phys. {\bf A623}, 559 (1997), hep-ph/9703248.

\bibitem{Williams:1998nj}
A.~G. Williams, H.~B. O'Connell, and A.~W. Thomas,
\newblock Nucl. Phys. {\bf A629}, 464c (1998), hep-ph/9707253.

\bibitem{Gardner:1998ta}
S.~Gardner and H.~B. O'Connell,
\newblock Phys. Rev. {\bf D59}, 076002 (1999), hep-ph/9809224.

\bibitem{Gardner:1998ie}
S.~Gardner and H.~B. O'Connell,
\newblock Phys. Rev. {\bf D57}, 2716 (1998), hep-ph/9707385.

\bibitem{Wolfenstein:1983yz}
L.~Wolfenstein,
\newblock Phys. Rev. Lett. {\bf 51}, 1945 (1983).

\bibitem{Wolfenstein:1964ks}
L.~Wolfenstein,
\newblock Phys. Rev. Lett. {\bf 13}, 562 (1964).

\bibitem{Hocker:2001xe}
A.~Hocker, H.~Lacker, S.~Laplace, and F.~Le~Diberder,
\newblock Eur. Phys. J. {\bf C21}, 225 (2001), hep-ph/0104062.

\bibitem{Chau:1984fp}
L.-L. Chau and W.-Y. Keung,
\newblock Phys. Rev. Lett. {\bf 53}, 1802 (1984).

\bibitem{Pakvasa:1990if}
S.~Pakvasa, S.~P. Rosen, and S.~F. Tuan,
\newblock Phys. Rev. {\bf D42}, 3746 (1990).

\bibitem{Jackson:1964zd}
J.~D. Jackson,
\newblock Nuovo Cim. {\bf 34}, 1644 (1964).

\bibitem{Byers:1967tw}
N.~Byers,
\newblock {\it Spin and Density Matrix of decaying states} , CERN 67 20,
  hep-ph/0412116.

\bibitem{Bensalem:2002pz}
W.~Bensalem, A.~Datta, and D.~London,
\newblock Phys. Lett. {\bf B538}, 309 (2002), hep-ph/0205009.

\bibitem{Aliev:2004yf}
T.~M. Aliev, V.~Bashiry, and M.~Savci,
\newblock Eur. Phys. J. {\bf C38}, 283 (2004), hep-ph/0409275.

\bibitem{Chen:2003}
C.~C. et~al,
\newblock Nucl.Phys.B(Proc Suppl) {\bf 115}, 263 (2003), hep-ph/0210067.

\bibitem{Sehgal:1999vg}
L.~M. Sehgal and J.~van Leusen,
\newblock Phys. Rev. Lett. {\bf 83}, 4933 (1999), hep-ph/9908426.

\bibitem{Wolfenstein:1999xb}
L.~Wolfenstein,
\newblock Phys. Rev. Lett. {\bf 83}, 911 (1999).

\bibitem{Kadeer:2005aq}
A.~Kadeer, J.~G. Korner, and U.~Moosbrugger,
\newblock (2005), hep-ph/0511019.

\bibitem{Binosi:2003yf}
D.~Binosi and L.~Theussl,
\newblock Comput. Phys. Commun. {\bf 161}, 76 (2004), hep-ph/0309015.
\end{thebibliography}
\end{document}